\documentclass[useAMS,usenatbib]{mnras}

\usepackage{setspace}
\usepackage{graphics}
\usepackage{graphicx}
\usepackage{amssymb}
\usepackage{longtable}
\usepackage{amsmath}
\usepackage{amsbsy,fixmath}
\usepackage{epstopdf}
\usepackage{color}
\usepackage[T1]{fontenc} 
\usepackage{aecompl}
\usepackage[none]{hyphenat}
\usepackage[normalem]{ulem}
\usepackage{physics}

\graphicspath{ {figs/}}

\defcitealias{Goicovic2016}{Paper~I}
\newcommand{\paperI}{\citetalias{Goicovic2016}}

\def\msun{\,{\rm M_\odot}}
\def\simlt{\lower.5ex\hbox{$\; \buildrel < \over \sim \;$}}
\def\simgt{\lower.5ex\hbox{$\; \buildrel > \over \sim \;$}}

\title[Binary evolution from infalling clouds]{Infalling clouds on to supermassive black hole binaries -- II. Binary evolution and the final parsec problem}
\author[Goicovic et al.]{Felipe G. Goicovic$^{1,2,3}$\thanks{E-mail:
fgarrido@astro.puc.cl}, Alberto Sesana$^{4,2}$, Jorge Cuadra$^{1,7}$ and Federico Stasyszyn$^{5,6}$\\
$^{1}$Instituto de Astrof\'isica, Pontificia Universidad Cat\'olica de Chile, Av. Vicu\~na Mackenna 4860, 7820436 Macul, Santiago, Chile\\
$^{2}$Max-Planck-Institut f\"ur Gravitationsphysik, Albert Einstein Institut (AEI), Am M\"ulenberg 1, 14476, Potsdam-Golm, Germany\\
$^{3}$Heidelberg Institute for Theoretical Studies (HITS), Schloss-Wolfsbrunnenweg 35, D-69118 Heidelberg, Germany\\
$^{4}$School of Physics and Astronomy, University of Birmingham, Edgbaston, Birmingham B15 2TT, United Kingdom\\
$^{5}$Instituto de Astronom\'ia T\'ecnica y Experimental, UNC-CONICET, Observatorio Astron\'omico, X5000BGR, C\'ordoba, Argentina\\
$^{6}$Leibniz-Institut f\"ur Astrophysik Potsdam (AIP), An der Sternwarte 16, 14482, Potsdam, Germany\\
$^{7}$Max Planck Institute f\"ur extraterrestriche Physik (MPE), D-85748 Garching, Germany\\
}
\begin{document}

\date{\today}

\pagerange{\pageref{firstpage}--\pageref{lastpage}} \pubyear{2016}

\maketitle

\label{firstpage}

\begin{abstract}
The formation of massive black hole binaries (MBHBs) is an unavoidable outcome of galaxy evolution via successive mergers. However, the mechanism that drives their orbital evolution from parsec separations down to the gravitational wave (GW) dominated regime is poorly understood, and their final fate is still unclear. 
If such binaries are embedded in gas-rich and turbulent environments, as observed in remnants of galaxy mergers, the interaction with gas clumps (such as molecular clouds) may efficiently drive their orbital evolution. Using numerical simulations, we test this hypothesis by studying the dynamical evolution of an equal-mass, circular MBHB accreting infalling molecular clouds. We investigate different orbital configurations, modelling a total of 13 systems to explore different possible impact parameters and relative inclinations of the cloud--binary encounter. 
{We focus our study on the prompt, transient phase during the first few orbits when the dynamical evolution of the binary is 
{fastest, finding that this evolution is}
dominated by the exchange of angular momentum through {gas capture by the individual black holes and accretion}.}
Building on these results, we construct a simple model for evolving a MBHB interacting with a sequence of clouds, which are randomly drawn from reasonable populations with different levels of anisotropy in their angular momenta distributions. We show that the binary efficiently evolves down to the GW emission regime within a few hundred million years, overcoming the `final parsec' problem regardless of the stellar distribution.
\end{abstract}

\begin{keywords}
 accretion, accretion discs -- black hole physics -- hydrodynamics -- galaxies: evolution -- galaxies: nuclei
\end{keywords}

\section{Introduction}

The important role of massive black holes (MBHs) in galaxy evolution has been established over the past decades. Hypothesised in the sixties to be the central engine of the then-discovered quasars \citep{1969Natur.223..690L}, a growing body of evidence confirmed the existence of MBHs in virtually every massive galaxy in the local Universe during the nineties \citep{1995ARA&A..33..581K}. The discovery of the tight correlations between the MBH masses and key properties of their host galaxies \citep{1998AJ....115.2285M,2000ApJ...539L...9F,2000ApJ...539L..13G} points towards a coevolution scenario, based on an interplay of gas accretion onto the MBH from the host, releasing a large amount of energy feeding back onto the host galaxy \citep[][and references therein]{Kormendy2013}. Since the majority of massive galaxies harbour a MBH at their nucleus, and since in the hierarchical model of structure formation galaxy mergers are frequent throughout cosmic history \citep{1978MNRAS.183..341W}, massive black hole binaries (MBHBs) are expected to be a common outcome of the galaxy evolution process \citep{Beg80}. Therefore, understanding their formation and dynamical evolution is of paramount importance in reconstructing the puzzle of the hierarchical growth of structures in the Universe.

Despite the fairly solid theoretical predictions, observational evidence of MBHB existence is sparse, although several candidates have been proposed in the literature \citep[see][for a review]{Dotti2012}. Unfortunately, direct imaging of these objects is currently impossible, as the small scales required to resolve them are beyond current capabilities on galactic nuclei beyond the local group. An exception to this is high frequency radio interferometry, that enables the resolution of parsec scales in low redshift galaxies. In fact, the most compelling MBHB candidate to date is a peculiar double radio core \citep{2006ApJ...646...49R}. Many more candidates have been proposed based on distinctive spectroscopic features \citep[see, e.g.][]{2011ApJ...738...20T,2012ApJS..201...23E}.  Most of the times, however, those observed signatures can be explained by alternative scenarios that do not require a MBHB \citep{Dotti2012,Bogdanovic2014,2015arXiv151201825L}.

From a theoretical perspective, the evolution of MBH pairs after a galaxy merger will be dominated by different processes depending on their separation, and on the  content and distribution of both gas and stars in the merger product \citep[e.g.,][]{Beg80, 2007MNRAS.379..956D, Sesana2010,Khan2013,Holley2015}.  At large separations, both MBHs sink towards the galactic nucleus on short timescales due to dynamical friction against  the surrounding background of dark matter, gas and stars \citep{1943ApJ....97..255C,MM01}. This process becomes inefficient when the mass enclosed in the orbit of the two MBHs is of the order of their total mass. At this point, when the pair forms a gravitationally bound binary, the relative velocity of the two black holes becomes larger than the velocity dispersion of the background medium, and the efficiency of dynamical friction drops sharply. 
In a gas-poor environment, the binary will continue shrinking by ejecting single stars via three-body encounters.
 Only once the separation is sufficiently low ($\lesssim 10^{-3}$ pc) gravitational radiation will efficiently extract angular momentum and energy from the binary and rapidly lead it to coalescence \citep{Beg80}. Early dynamical studies found that at parsec separations there are too few MBHB--star scattering events to reach the coalescence within a Hubble time; an issue called the ``final parsec problem'' \citep[see, e.g.][]{MM01,Yu02}. However, in the past decade, several semi-analytical and numerical works have shown that in the non-relaxed, triaxial rotating remnant of a galaxy merger, the supply of stars to the MBHB should be large enough to lead to a final coalescence on a Gyr timescale \citep{2006ApJ...642L..21B,Sesana2010,2011ApJ...732L..26P,2011ApJ...732...89K,Khan2013,Holley2015,2015ApJ...810...49V,2015MNRAS.454L..66S}.

The picture can be quite different for MBH pairs within gas-rich environments.  In this case, interaction with gas can be very efficient in absorbing and transporting outwards the angular momentum of the pair, leading to a more rapid evolution and eventual coalescence.  Different numerical studies have shown an orbital decay, driven by global disc torques, on time-scales of only $\sim 10^7\,$yr within the massive gaseous nuclear disc that forms after a gas-rich galaxy merger \citep{2004ApJ...607..765E, 2005ApJ...630..152E, 2007Sci...316.1874M, Fiacconi2013, Roskar2015, DelValle2015}.  Once the MBHs form a compact enough, bound binary, its torque will open up a cavity in the gas distribution \citep[e.g.,][]{Artymowicz1994, DelValle2012, DelValle2014}.
Many theoretical and numerical studies have focused on this phase of a sub-parsec binary surrounded by a gaseous circumbinary disc \citep[e.g.,][]{Ivanov1999,ArmNat05,C09,Haiman2009,Lod09,Nix11a,Roedig2011,Roedig2012,Kocsis2012,Pau2013,Roedig2014, Dunhill2015}, typically finding a much slower orbital evolution. However, most of these studies assume the existence of an extended, stable circumbinary disc, generally corotating with the binary. Notable exceptions are \cite {Roedig2014} and the work of Nixon and collaborators \citep[see, e.g.][]{Nix11a,2013MNRAS.434.1946N} that extensively investigated the interaction with misaligned or counter-rotating (with respect to the MBHB angular momentum) discs. However, also in this case, relaxed steady discs are assumed as initial condition. In all the aforementioned studies, the evolution of the MBHB is `decoupled' from the host galaxy, in the sense that there is no attempt to model the formation of the circumbinary disc structure, nor to link it to the fuelling mechanisms that transport gas from galactic scales down to the binary.

In the aftermath of a galaxy merger, gas is efficiently fuelled to the centre of the remnant \citep{1996ApJ...471..115B}, and the formation of massive circumnuclear discs is observed on a scale of hundreds of parsec \citep{2007Sci...316.1874M}, consistent with observations of ultraluminous infrared galaxies \citep{1988ApJ...325...74S}. Those discs however are extremely thick compared to the size of a putative sub-parsec binary. Turbulence and gravitational instabilities in this large scale disc can trigger the formation of gas clumps, which will be the seeds for molecular clouds \citep{Agertz2009}. These clouds can travel through the interstellar medium almost unaffected by hydrodynamical drag, and might produce discrete accretion events onto a central MBH \citep{Hobbs11}, possibly coming with a wide range of angular momenta directions with respect to that of the binary orbit. In the context of MBHBs, the recent studies of \citet{Dunhill2014} and \citet[][hereafter Paper I]{Goicovic2016} have explored the hypothesis of infalling clouds as the source of gas for MBHBs, investigating the formation of discs, either around the binary or each individual MBH, for a wide range of orbital configurations. In particular, \paperI~modelled the interaction of near-radial infall onto equal-mass MBHBs with 12 different configurations, exploring different inclinations and impact parameters.

Based on the simulations performed in \paperI, we model here the dynamical effects that the interaction with these clouds have on the MBHB orbit. We focus on the angular momentum transfer and on the evolution of the binary orbital elements, paying particular attention to the early phases of the interaction, when most of the angular momentum transfer occurs. The paper is organised as follows. In Section~\ref{sec:simulation} we briefly describe our numerical model, explaining the modifications respect to the simulations presented in \paperI. In Section~\ref{sec:evolution} we present the evolution of the binary angular momentum and orbital parameters measured with our simulations. In Section~\ref{sec:angmomexch} we present a simple model where the evolution of the MBHBs is driven only by the exchange of angular momentum with the accreted material, and show that it roughly captures the behaviour observed in the simulations. We explore the implication of this model for the long-term evolution of MBHBs in Section~\ref{sec:application}, and summarise our findings in Section~\ref{sec:summary}.

\section{The numerical model}
\label{sec:simulation}

We model the interaction between the gas clouds and the MBHBs using the smoothed particle hydrodynamics (SPH) technique, as described in \paperI. The binary consists of two sink particles, initially having equal masses and a circular orbit. On the other hand, the cloud is initially spherical with uniform density, a turbulent velocity field, and a total mass 100 times smaller than the binary.

By changing the initial orbit of the cloud, we model a total of 13 systems. The first 12 systems are the same described in \paperI, which correspond to the combination of 4 different orientations relative to the binary orbit (A: aligned, CA: counter-aligned, PE: perpendicular edge-on, and PF: perpendicular face-on) and three pericentre distances ($r_{\rm p}=0.7, 1.5, 3~R_{\rm bin}$, where $R_{\rm bin}=0.5a$ is the initial binary radius). We model an additional impact parameter ($r_{\rm p}=6R_{\rm bin}$) for the aligned configuration, needed to obtain a distribution of changes of both the accreted mass and binary orbit (more details on this choice are given in \S~\ref{sec:montecarlo}).  Throughout the paper we refer to our different models using the letters that indicate the orbit orientation and the number that gives the pericentre distance in $R_{\rm bin}$.

As described in \paperI, we modified the standard version of \textsc{gadget}-3 \citep[see][]{Springel2005} to include a deterministic accretion recipe, where the sink particles representing the MBHs accrete all bound SPH particles within a fixed radius ($r_{\rm sink}=0.1a_0$), as implemented by \cite{Cuadra2006}.  
Additionally, we treat the thermodynamics of the gas using a barotropic equation of state (i.e. pressure is a function of density only) with two regimes: for densities below a critical value, the gas evolves isothermally, while for higher densities the gas is adiabatic. This prescription captures the expected temperature dependence with density of the gas without an explicit implementation of cooling, and has the effect of halting the collapse of the densest gas regions, avoiding excessively small time-steps that would stall the simulations.

Additionally, we made two extra changes to the set up from \paperI\ in order to follow properly the evolution of the binary orbit.
We first set a fixed time-step for the MBHs equal to $10^{-4} P_0$, where $P_0$ is the initial binary orbital period.  
We also decreased the Courant factor from $0.1$ to $0.03$. This factor determines the size of the hydrodynamical time-step for each gas particle.
With these two changes, we can measure the small changes in the binary orbit  that are produced by the interaction with the gas clouds.
 We are thus able to disentangle the binary evolution from the numerical noise (\S~\ref{sec:angmom})  
without the need to remove the MBHs from the tree calculation of the gravitational forces \citep[cf.][for \textsc{gadget}-2 models]{C09}.

\section{Dynamical evolution of the system}
\label{sec:evolution}

\subsection{Angular momentum conservation}
\label{sec:angmom}
\begin{figure*}
\centering
 \begin{picture}(500,500)
  \put(0,0){\includegraphics[width=0.3333\textwidth]{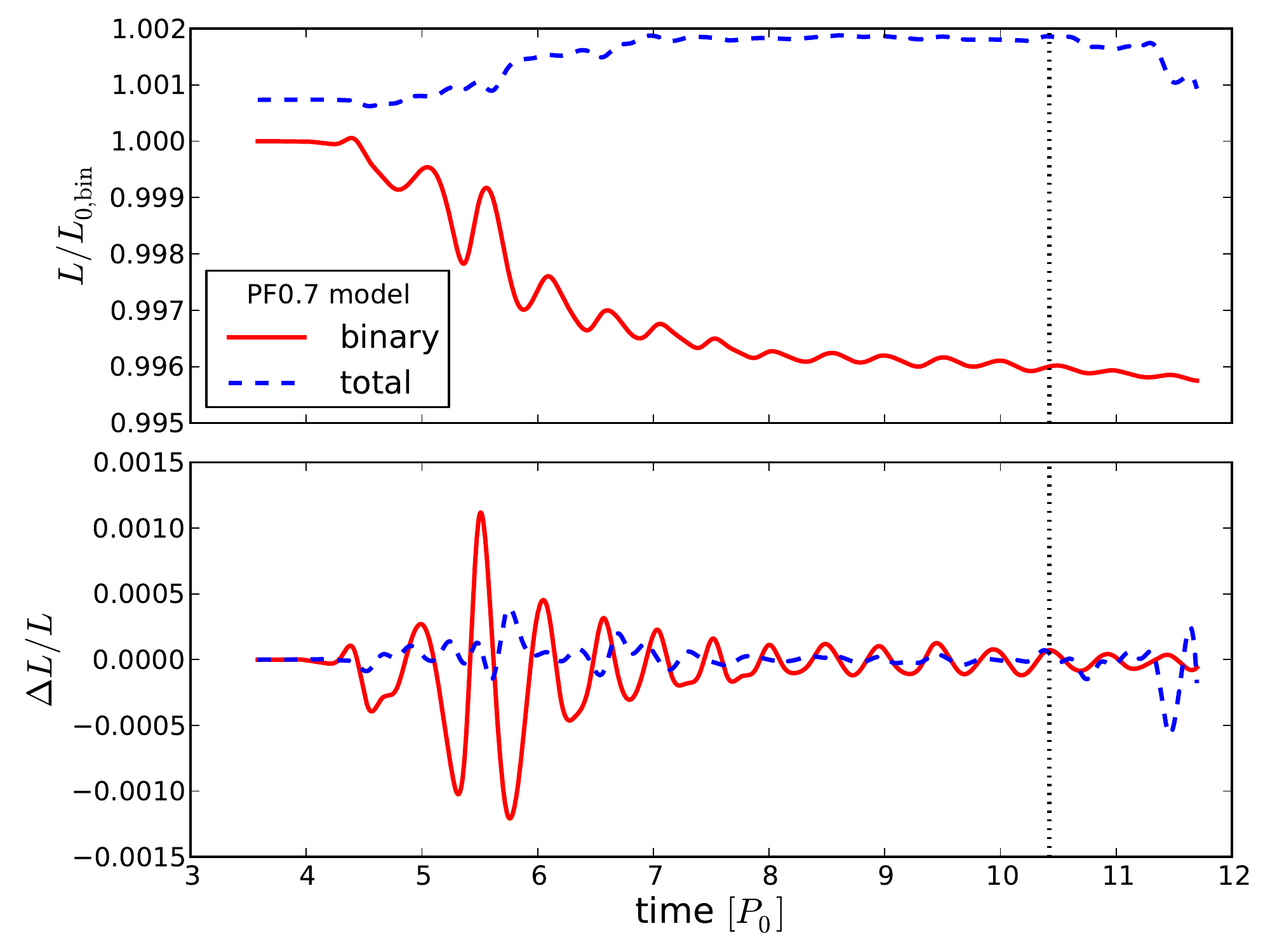}}
  \put(167,0){\includegraphics[width=0.3333\textwidth]{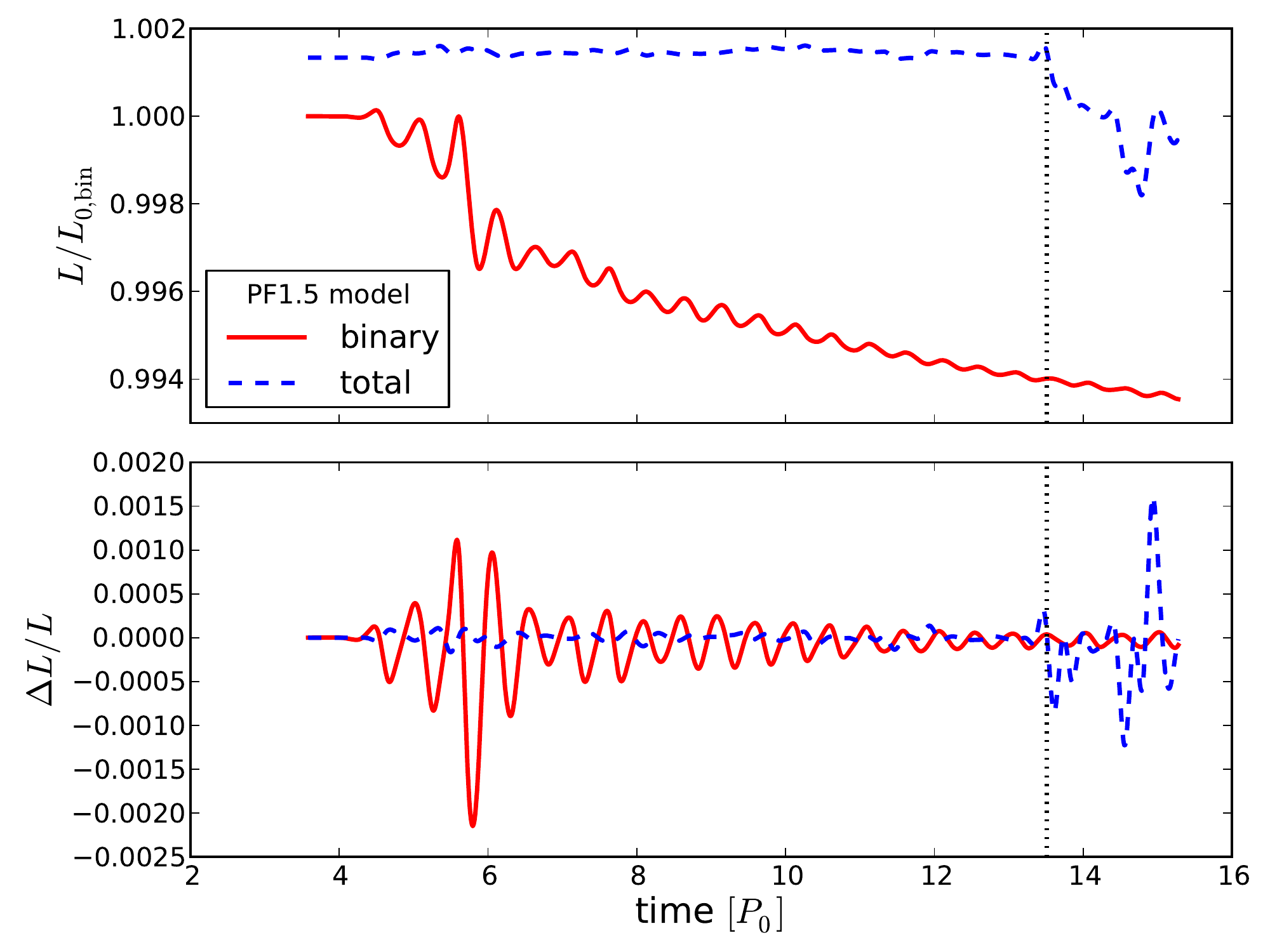}}
  \put(333,0){\includegraphics[width=0.3333\textwidth]{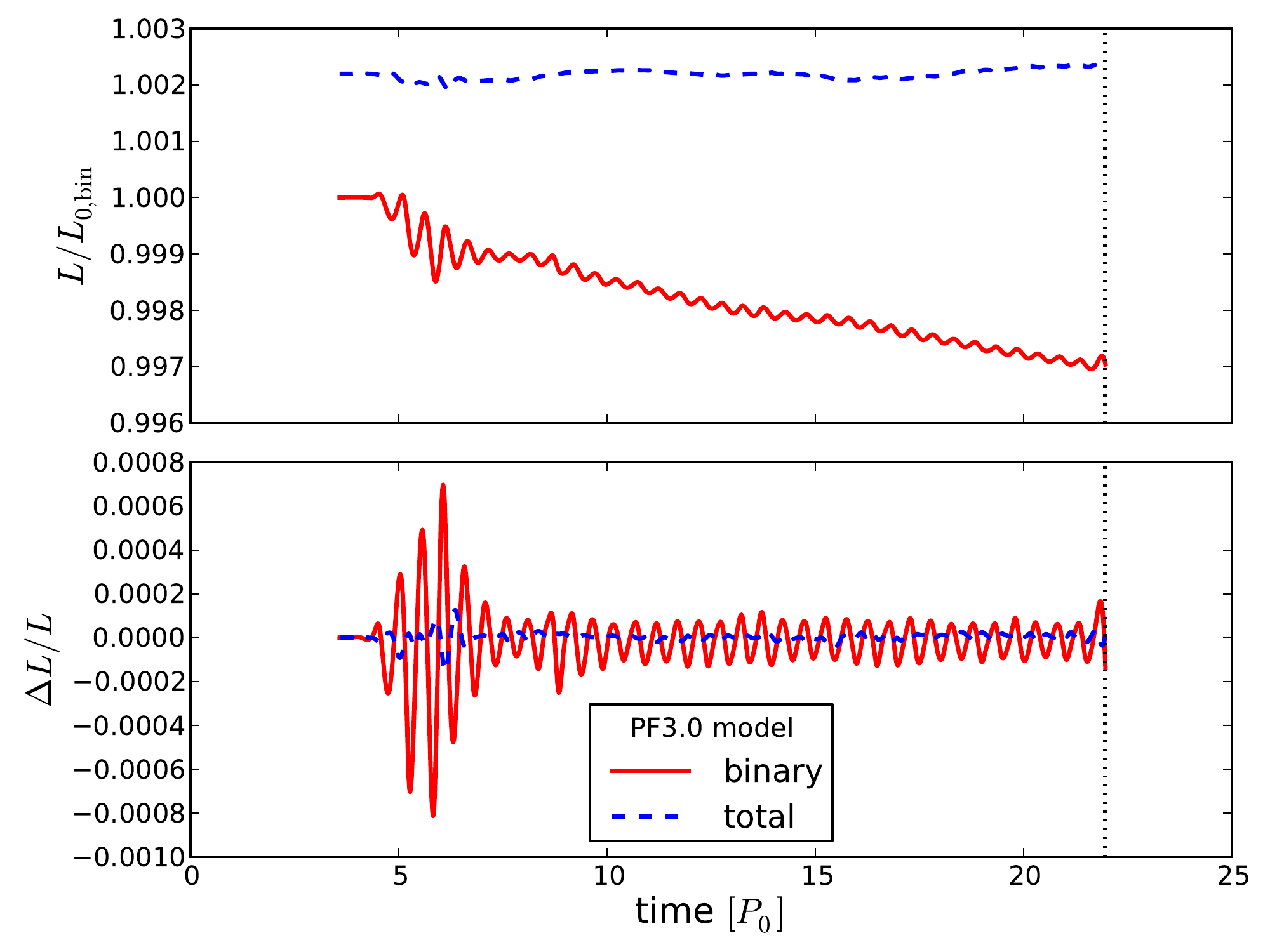}}
  \put(0,125){\includegraphics[width=0.3333\textwidth]{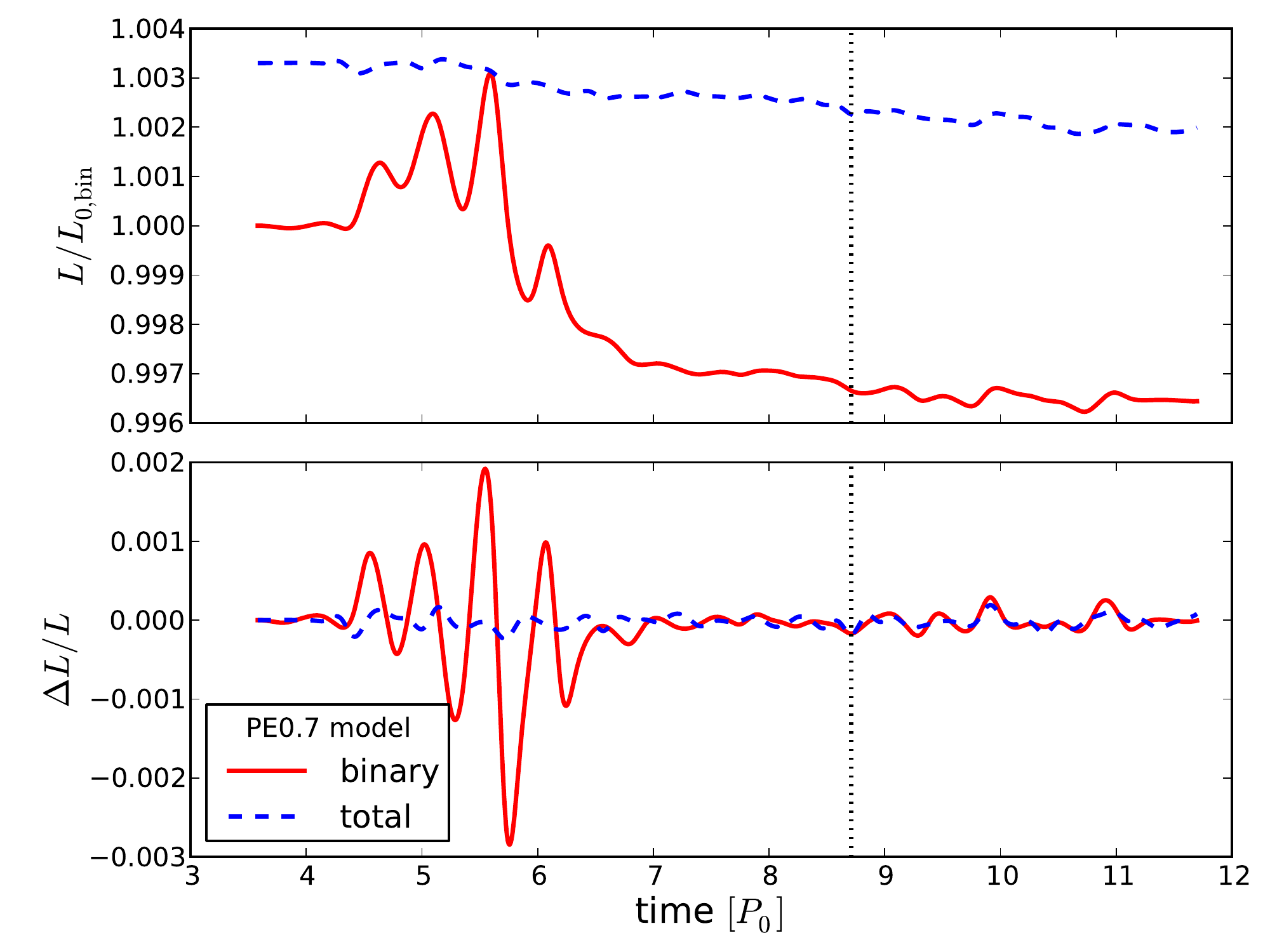}}
  \put(167,125){\includegraphics[width=0.3333\textwidth]{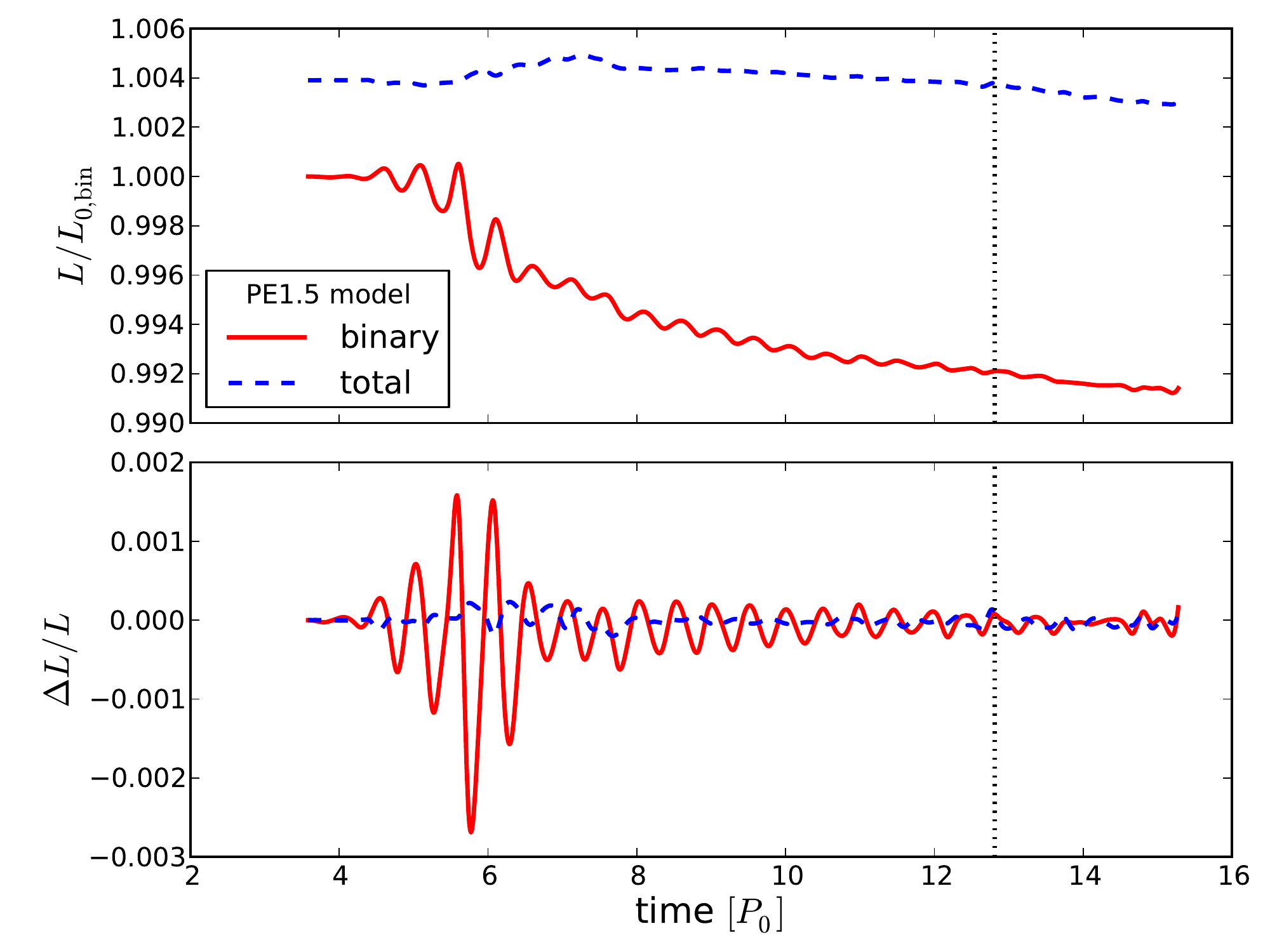}}
  \put(333,125){\includegraphics[width=0.3333\textwidth]{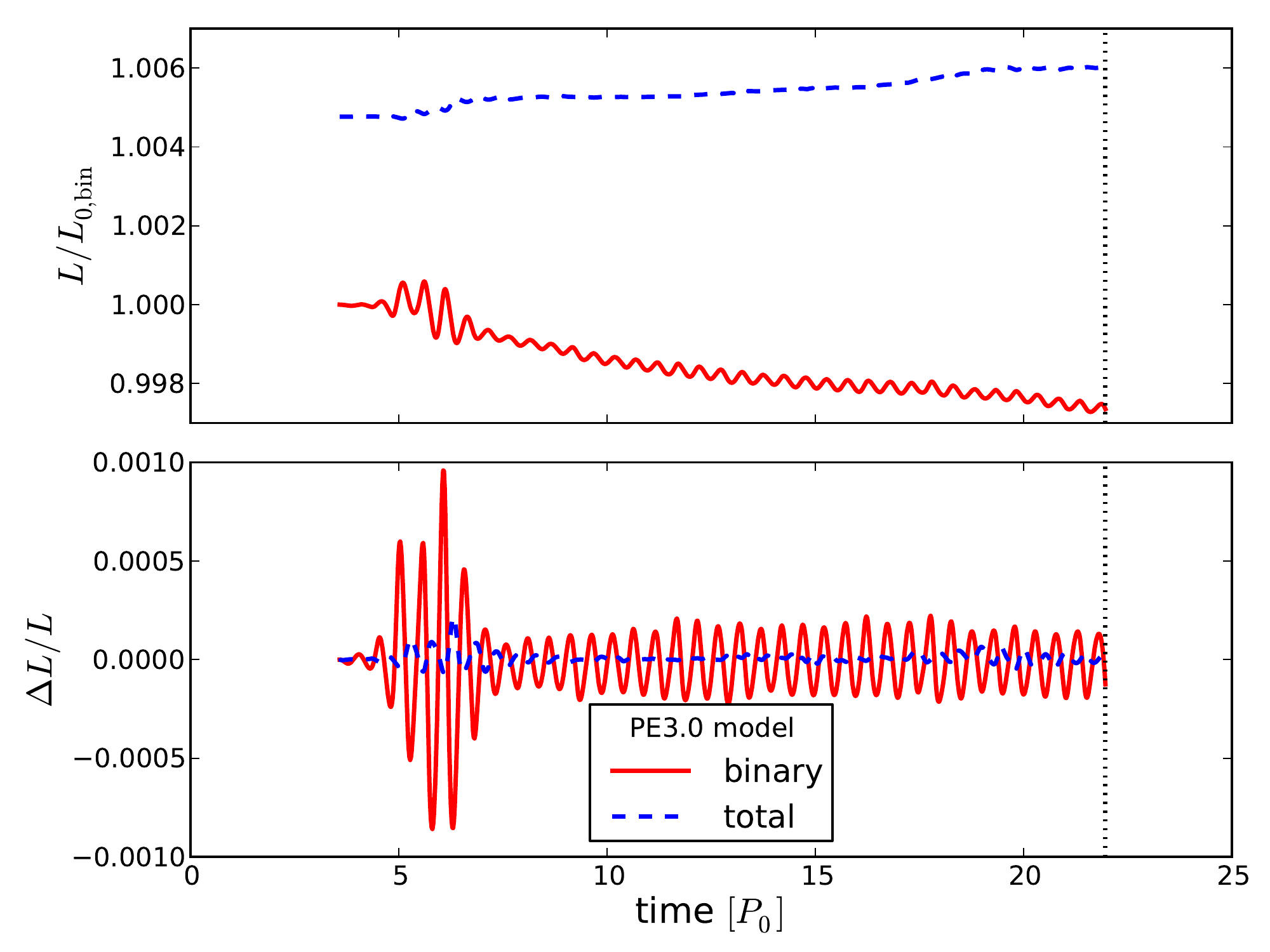}}
  \put(0,250){\includegraphics[width=0.3333\textwidth]{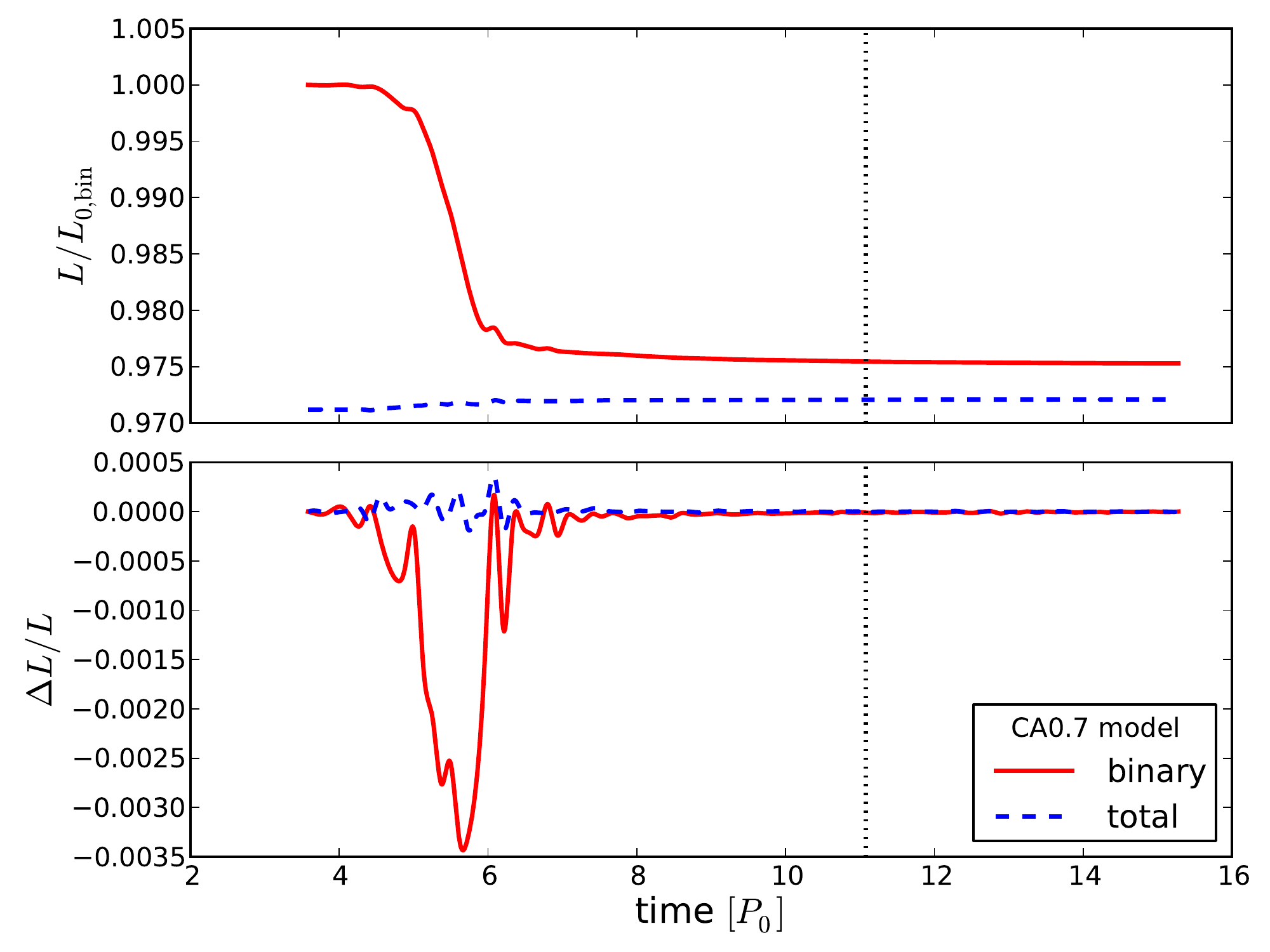}}
  \put(167,250){\includegraphics[width=0.3333\textwidth]{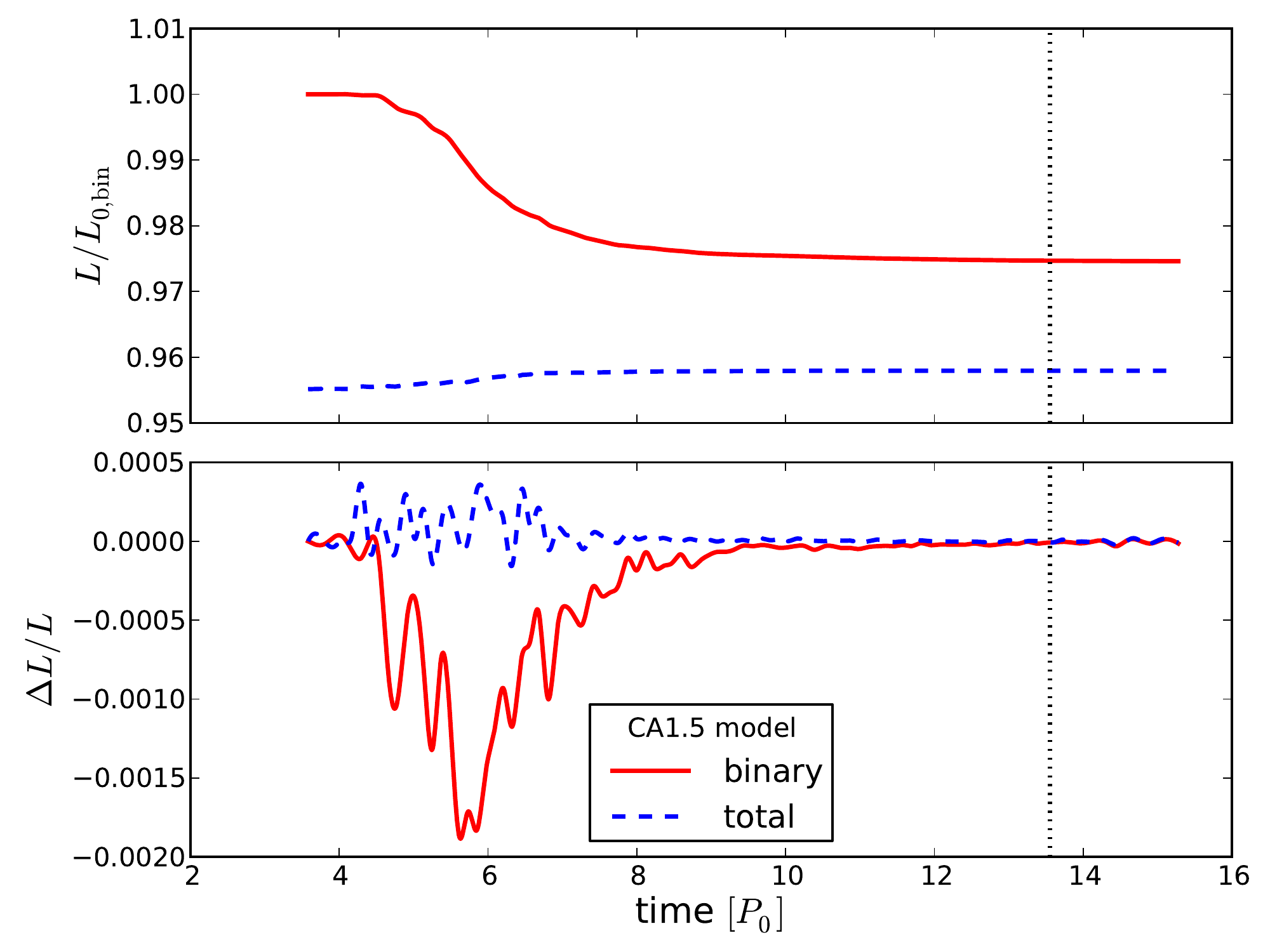}}
  \put(333,250){\includegraphics[width=0.3333\textwidth]{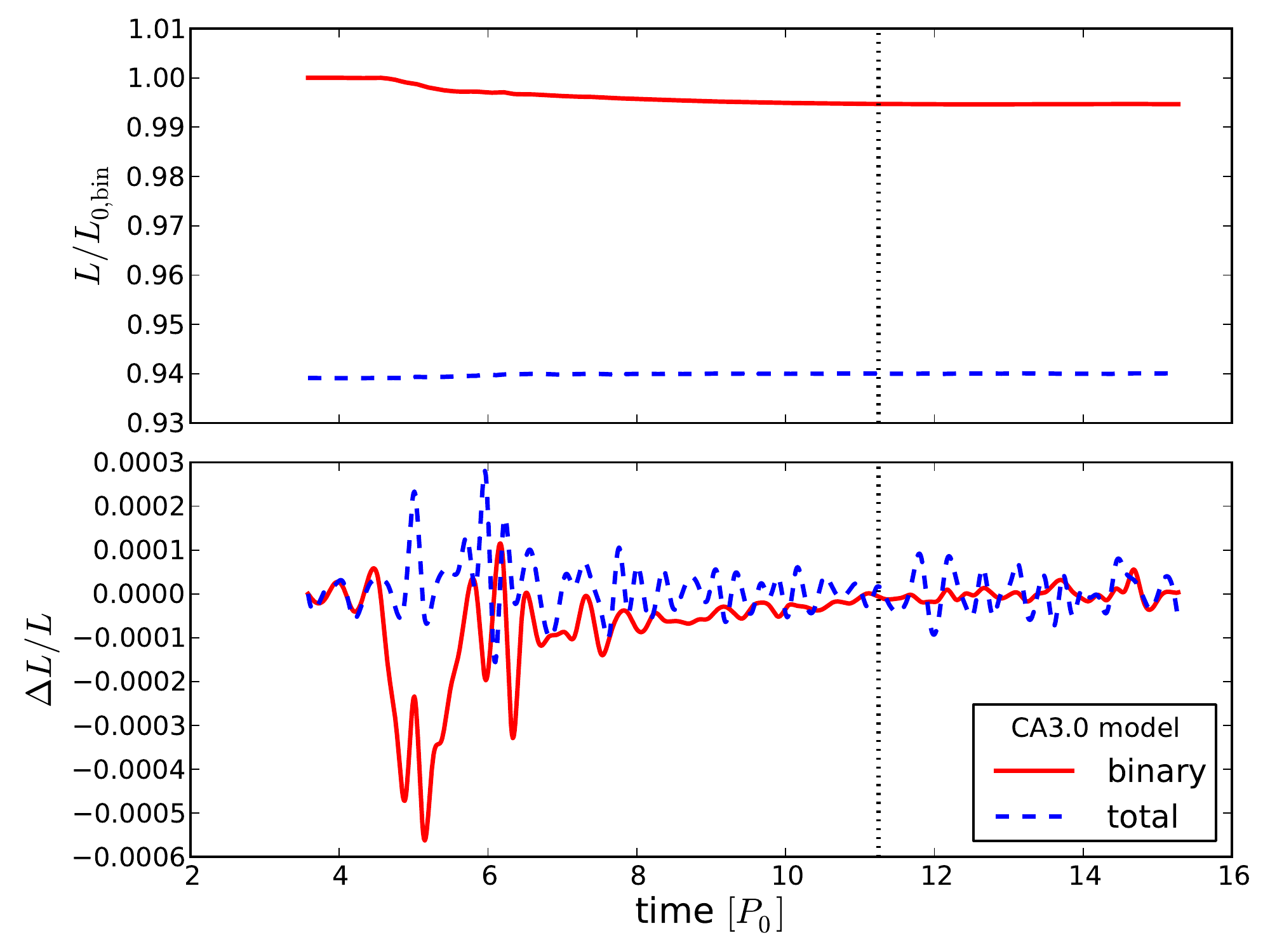}}
  \put(0,375){\includegraphics[width=0.3333\textwidth]{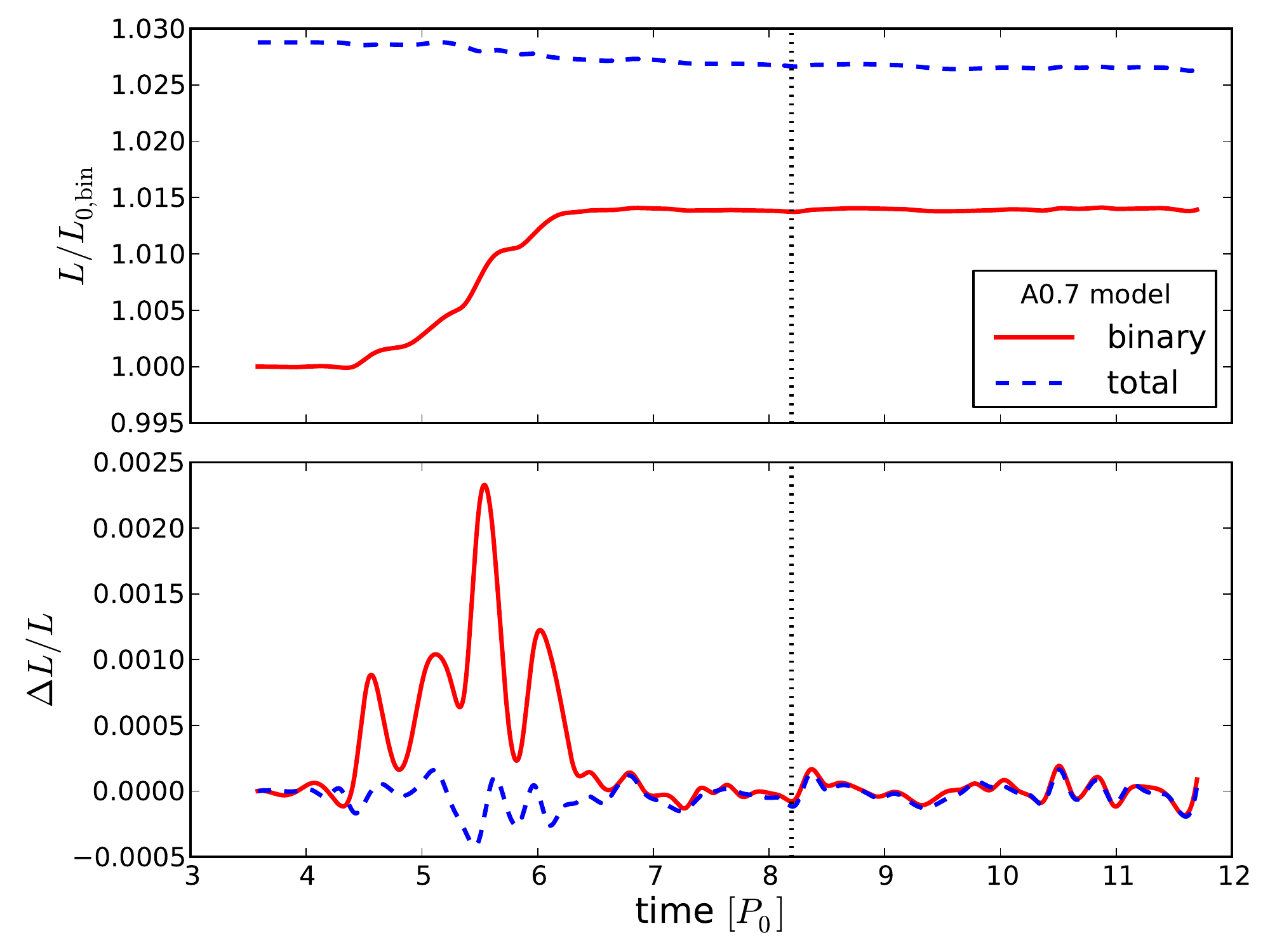}}
  \put(167,375){\includegraphics[width=0.3333\textwidth]{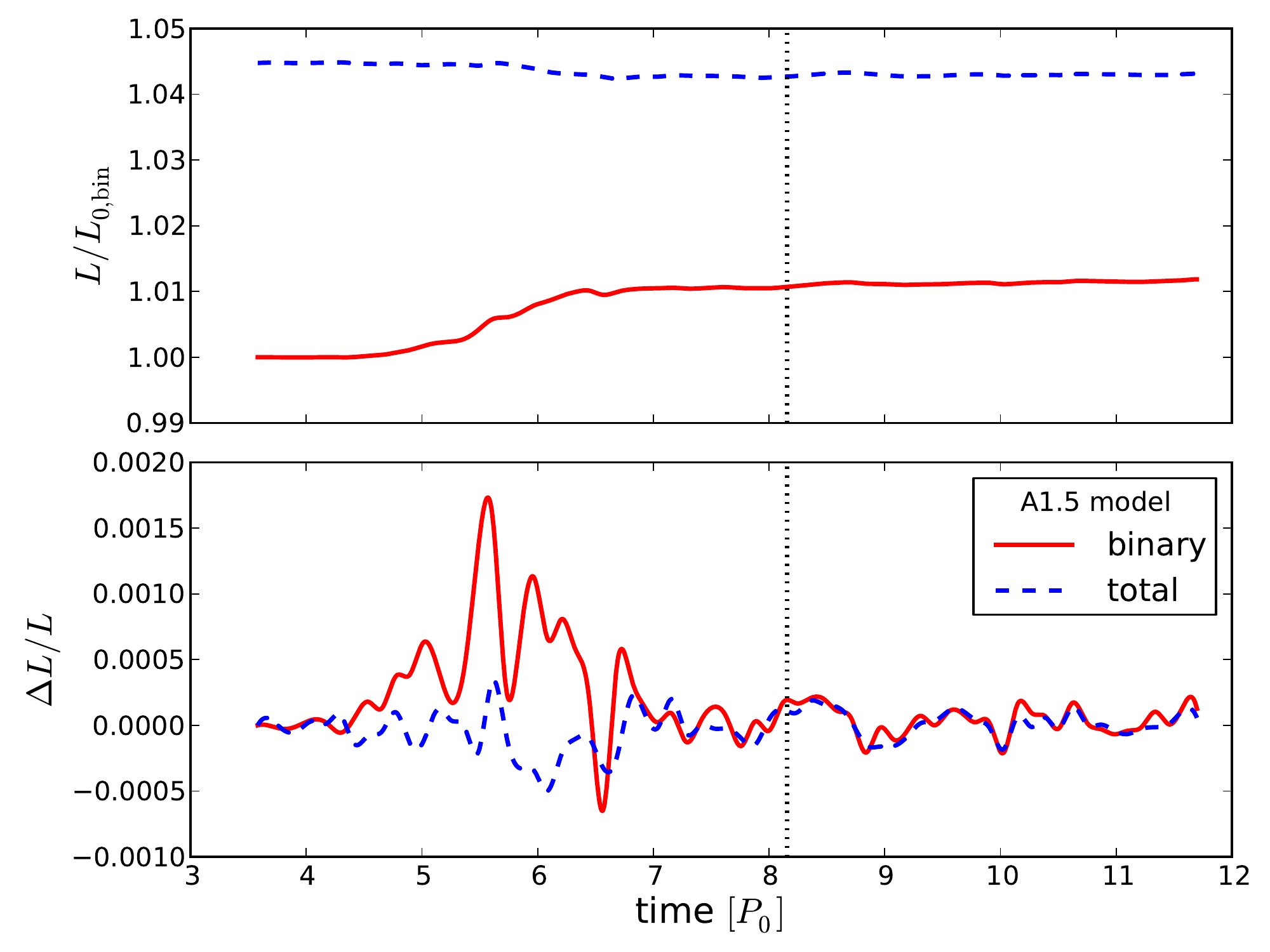}}
  \put(333,375){\includegraphics[width=0.3333\textwidth]{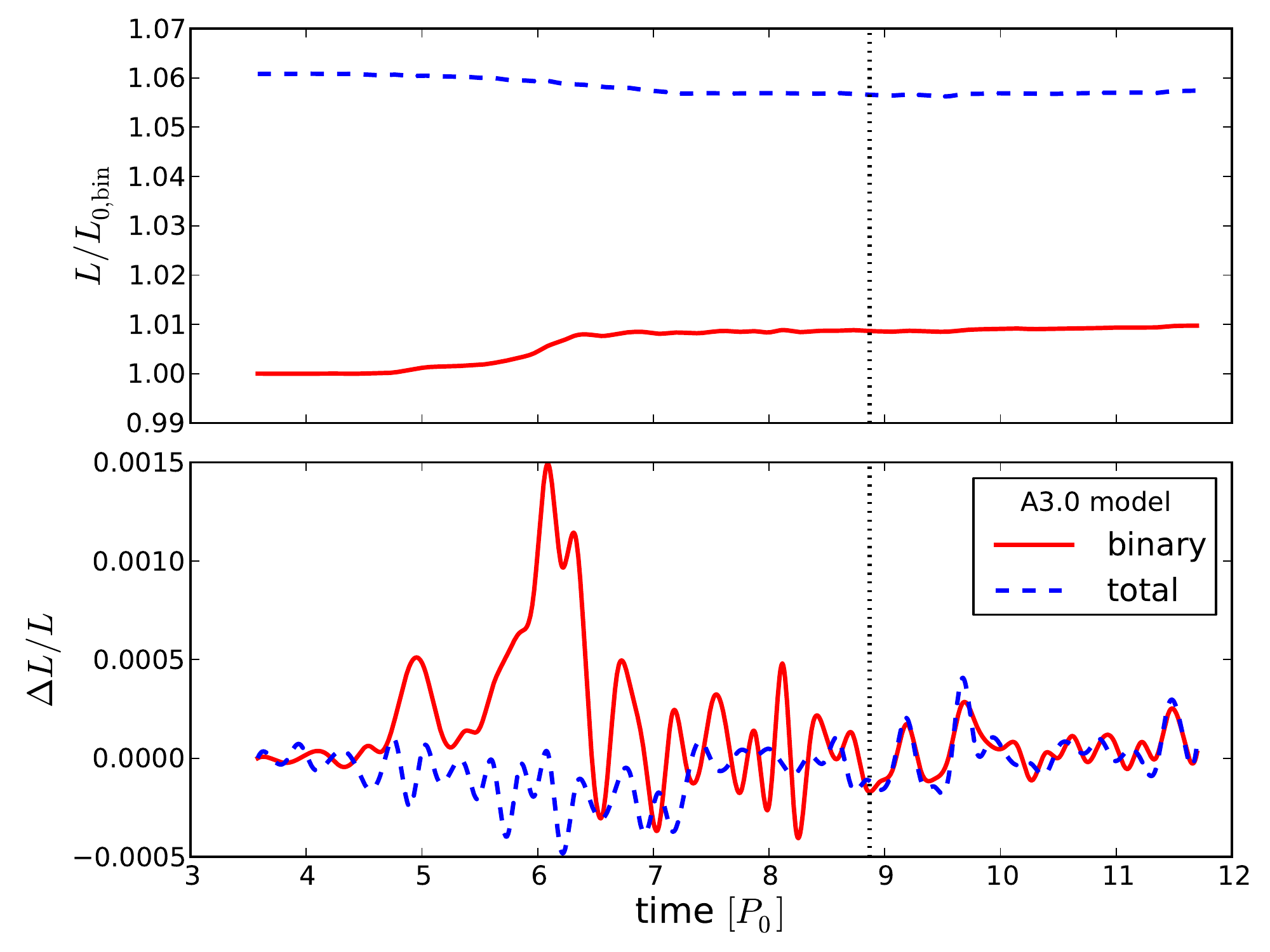}}
 \end{picture}
\caption{Evolution of the angular momentum of the entire system (blue) and the binary (red) for twelve of our models. The upper plots of each panel show the total magnitude of the angular momentum vector normalised by the binary initial value, while the lower plots show the relative change between successive snapshots. We increase the impact parameter from left to right, and each row represent a different inclination, as indicated in the legends. The vertical dashed line indicates the time when the fluctuations of the total angular momentum are larger than half of the changes of the binary angular momentum. Note the different vertical scales in each panel.}
\label{ang_mom}
\end{figure*}

We aim to measure the small changes in the binary orbit caused by the interaction with a much less massive cloud.  Therefore, we first need to establish whether our simulations have the accuracy to resolve these effects. With that aim, we study the conservation of angular momentum $L$  in our different models in Fig.~\ref{ang_mom}. 
The upper plots on each panel show the evolution of the angular momentum of the binary and that of the whole system.  The latter should be constant, but due to numerical issues we see it does vary in the models.  Nevertheless, the total change in angular momentum experienced by the binary during the interaction is always noticeably larger than the change in the total angular momentum of the system, which means the binary evolution obtained in the simulations is still meaningful.
The lower panels of the figure allow us to quantify {\em until when} we can trust the binary evolution, as explained below. 
 
Notice that due to the finite size of the sink radius, we do expect some loss of angular momentum throughout the binary evolution as the gas is accreted by each MBH. However, we estimate that the cumulative effect of this loss is at most of the order of $~10^{-4}L_0$, which is significantly smaller that the changes we observe for the total angular momentum of the system in Fig.~\ref{ang_mom}. Therefore, these deviations are due to numerical inaccuracies of the code. 

The Lagrangian formulation used by SPH codes to solve the hydrodynamical equations conserve angular momentum exactly. However, the numerical integration of these equations, { {\em using individual particle timesteps,} breaks the time symmetry of the model, preventing exact conservation. More importantly, the approximation of gravitational forces through a tree algorithm introduces numerical noise on the accelerations computed by the code, which translates in oscillations in the angular momentum \citep[for a discussion of these two issues see][]{Springel2005}.}
Given these inaccuracies, we can trust our simulations only up to the point where the numerical noise starts to dominate the evolution of the system. This corresponds to the time at which the fluctuations of the total angular momentum are of the order of the binary angular momentum changes.  More precisely, we compute the amplitude of the fluctuations in $\Delta L/L$ for both the entire system and the binary in chunks of half binary orbital periods, and show them in the lower plots of each panel of Fig.~\ref{ang_mom}. When the ratio of the former to the latter becomes larger than $1/2$, we discard the subsequent binary evolution.  That point is indicated in Fig.~\ref{ang_mom} by the dashed vertical lines, and it always happens at $T\gtrsim8P_0$, being $P_0$ the initial binary period. Note that most, if not all, the evolution of $L_{\rm bin}$ occurs at $T<8P_0$, corresponding to the phase in which most of the mass is accreted by the MBHB (as shown in \paperI). For the PE and PF simulations this period is actually longer because of the more prolonged accretion, but angular momentum fluctuations are correspondingly smaller. This indicates that significant $\Delta{L_{\rm bin}}$ is driven by torques exerted by accreting particles, closely interacting with the MBHB. When accretion stops, also the binary evolution is dumped to a level consistent with the numerical accuracy of the code. This is because the cloud is light, and gas that does not interact strongly with the binary hardly modifies its dynamics. {A robust measurement of the secular evolution of the system would require refining some parameters of the simulations (e.g., the opening angle of the gravity-tree nodes), which translates in considerable longer computing times not affordable for our standard configuration.}

\subsection{Binary evolution}

\begin{figure}
\begin{picture}(230,520)
\put(0,0){\includegraphics[width=0.46\textwidth]{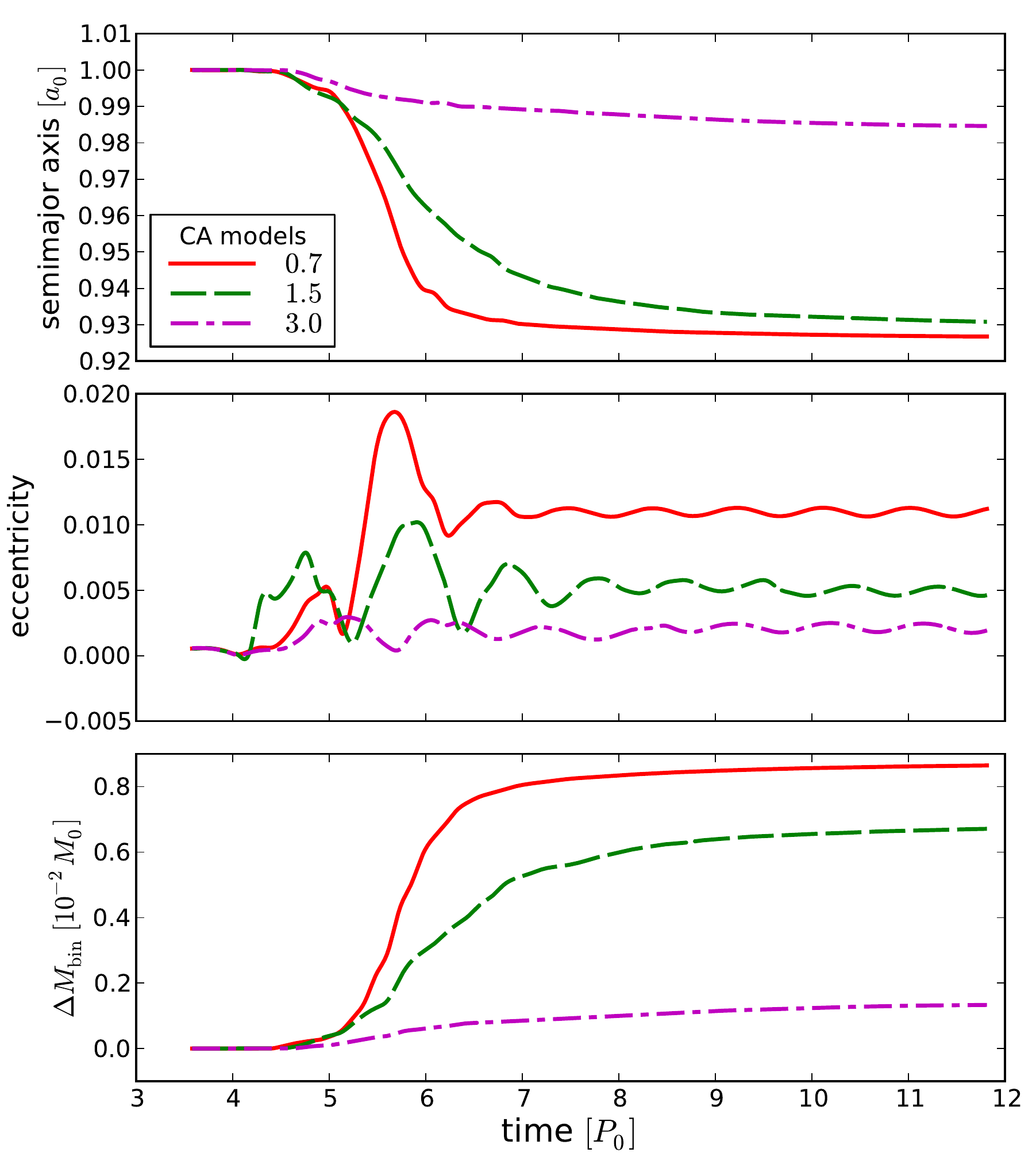}}
\put(0,260){\includegraphics[width=0.46\textwidth]{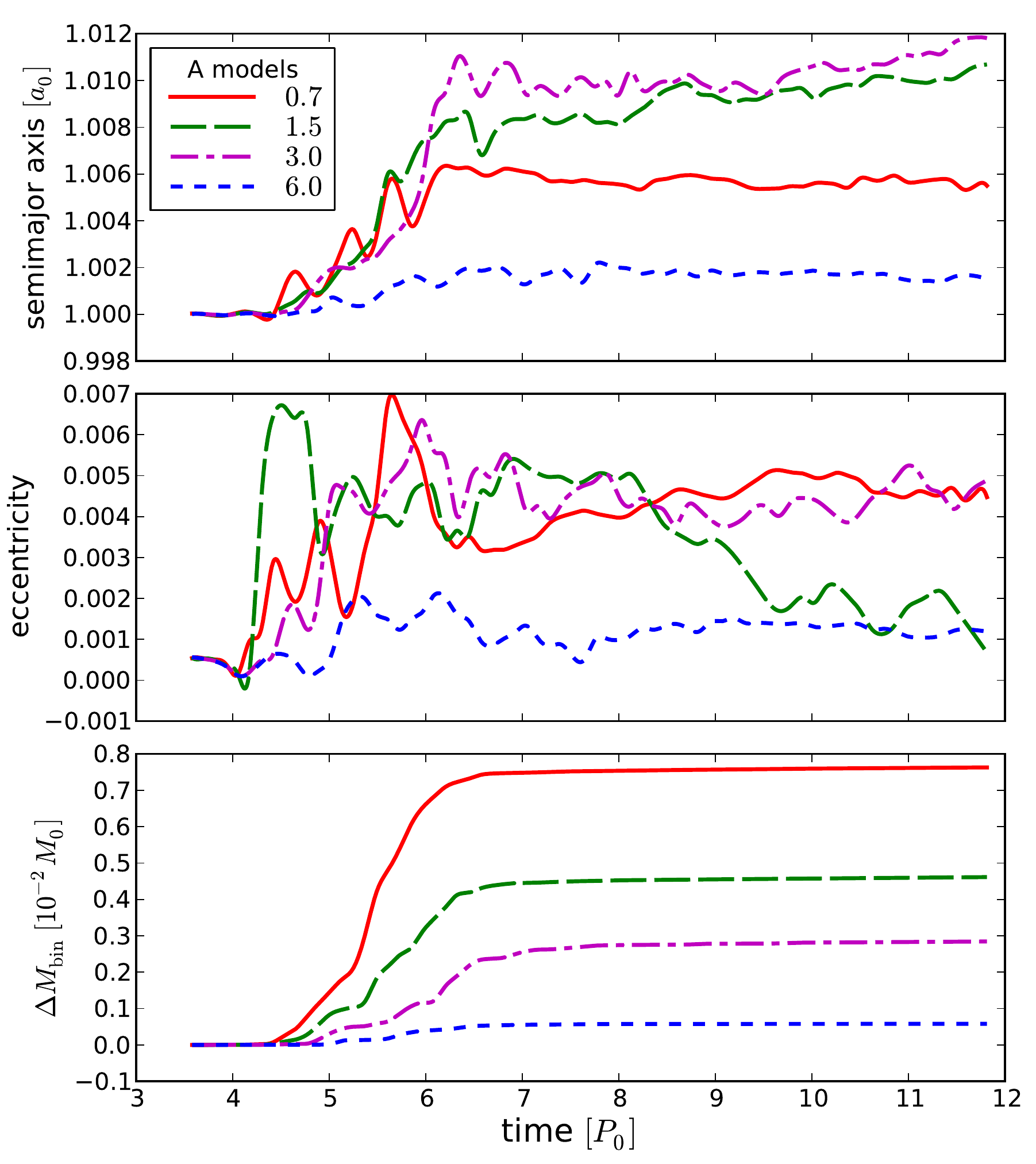}}
\end{picture}
\caption{Evolution of the binary semimajor axis (upper panels), eccentricity (middle panels) and accreted mass (lower panels) for the A and CA models. The different lines in each panel represent the different pericentre distances, as indicated in the legend. 
}
\label{orbit1}
\end{figure}

\begin{figure}
\begin{picture}(230,520)
\put(0,0){\includegraphics[width=0.46\textwidth]{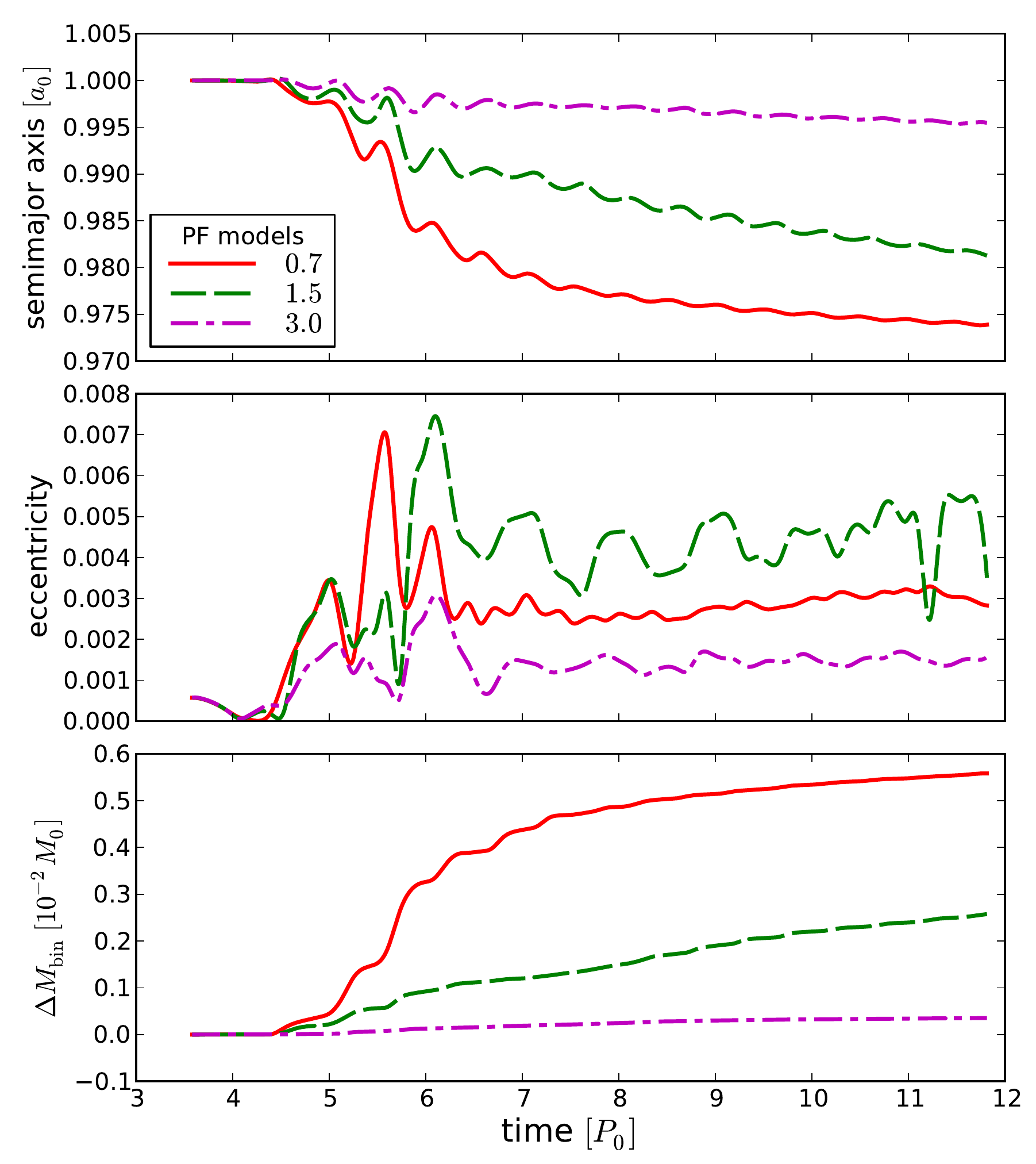}}
\put(0,260){\includegraphics[width=0.46\textwidth]{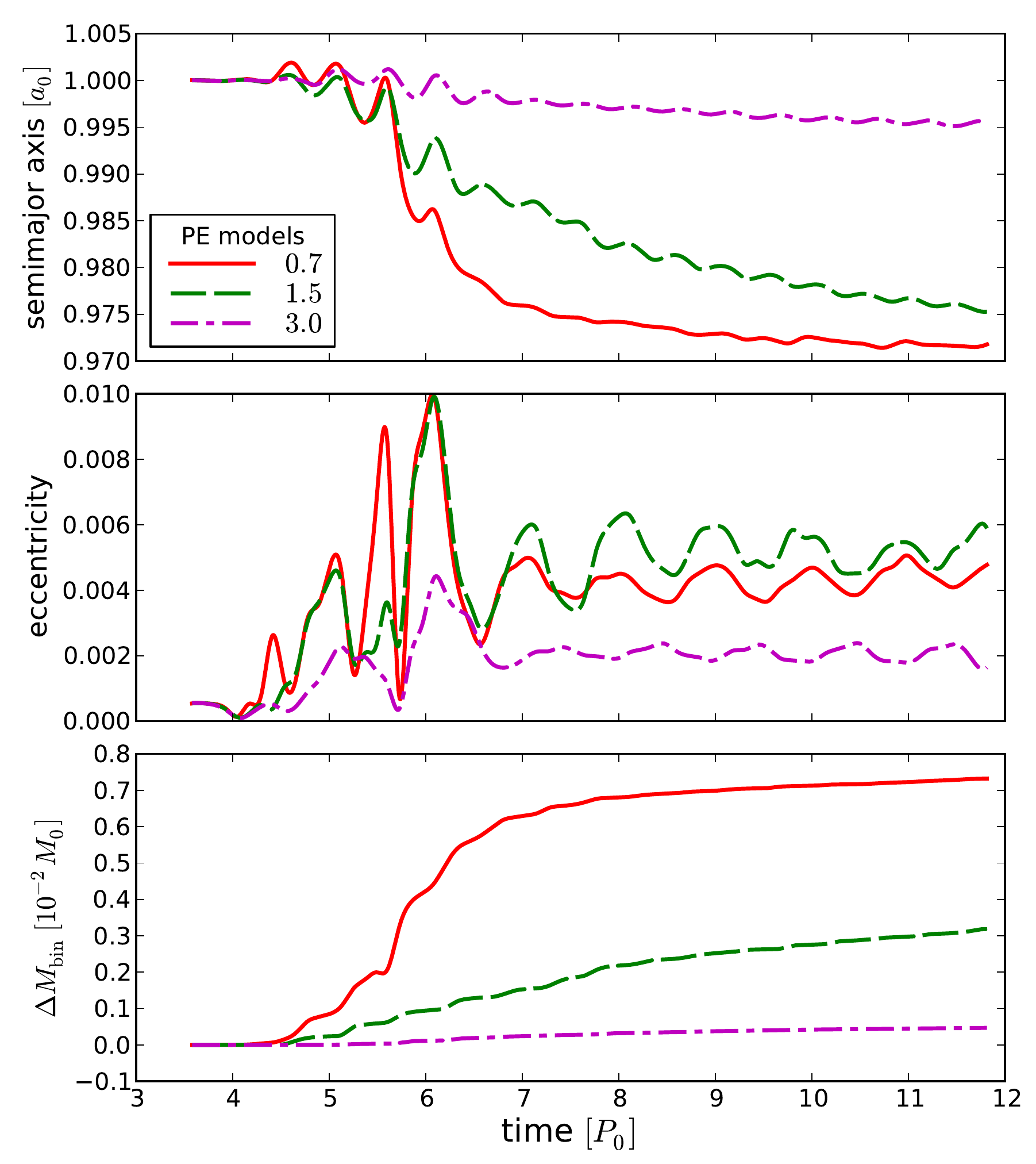}}
\end{picture}
\caption{Same as Fig.~\ref{orbit1}, but for the perpendicular simulations (PE and PF models).}
\label{orbit2}
\end{figure}

We present the results of the transient evolution of the binary orbital components and mass in Figs.~\ref{orbit1} and~\ref{orbit2}. We observe that the bulk of accretion occurs during the first few orbits ($\approx 4 - 8$), which correspond to the first passage of the cloud, as discussed in \paperI. This is the same period where we observe a significant change of the orbital parameters, especially the semimajor axis, which implies that the dynamical evolution of the binary is intimately related to the accretion. The eccentricity evolution for every system is extremely small, typically less than 1\%, which means that the binary remains roughly circular during the transient interaction with the gas cloud. For this reason we restrict our analysis to the semimajor axis evolution.

\begin{table}
\centering
\caption{Total change of the binary angular momentum $(\Delta L)$, mass $(\Delta M)$ and semimajor axis $(\Delta a)$ for the different orbits modelled, from the beginning of the simulation up to $t_{\rm conf}$, which is the time where we no longer disentangle the physical interaction from the numerical noise. }
\begin{tabular}{l*{3}{c}r}
\hline\hline
 Model & $t_{\rm conf}$ & $\Delta L$ & $\Delta M$ & $\Delta a$\\
 & $(P_{0})$ & $(L_{0})$ & $(M_{0})$ & $(a_{0})$\\
\hline
A0.7 & $8.1$ & $0.01380$ & $0.00754$ & $0.00545$  \\
A1.5 & $8.1$ & $0.01058$ & $0.00453$ & $0.00829$  \\
A3.0 & $8.9$ & $0.00864$ & $0.00278$ & $0.00982$  \\ 
A6.0 & $8.8$ & $0.00137$ & $0.00058$ & $0.00160$  \\ 
\hline
CA0.7 & $11.1$ & $-0.02456$ & $0.00862$ & $-0.07312$  \\
CA1.5 & $13.5$ & $-0.02532$ & $0.00678$ & $-0.06988$  \\
CA3.0 & $11.3$ & $-0.00529$ & $ 0.00132$ & $-0.01521$  \\ 
\hline
PE0.7 & $8.7$ & $-0.00335$ & $0.00695$ & $-0.02707$  \\
PE1.5 & $12.8$ & $-0.00790$ & $0.00334$ & $-0.02547$  \\
PE3.0 & $22.0$ & $-0.00265$ & $0.00080$ & $-0.00764$  \\ 
\hline
PF0.7 & $10.4$ & $-0.00399$ & $0.00542$ & $-0.02389$  \\
PF1.5 & $13.5$ & $-0.00600$ & $0.00281$ & $-0.02013$  \\
PF3.0 & $21.8$ & $-0.00283$ & $ 0.00076$ & $-0.00796$  \\ \hline
\hline
\end{tabular}
\label{delta}
\end{table}
Finally, we summarise the total change of the binary angular momentum, mass and semimajor axis for every configuration in Table~\ref{delta}, up to the time where we trust the numerics, as described in the previous Section.

\section{Angular momentum exchange}
\label{sec:angmomexch}

In order to link the evolution of the binary angular momentum to that of the semimajor axis and mass, we write the magnitude of the angular momentum as a combination of the binary properties, similar to what is shown in \citet{Roedig2012}.
However, instead of writing it as a function of the total binary mass $M$ and the reduced mass $\mu$, we consider $M$ and the mass ratio $q$.  This way, we can separate better the effect of total accretion (changing $M$) from that of differential accretion on to the two masses (changing $q$).

The magnitude of the binary angular momentum is
\begin{equation}
  L_{\rm bin}=\frac{q}{(1+q)^2}M^{3/2}\sqrt{Ga(1-e^2)},
\end{equation}
where $G$ is the gravitational constant. Differentiating with respect to all the parameters we get
\begin{equation}
  \frac{\Delta{L_{\rm bin}}}{L_{\rm bin}}=\frac{1-q}{q(1+q)}\Delta{q}+\frac{3}{2}\frac{\Delta{M}}{M}+\frac{1}{2}\frac{\Delta{a}}{a}-\frac{e}{1-e^2}\Delta{e}.
  \label{deltaL}
\end{equation}

\begin{figure*}
\centering
 \begin{picture}(500,500)
  \put(0,0){\includegraphics[width=0.3333\textwidth]{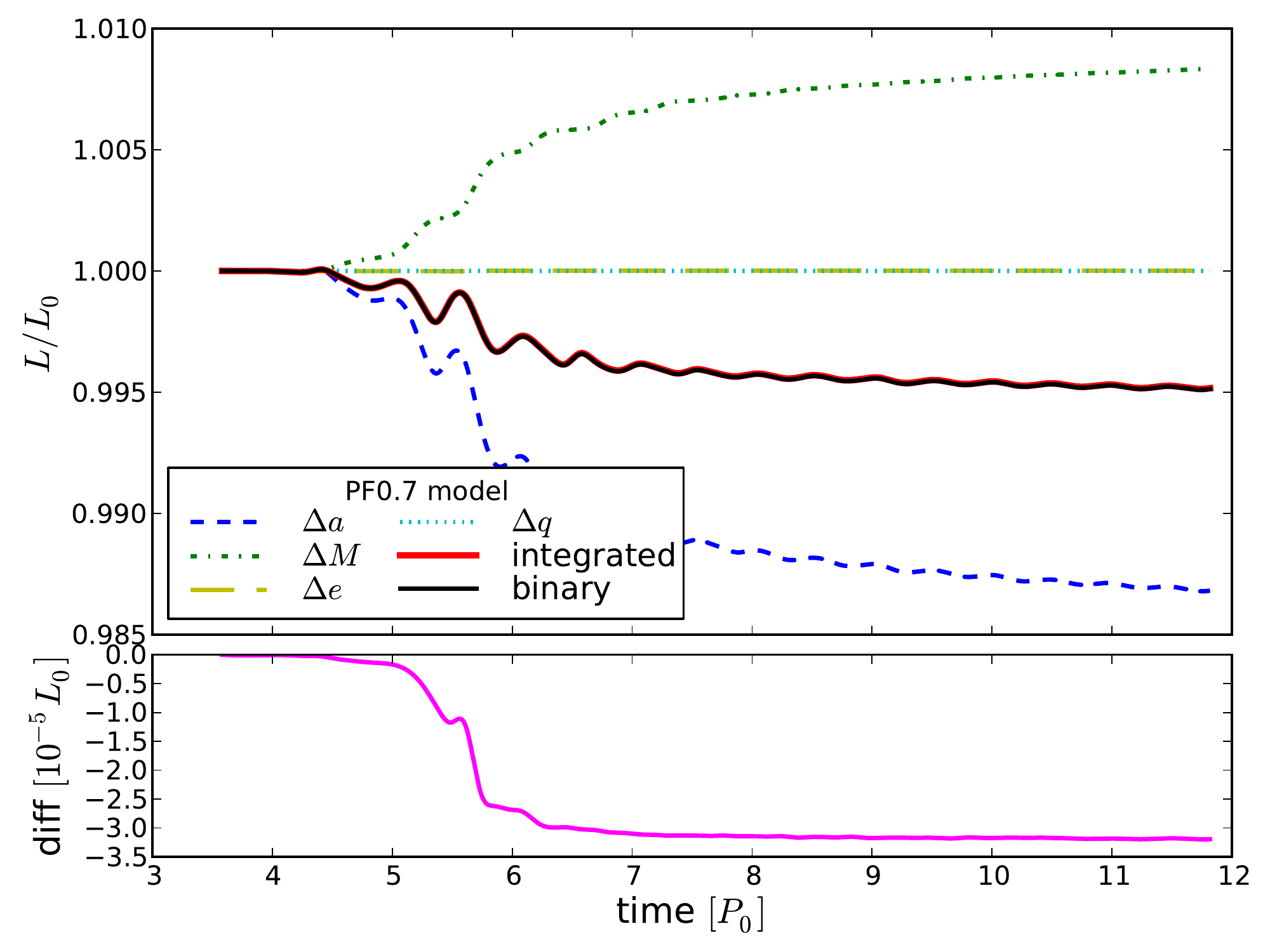}}
  \put(167,0){\includegraphics[width=0.3333\textwidth]{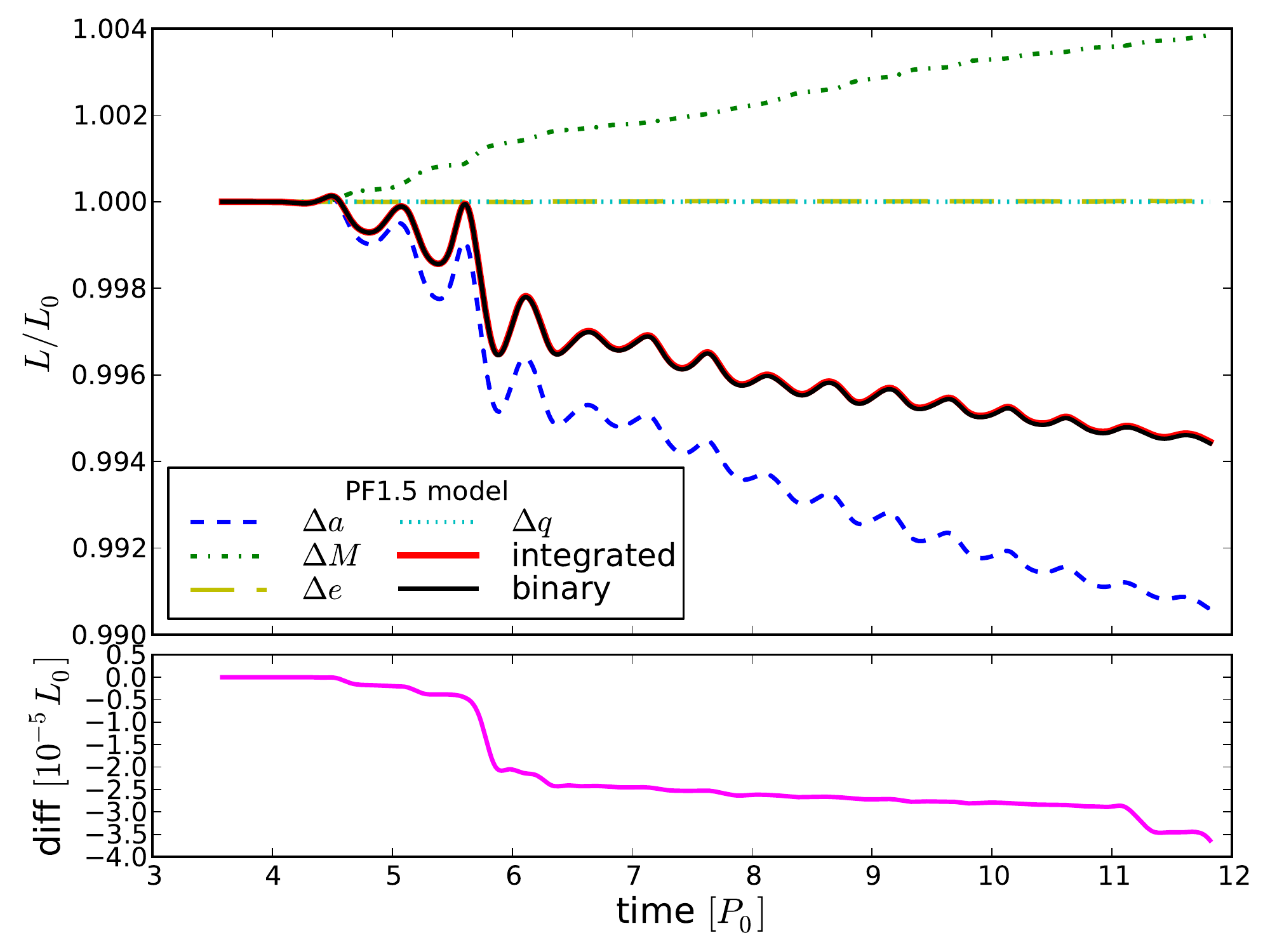}}
  \put(333,0){\includegraphics[width=0.3333\textwidth]{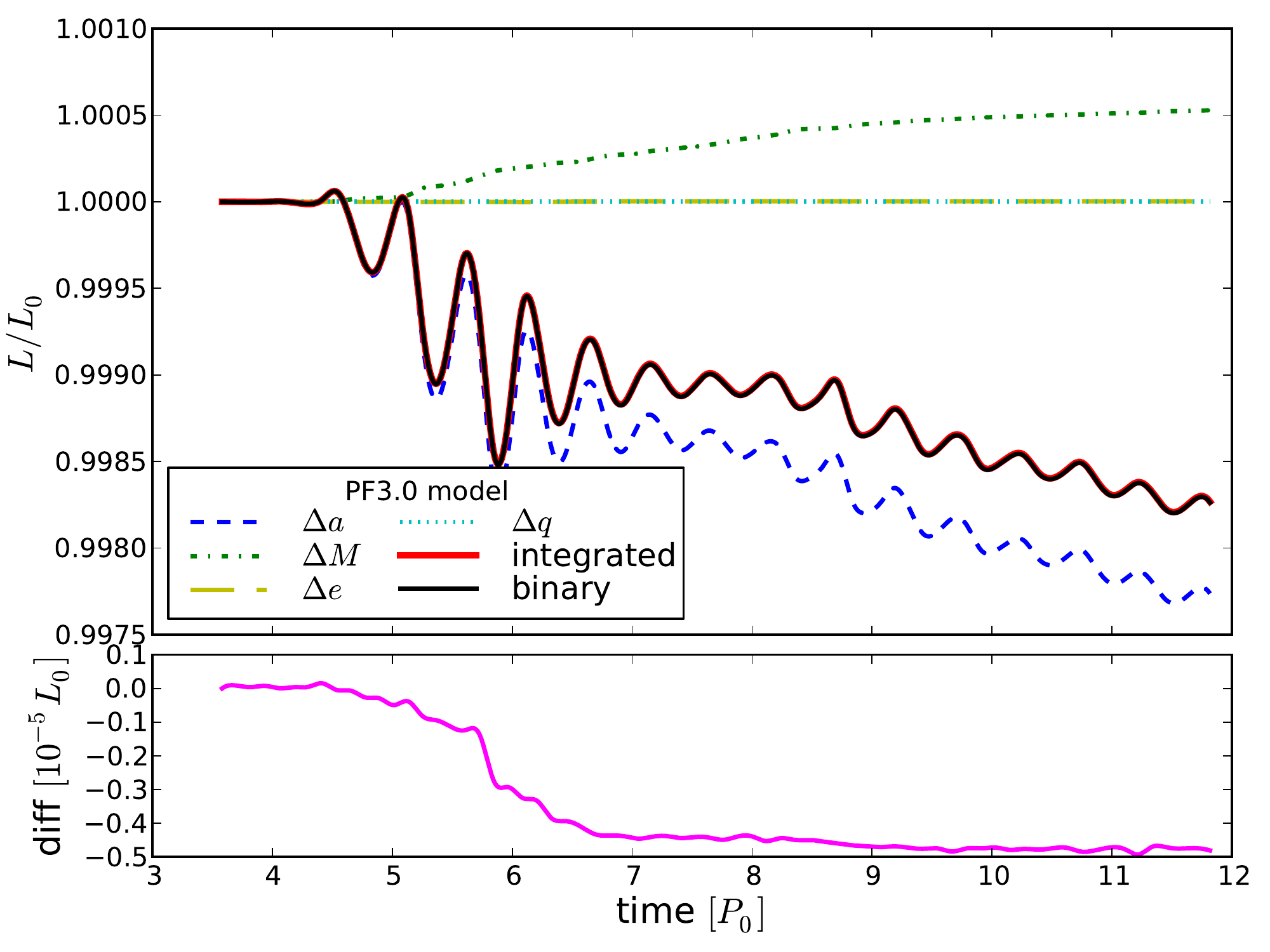}}
  \put(0,125){\includegraphics[width=0.3333\textwidth]{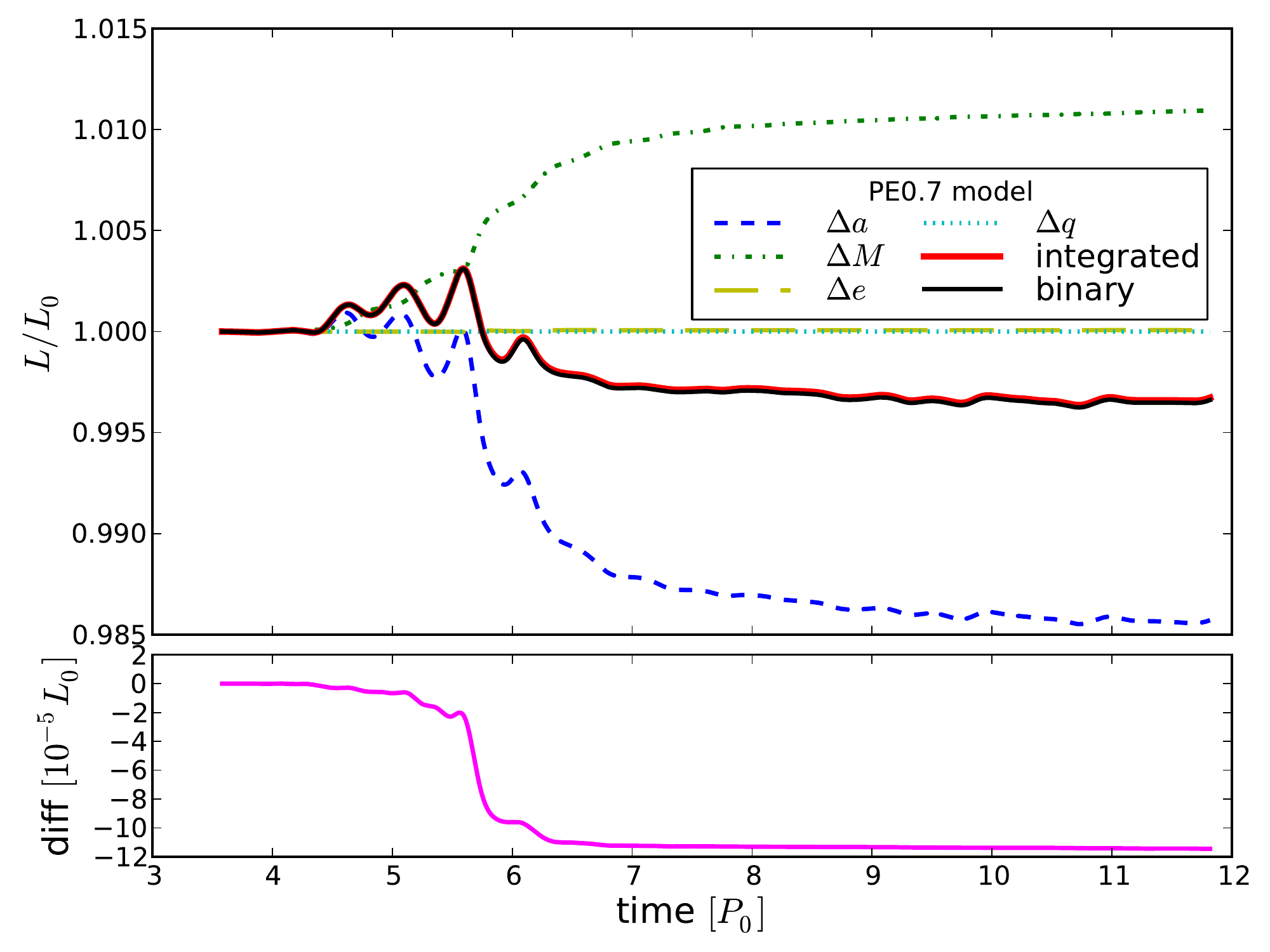}}
  \put(167,125){\includegraphics[width=0.3333\textwidth]{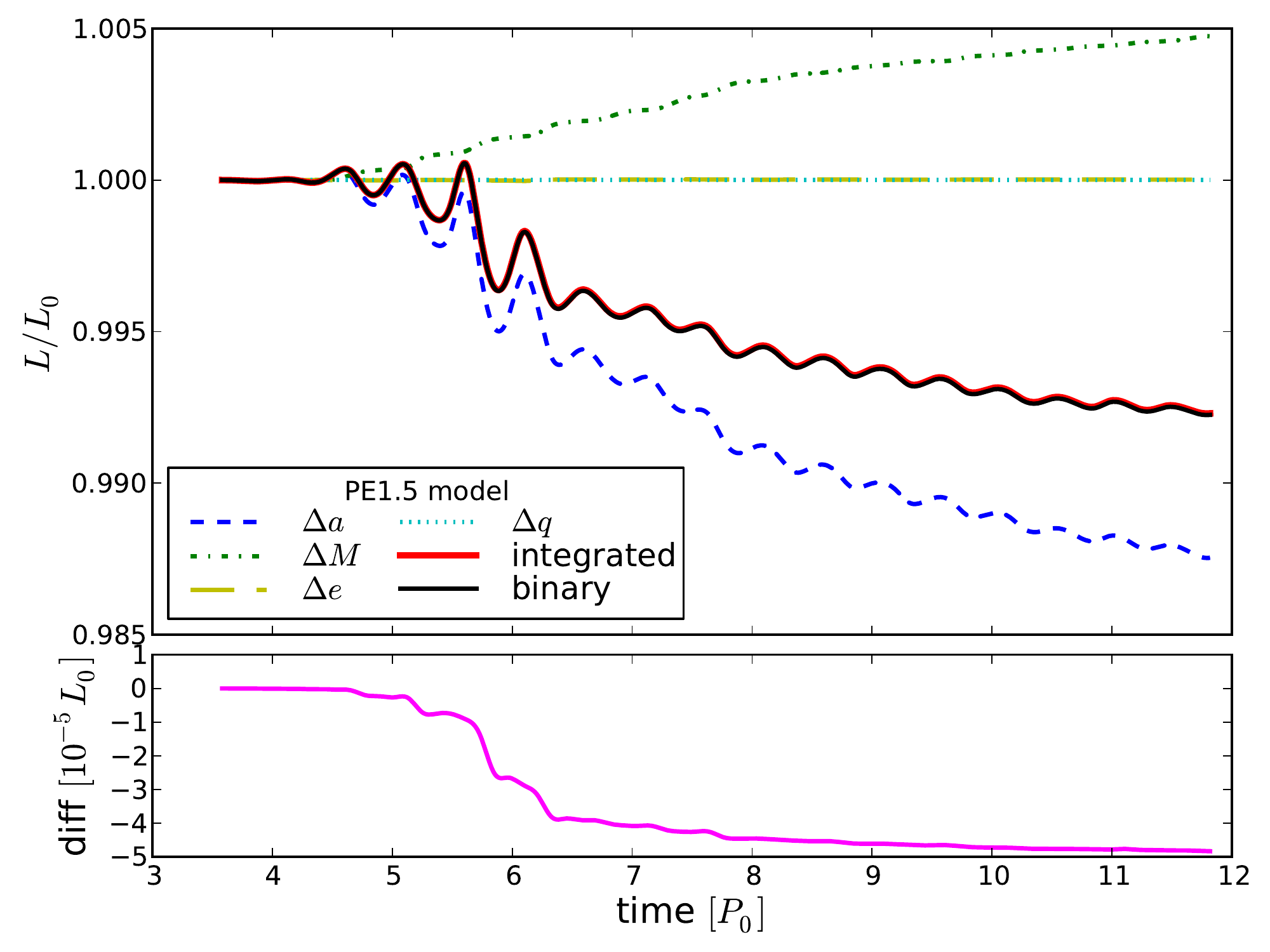}}
  \put(333,125){\includegraphics[width=0.3333\textwidth]{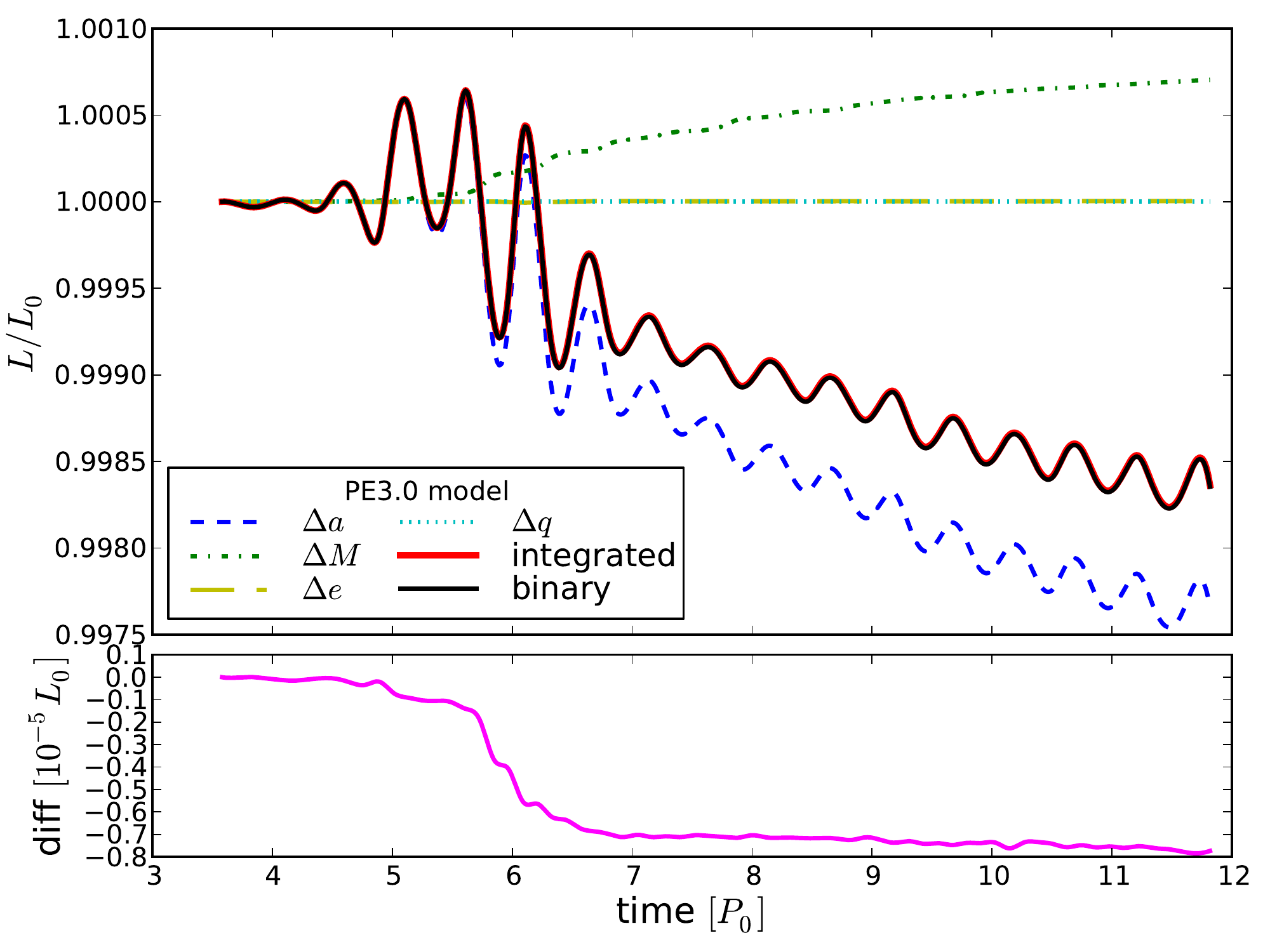}}
  \put(0,250){\includegraphics[width=0.3333\textwidth]{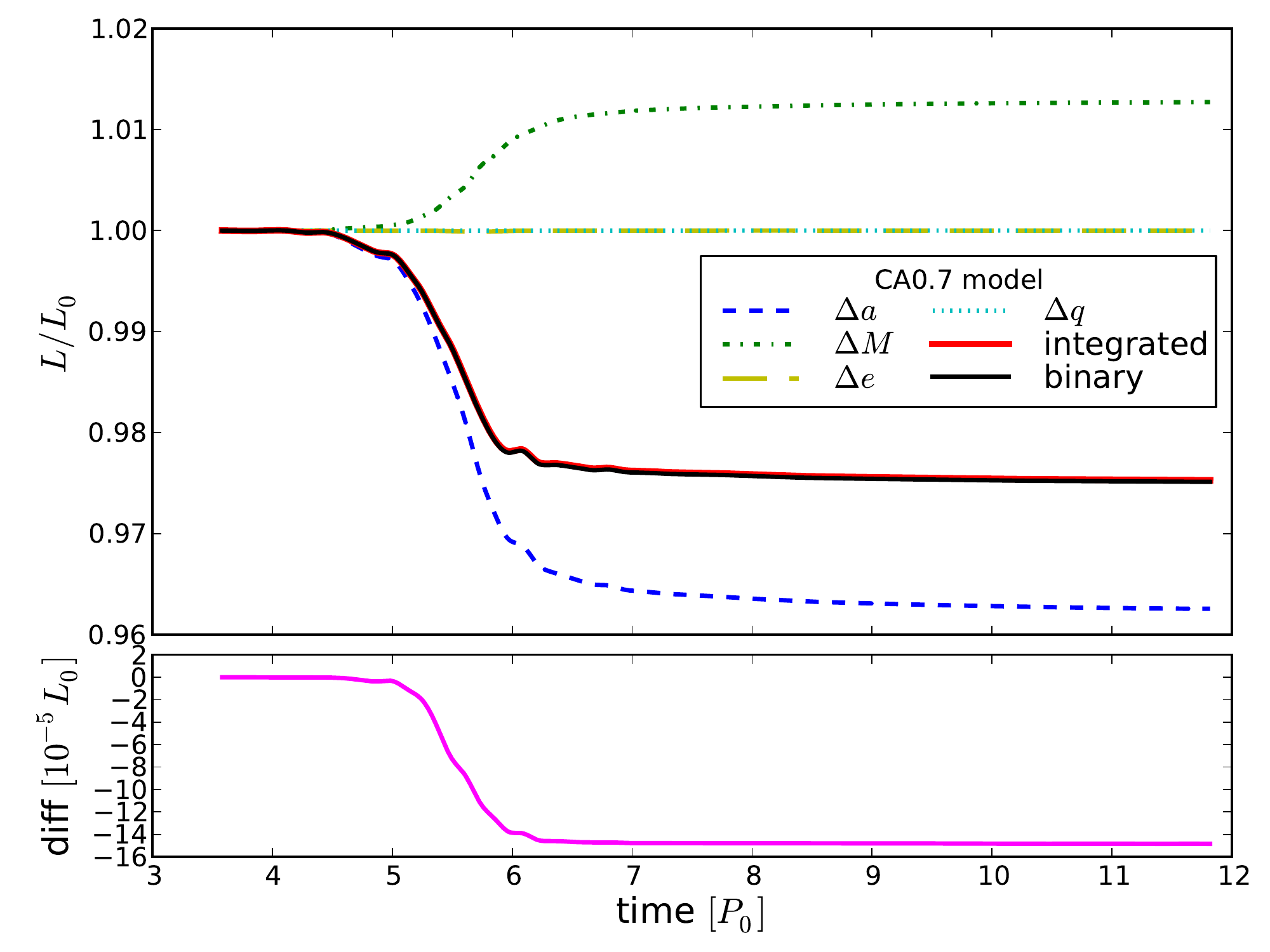}}
  \put(167,250){\includegraphics[width=0.3333\textwidth]{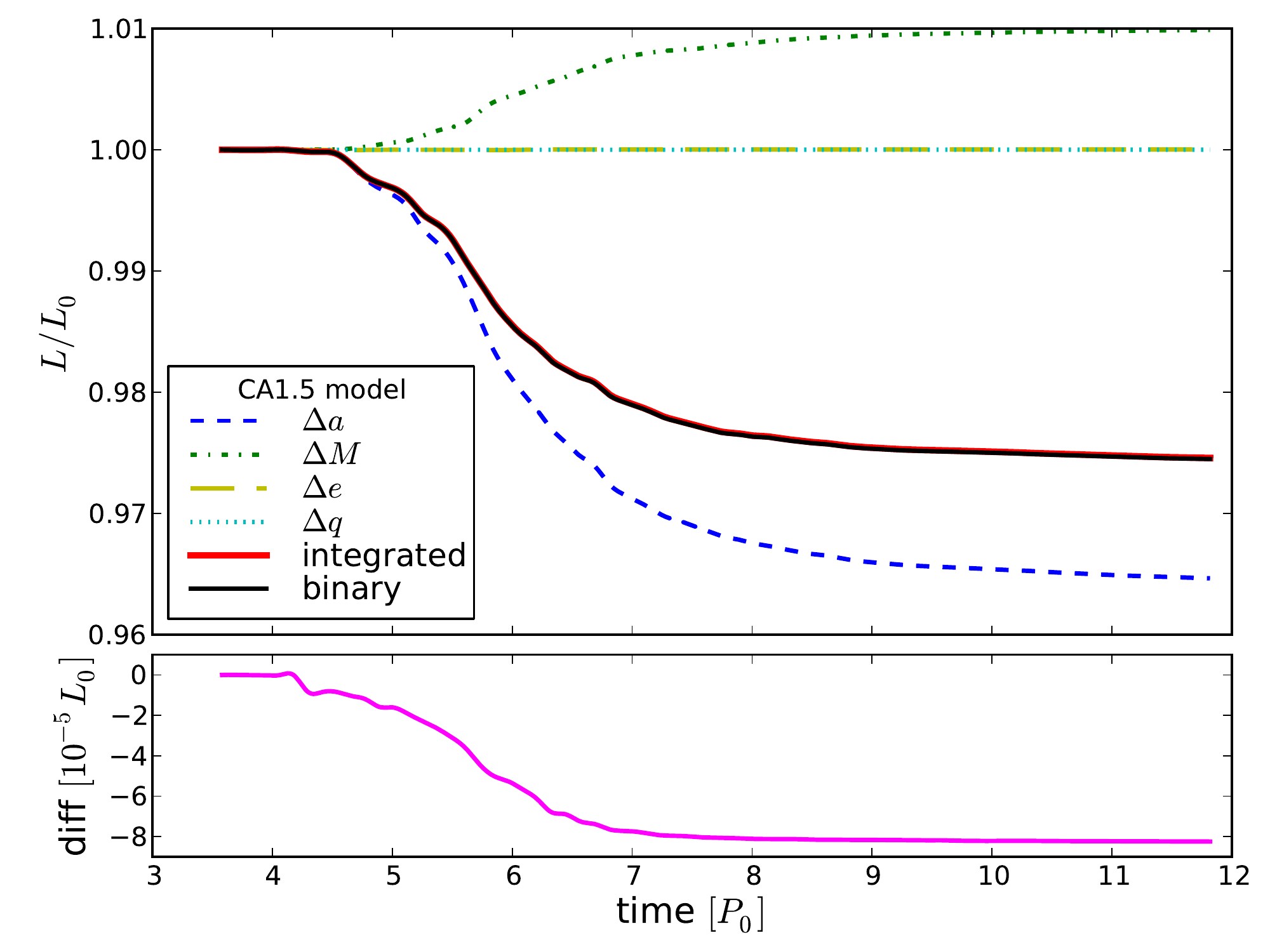}}
  \put(333,250){\includegraphics[width=0.3333\textwidth]{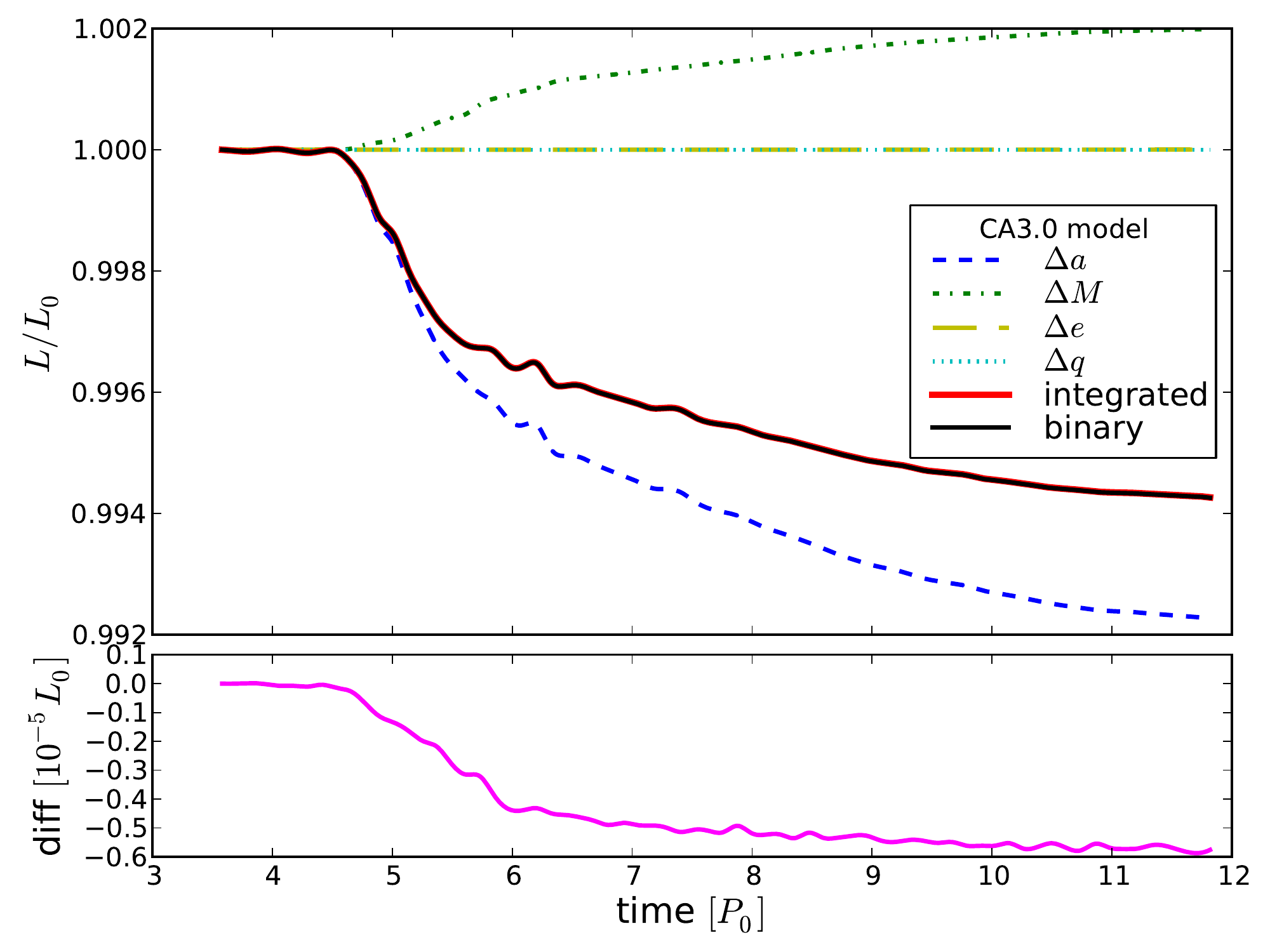}}
  \put(0,375){\includegraphics[width=0.3333\textwidth]{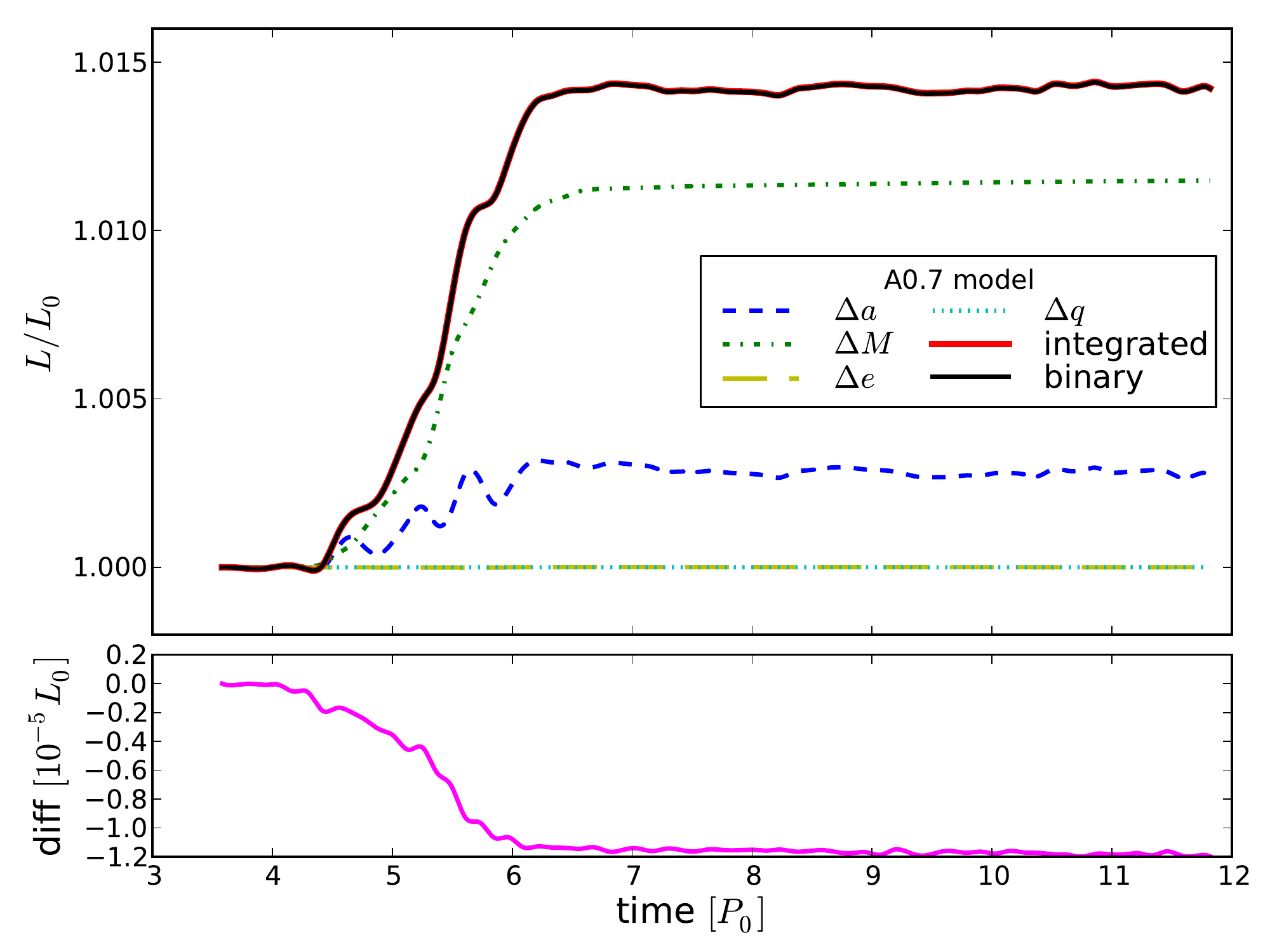}}
  \put(167,375){\includegraphics[width=0.3333\textwidth]{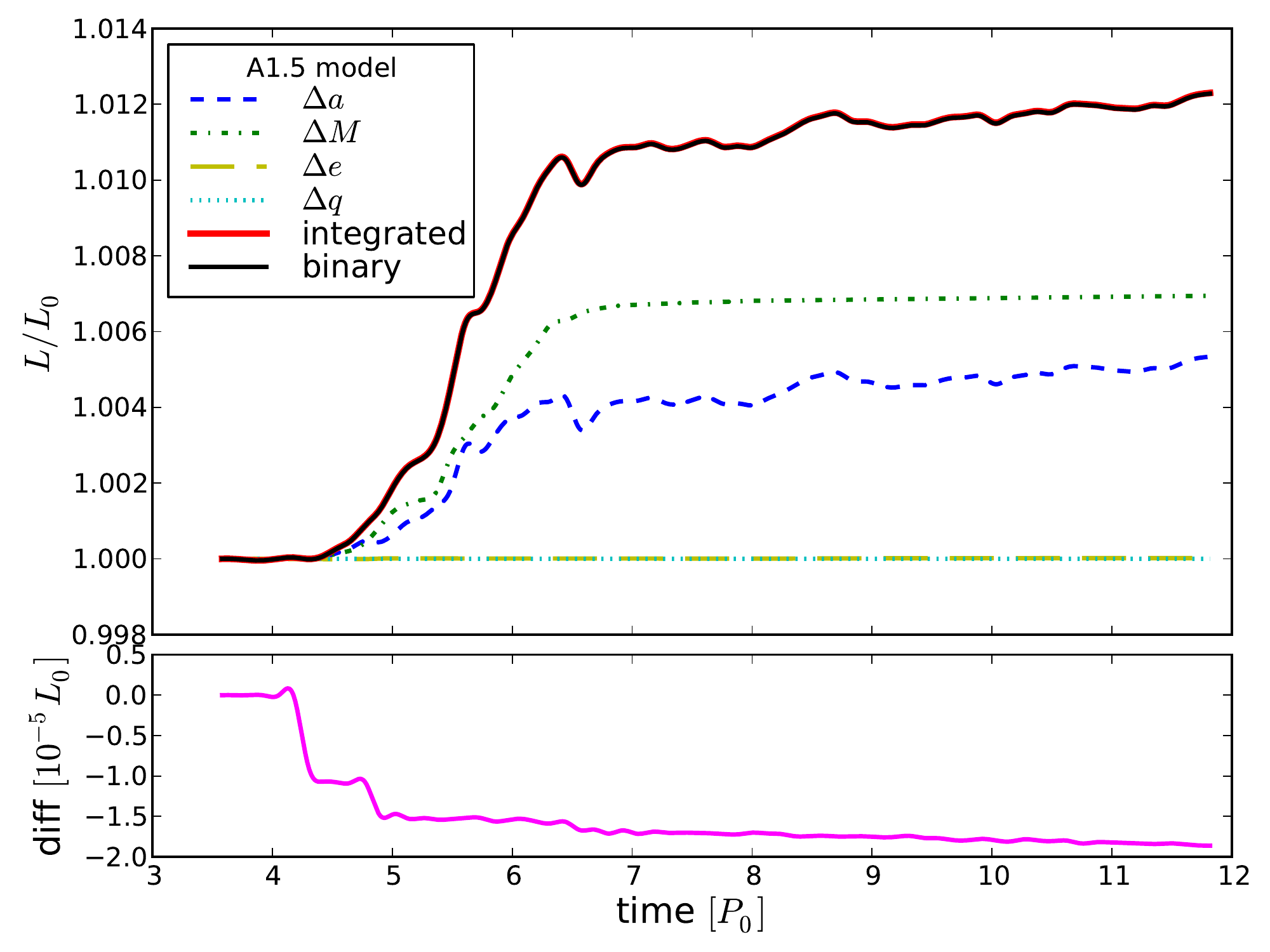}}
  \put(333,375){\includegraphics[width=0.3333\textwidth]{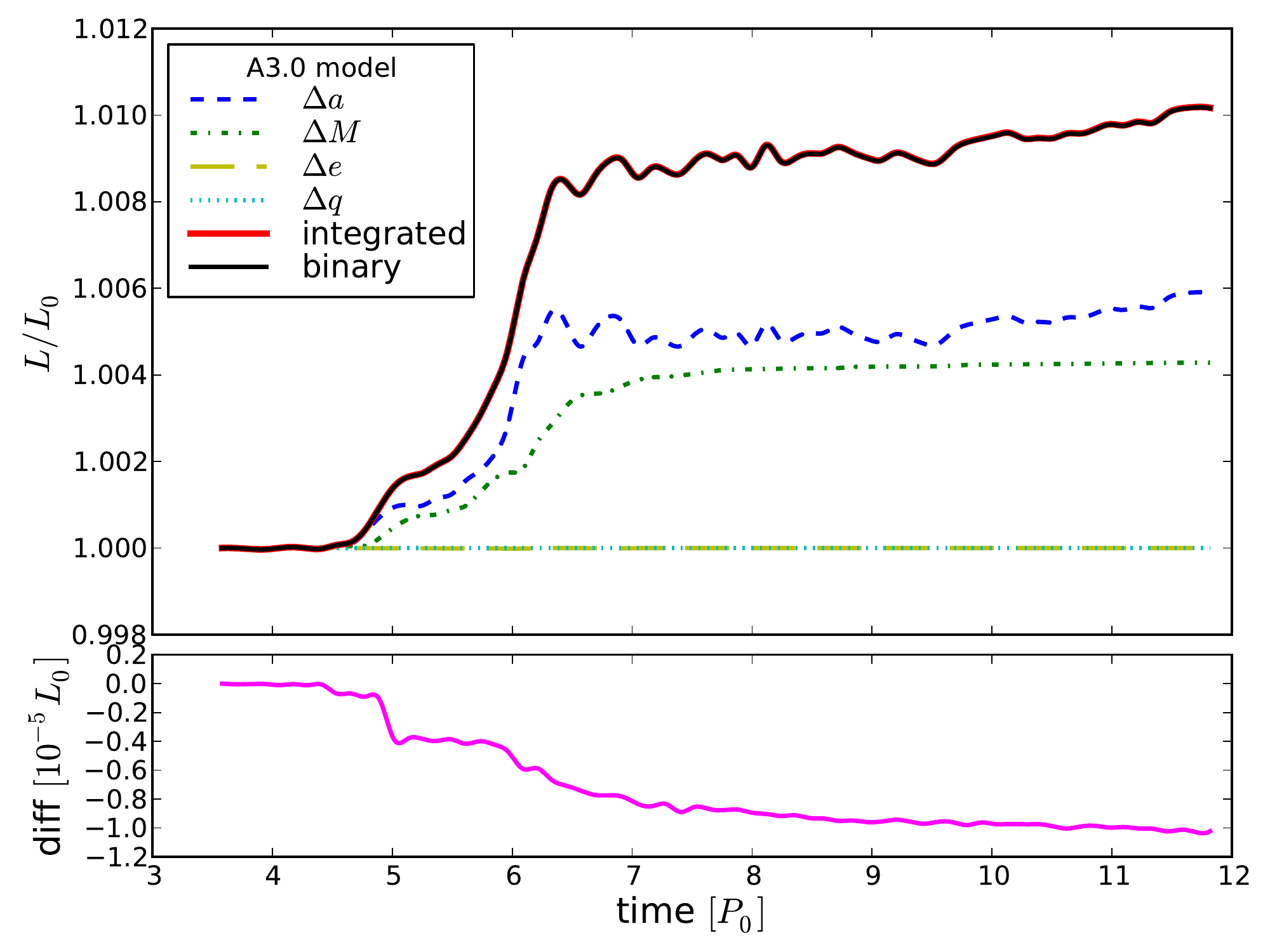}}
 \end{picture}
\caption{{\em Upper plots:} Decomposition of the binary total angular momentum following eq.~\ref{deltaL} (non-solid lines), the sum of all components (solid red line) and the angular momentum of the binary computed directly from the snapshots (solid black line). We increase the impact parameter from left to right, while the different inclinations are shown from top to bottom, as indicated on each legend.  Notice that the $\Delta{q}$ and $\Delta{e}$ components of the angular momentum are indistinguishable from unity in these plots. 
{\em Lower plots:} Difference between the momentum measured from the snapshots and the value recovered from the contribution of all components.}
\label{Ldecomp}
\end{figure*}

In all of our models, the binary remains approximately equal-mass and circular, $(1-q) \ll 1$ and $e \ll 1$ (see Figs.~\ref{orbit1} and~\ref{orbit2}).
This means that the factors in front of $\Delta{q}$ and $\Delta{e}$ in eq.~\ref{deltaL} are negligible compared to the $3/2$ and $1/2$ in front of $\Delta{M}/M$ and $\Delta{a}/a$, respectively.  We can therefore calculate approximately the total change in angular momentum based on the change in $a$ and $M$ only,
\begin{equation}
  \frac{\Delta{L_{\rm bin}}}{L_{\rm bin}}\approx\frac{3}{2}\frac{\Delta{M}}{M}+\frac{1}{2}\frac{\Delta{a}}{a}.
  \label{deltaL2}
\end{equation}
This approximation is confirmed by the upper panels of Fig.~\ref{Ldecomp}, where we show the decomposition of the binary total angular momentum into its individual components. Here we observe that the contributions due to the evolution in mass ratio (dotted red lines) and eccentricity (long dashed cyan lines) are negligible compared to those due to the change in total mass (dotted-dashed green lines) and semimajor axis (dashed blue lines). The solid lines in each panel represent the binary angular momentum, black is the value measured directly form the simulations, while red is the one recovered by integrating the individual components of equation \eqref{deltaL}. Note that the black and red solid lines are indistinguishable from each other on this scale. In order to establish the resolution of our simulations, we compute the difference between these two lines, shown in the lower panels of Fig.~\ref{Ldecomp}.
The differences displayed here are usually within the range $\sim 10^{-5}-10^{-4}L_{0}$, which {implies that the binary orbital elements are related to the angular momentum through our first order expansion (eq.~\ref{deltaL}) very accurately.} {These results, together with what we show in Fig.~\ref{ang_mom}, confirm that the evolution of our systems is dominated by the physical interaction with the gas rather than numerical noise, at least during the prompt accretion phase.}

\subsection{Analytical estimate of the binary evolution}

In order to estimate the expected evolution of $L$ and $a$, we develop a simple analytical model based on the exchange of angular momentum through accretion only (i.e., ignoring the non-accreted gas). The initial angular momentum of the gas is determined by the initial conditions that we impose for the cloud, described in detail in \paperI. Then, the average angular momentum of a portion of gas with mass $M_{\rm gas}$ is
\begin{equation}
L_{\rm gas}=d M_{\rm gas} v_{\rm ini}\sin\theta_{\rm vel},
\end{equation}
where $d=15a$ is the initial distance of the cloud, $v_{\rm ini}=0.25\sqrt{GM/a}$ is its initial orbital velocity, and $\theta_{\rm vel}$ is the angle between its velocity vector and the binary plane. Replacing the different values we obtain
\begin{equation}
  L_{\rm gas}=\alpha M_{\rm gas}\sqrt{GMa},
  \label{eq_Lgas}
\end{equation}
where 
\begin{equation}
\alpha=\frac{15}{4} \sin\theta_{\rm vel} \approx 0.7,1.1,1.4,1.8
\label{alpha_est}
\end{equation} 
for increasing impact parameter. Expressed in terms of the initial angular momentum of the binary $L_{0}=\mu\sqrt{GMa}$
\begin{equation}
  \frac{L_{\rm gas}}{L_{0}}=\alpha\frac{M_{\rm gas}}{\mu}=4\alpha\frac{M_{\rm gas}}{M}.
\label{Lgas}
\end{equation}

In order to check if this approximation is correct we compute the equivalent $\alpha$ of the cloud as a whole at the beginning of the simulation. We obtain
\begin{equation}
\alpha_{\rm sim}=0.73,1.14,1.55,1.9
\end{equation}
which are very close to the values derived from the initial conditions, as expected.  The small differences are due to the initial random turbulent velocity field.

\subsubsection{Angular momentum evolution}

We take the simple assumption that each accreted gas particle brings its initial angular momentum to the MBHB. For the aligned (A) and counter-aligned (CA) cases, where the majority of the accretion occurs on the same plane to that of the binary, we simply add or subtract the angular momentum estimated using equation \eqref{Lgas} by considering the appropriate $M_{\rm gas}$ (i.e., the total mass accreted in each case, see Table~\ref{delta}). The values obtained using this approximation and the comparison with the actual values measured from the simulations are shown in Table~\ref{deltaL_par}.

\begin{table*}
\centering
\caption{Total evolution of the angular momentum magnitude ($\Delta L$) and semimajor axis ($\Delta a$) for the A and CA models. The subscript `est' corresponds to a value estimated using our initial simple model, while `corr' is the  corrected estimation using the appropriate $\alpha$ and including the slingshot, and `meas' means that is measured directly from the simulations.} 
\begin{tabular}{l*{9}{c}r}
\hline\hline
 Model & $\Delta L_{\rm meas}$  & $\Delta L_{\rm est}$  & $\frac{\Delta L_{\rm est}}{\Delta L_{\rm meas}}$ & $\Delta L_{\rm corr}$ & $\frac{\Delta L_{\rm corr}}{\Delta L_{\rm meas}}$ & $\Delta a_{\rm meas}$ & $\Delta a_{\rm est}$ & $\frac{\Delta a_{\rm est}}{\Delta a_{\rm meas}}$ & $\Delta a_{\rm corr}$ & $\frac{\Delta a_{\rm corr}}{\Delta a_{\rm meas}}$\\
 & ($L_{0}$) & ($L_{0}$) & & ($L_{0}$) & & ($a_{0}$) & ($a_{0}$) & & ($a_{0}$) &\\
\hline
A0.7 & 0.0138 & 0.0211 & 1.53 & 0.0145 & 1.06 & 0.0053 & 0.0196 & 3.70 & 0.0064 & 1.20 \\
A1.5 & 0.0106 & 0.0199 & 1.88 & 0.0114 & 1.06 & 0.0085 & 0.0263 & 3.09 & 0.0092 & 1.08 \\
A3.0 & 0.0086 & 0.0156 & 1.81 & 0.0105 & 1.21 & 0.0098 & 0.0228 & 2.33 & 0.0127 & 1.29 \\ 
A6.0 & 0.0014 & 0.0042 & 2.98 & 0.0013 & 0.93 & 0.0016 & 0.0066 & 4.13 & 0.0019 & 1.19\\ 
CA0.7 & -0.0246 & -0.0241 & 0.98 & -0.0239 & 0.97 & -0.0731 & -0.0741 & 1.01 & -0.0737 & 1.01\\
CA1.5 & -0.0253 & -0.0298 & 1.18 & -0.0261 & 1.03 & -0.0699 & -0.0800 & 1.14 & -0.0725 & 1.04\\
CA3.0 & -0.0053 & -0.0071 & 1.34 & -0.0052 & 0.98 & -0.0148 & -0.0180 & 1.22 & -0.0144 &  0.97\\ \hline
\hline
\end{tabular}
\label{deltaL_par}
\end{table*}

Comparing to the observed $L$ evolution, those numbers are a factor of $\simlt 3$ too large in the A cases and a $\le 40\%$ overestimate in the CA cases. 
The first source for this discrepancy could be the non-accreted material, which interacts with the MBHB taking away some extra angular momentum. This remaining gas is clearly seen in the form of minidiscs and circumbinary discs (see Fig.~3 in \paperI). In order to compute the amount of angular momentum acquired by the non-accreted gas, we identify directly from the simulations the remaining particles after the first passage of the cloud. Only for these particles we then compute the angular momentum difference respect to the initial state, which comes from the interaction with the binary. For the A models we obtain
\begin{equation}
\left(\frac{\Delta L_{\rm out}}{L_0}\right)_{\rm A}\approx 0.0021,\, 0.0024,\, 0.0011,\, 0.0016,
\end{equation}
for increasing impact parameters. Similarly, for the CA models,
\begin{equation}
\left(\frac{\Delta L_{\rm out}}{L_0}\right)_{\rm CA}\approx 0.0029,\, 0.0036,\, 0.0031.
\end{equation}
These values are around one order of magnitude too small to explain our overestimate of the angular momentum evolution.

The second effect that could account for these differences is that the accreted particles might have in average {less} angular momentum that their non accreted counterparts. In order to compute this, we estimate the average $\alpha$ of the accreted particles, which is an indication of their angular momentum budget. For the A models we obtain
\begin{equation}
\alpha_{\rm accr,A}=0.55,0.76,1.04, 1.25
\end{equation}
while for the CA models
\begin{equation}
\alpha_{\rm accr,CA}=0.61,0.83,0.98.
\end{equation}
All these values are smaller than the ones we use based on the initial conditions (equation \ref{alpha_est}), which implies that we are overestimating the average angular momentum of the gas accreted by the binary. 
This is because the gas with larger angular momentum will typically have orbits with a periapsis further away from the MBHB, avoiding being captured.

We estimate again the angular momentum change using a modified version of equation \eqref{Lgas}, where we implement the corrections we previously described as follows:
\begin{equation}
\frac{\Delta L_{\rm bin}}{L_{0}}=4\alpha_{\rm accr}\frac{\Delta M}{M}-\frac{\Delta L_{\rm out}}{L_0}.
\label{Lcorr}
\end{equation}
The values obtained with this equation are shown in Table~\ref{deltaL_par}. These numbers are now in remarkable agreement with the actual binary evolution. The largest difference we obtain with this approximation is 20\% for the A3.0 model -- for all the other cases the discrepancies are not larger than 7\%.

\begin{table*}
\centering
\caption{Total evolution of the angular momentum inclination angle ($\Delta\theta$), semimajor axis ($\Delta a$) for the perpendicular models. The definition of the subscripts is the same as Table~\ref{deltaL_par}.} 
\begin{tabular}{l*{7}{c}r}
\hline\hline
 Model & $\Delta\theta_{\rm meas}$ & $\Delta\theta_{\rm est}$  & $\Delta\theta_{\rm corr}$ & $\Delta a_{\rm meas}$ & $\Delta a_{\rm est}$  & $\Delta a_{\rm meas}/\Delta a_{\rm meas}$ &  $\Delta a_{\rm corr}$  & $\Delta a_{\rm corr}/\Delta a_{\rm meas}$\\
 &(deg)&(deg)& (deg) & ($a_0$) & ($a_0$) & & ($a_0$)& \\
\hline
PE0.7 & 0.88 & 1.11 & 0.81 & -0.0271 & -0.0208 & 0.77 & -0.0276 & 1.02 \\ 
PE1.5 & 0.67 & 0.84 & 0.54 & -0.0255 & -0.0101 & 0.40 & -0.0258 & 1.01 \\ 
PE3.0 & 0.23 & 0.26 & 0.18 & -0.0076 & -0.0024 & 0.32 & -0.0077 & 1.01 \\ 
PF0.7 & 0.53 & 0.87 & 0.53 & -0.0239 & -0.0162 & 0.68 & -0.0242 & 1.01 \\ 
PF1.5 & 0.49 & 0.71 & 0.46 & -0.0201 & -0.0084 & 0.42 & -0.0204 & 1.02 \\ 
PF3.0 & 0.15 & 0.24 & 0.17 & -0.0080 & -0.0028 & 0.28 & -0.0079 & 0.99 \\ 
\hline
\hline
\end{tabular}
\label{deltaL_perp}
\end{table*}

\begin{figure}
\centering
\includegraphics[width=0.46\textwidth]{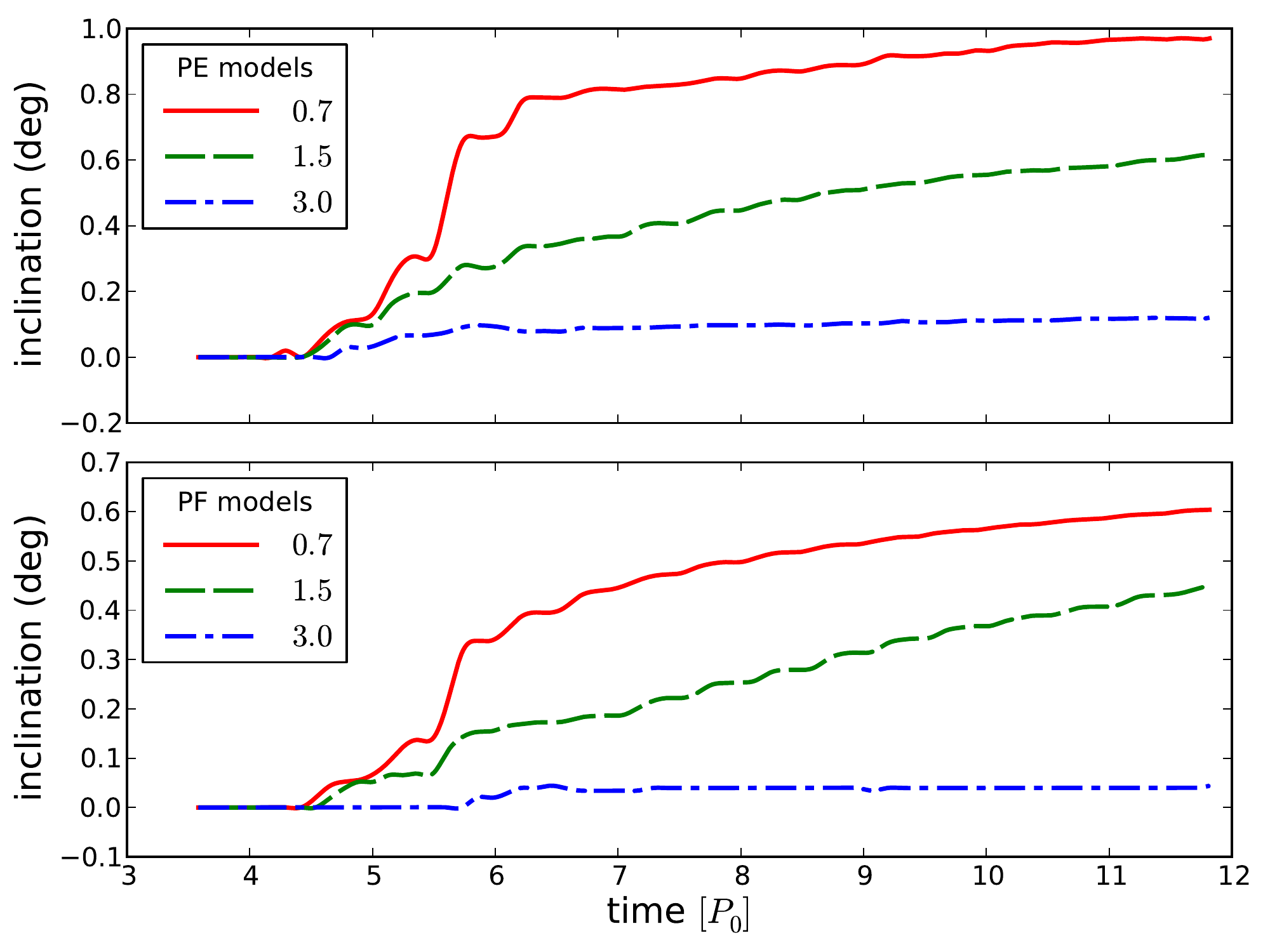}
\caption{Evolution of the binary inclination with respect to its initial orientation for the perpendicular models, edge-on (upper panel) and face-on (lower panel). The colours represent the different impact parameters, as indicated in the legend.}
\label{perp_angle}
\end{figure}

In the case of the perpendicular configurations, the main effect of the accreted gas will be to tilt the binary in the direction of the orbital angular momentum of the cloud, because the typical gas velocity will be perpendicular to that of the MBHB. This is clearly seen in Fig.~\ref{perp_angle}, where we show the time evolution of the angular momentum inclination angle respect to its initial orientation. We can quantify the tilt angle $\Delta\theta$ by applying the same approximation as before:
\begin{equation}
 \sin{(\Delta\theta)}=\frac{\Delta{L_{\rm bin}}}{L_{\rm bin}}=4\alpha\frac{\Delta{M}}{M}.
 \label{theta}
\end{equation}
The values estimated using this expression are shown in Table~\ref{deltaL_perp}, and are an overestimate of the measured ones. Similar as the other configurations, this is due to the mean $\alpha$ value of the accreted particles. For the PE models those are
\begin{equation}
\alpha_{\rm accr,PE}=0.51,0.71,0.98,
\end{equation}
while for the PF models we get
\begin{equation}
  \alpha_{\rm accr,PF}=0.44,0.71,0.98,
\end{equation}
which appear to be consistent with the overestimations for both configurations.
We estimate the inclination angles with equation~\eqref{theta}, but using the values of $\alpha_{\rm accr}$, and present them in Table~\ref{deltaL_perp} ($\Delta\theta_{\rm corr}$). With this correction, we obtain values much closer to the measured ones, especially for the PF configurations. For the PE models, the corrected values are only a slight underestimate of the simulated ones.

In summary, using a simple analytical model based only on the angular momentum exchange between the accreted material and the binary we are able to reproduce all the trends for the total evolution of the angular momentum vector (either magnitude or inclination). 
This confirms our hypothesis that the transient evolution of a binary during the near-radial infall of gaseous clouds is dominated by the accretion onto the MBHB.

\subsubsection{Semimajor axis evolution}

We can now use equation \eqref{deltaL2} to estimate, in the same manner, the evolution of the binary semimajor axis, and compare it to what it is found in the simulation.

We start with the prograde and retrograde cases. If we equate the RHS of equation~\eqref{deltaL2} to the RHS of equation~\eqref{Lgas}, and we keep in mind that we should add the latter to the $L$ budget in the A case and subtract it in the CA case, we get
\begin{equation}
  \left(\frac{\Delta{a}}{a}\right)_{\rm A, CA}=-8\left(\frac{3}{8} \mp \alpha\right)\frac{\Delta{M}}{M},
  \label{deltaa_ACA}
\end{equation}
where the `$-$' sign corresponds to the A case and the `$+$' to the CA case. Comparing these estimates with what was measured in the simulation (Table~\ref{deltaL_perp}), we see that we overestimate the binary shrinking in the corotating case, but we get close to the observed values in the retrograde case. All the discrepancies are consistent with the overestimation of the angular momentum change. If we compute $\Delta a$ using an appropriate angular momentum change (i.e.,~$\Delta L_{\rm corr}$ from equation~\ref{Lcorr}), we obtain values much closer to the measured ones, typically within a few percent. These values are presented in Table~\ref{deltaL_par} denoted with $\Delta a_{\rm corr}$.

We can apply the same reasoning to the PE and PF simulations, assuming now that the $L_{\rm gas}$ brought by the accreted gas is perpendicular to $L_{\rm bin}$, and so it does not change its magnitude but only its direction. We therefore have:

\begin{equation}
 0\approx\frac{3}{2}\frac{\Delta{M}}{M}+\frac{1}{2}\frac{\Delta{a}}{a},
  \label{deltaL2P}
\end{equation}
simply meaning
\begin{equation}
  \frac{\Delta{a}}{a}=-3\frac{\Delta{M}}{M},
  \label{deltaa_PEPF}
\end{equation}
regardless of the impact parameter. The estimations for the semimajor axis evolution in the perpendicular models are shown in Table~\ref{deltaL_perp}. These numbers show that with our simple model we underestimate the evolution. A likely reason is that the binary slingshots away some of the non-accreted material, preferentially in its direction of motion. The gas therefore takes away further angular momentum from the binary increasing the shrinking which is not included with our simplifications. This effect increases with the impact parameter, as less mass is accreted and much of it is subject to this slingshot. 
To compute the angular momentum taken away by the remaining gas, we use the fact that the angular momentum magnitude does change in the perpendicular configurations (see Fig.~\ref{ang_mom}). We assume that this total change comes from the slingshot.
Similar to what we did in the A and CA models, we include this $\Delta L_{\rm out}$ in the LHS of equation~\eqref{deltaL2P} to estimate the corrected semimajor axis evolution. We show these estimations in Table~\ref{deltaL_perp}. Note the remarkable accuracy with what we can now reproduce the measured values from the simulations, with differences no larger than 2\%.

In summary, as for the angular momentum evolution, with our analytical model considering only the accreted particles we are able to reproduce the trends of the semimajor axis evolution for all configurations. All the discrepancies found with respect to the simulated values are consistent with the simplifications we impose. When we implement the effect of the non-accreted material and the appropriate angular momentum budget of the accreted counterparts, we can reproduce the values with much better accuracy.

\subsubsection{Caveat: outflows}

The peak accretion rates found in our simulations, when scaled to physical units, are usually highly super-Eddington. 
{In \paperI~we found that the accretion rates vary between $\approx 1-50\,\dot M_{\rm edd}$ for a $10^6M_\odot$ binary during the prompt accretion phase. Naturally this will depend on the orbital configuration, with the smaller impact parameters having the highest peaks, but typically the accretion rates will be super-Eddington during the first few orbits.} 
Certainly the material could be accreted through a slim disc \citep[e.g.,][]{Abramowicz1988}. However, if we take the conservative approach that accretion is capped to the Eddington rate, the rest of the material will likely be ejected by radiation-pressure driven outflows \citep[e.g.,][]{KingPounds2003}. Therefore, part (possibly most) of the material which is accreted in our simulation, will be instead ejected in an outflow, making our description of the dynamics inaccurate. To estimate how much this can affect the MBHB dynamics we take for simplicity the (reasonable) working hypothesis of an isotropic outflow {\it in each of the BHs' reference frame}. It is easy to show that if a mass $\Delta{M_{\rm out}}$ is ejected, the angular momentum loss for the binary (assumed to be equal mass, circular) is $\Delta{L_{\rm out}}=(\Delta{M_{\rm out}}/4)\sqrt{GMa}$, that can be also written as $\Delta{L_{\rm out}}/L=\Delta{M_{\rm out}}/M$. This is the same as in equation (\ref{Lgas}), with $\alpha=0.25$. So even if all the captured mass is ejected in an outflow instead of being accreted, this amounts to including a factor 0.25 into the parenthesis of equation (\ref{deltaa_ACA}), which does not change the evolution significantly. For the PE and PF cases, equation (\ref{deltaa_PEPF}) becomes $\Delta{a}/a=-2\Delta{M}/M$, i.e. the shrinkage of $a$ is 33\% less. Hence, even if all the captured mass is ejected in an outflow instead of being accreted, the evolution of the semimajor axis is only mildly affected.
{This is confirmed by the results shown in Appendix~\ref{sec:Ap1},  where we re-simulate the A0.7 model with different accretion radii. By shrinking the accretion radius by a factor of eight, the mass accretion decreses by about 25\%, however there seems to be no correlation with the semimajor-axis evolution. This is particularly true for all runs with $r_{\rm sink}\leq 0.1$, for which the binary evolution is essentially identical. This is because the relevant exchange of angular momentum occurred during the capture of the material, which we can resolve quite well, instead of the accretion itself.}

Finally, radiation driven outflows could also affect the dynamics of the rest of the infalling cloud, inhibiting further accretion, but we cannot quantify this possible effect with our current approach.

\section{Application: long term evolution of binaries via accretion of discrete gas clouds}
\label{sec:application}

\begin{figure}
\centering 
\includegraphics[width=0.45\textwidth]{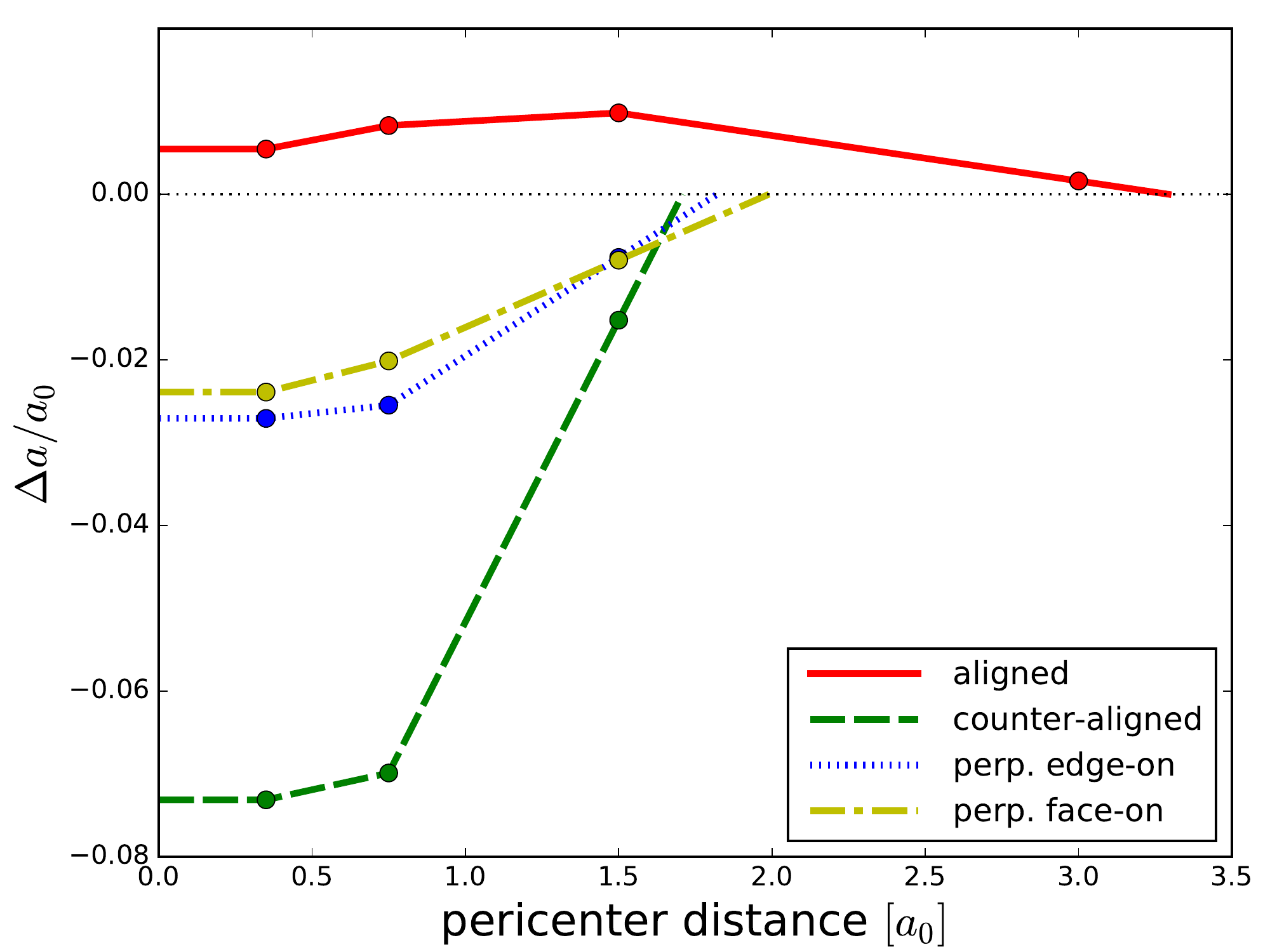}
\caption{Total change of the binary semimajor axis as a function of the cloud's pericentre distance. The filled circles are the values measured from the simulations, while the lines are the linear interpolation/extrapolation of those points. The extrapolation after the largest impact parameters define the maximum value of the pericentre distance for the different inclinations, where the evolution of the semimajor axis is zero (dotted black line). For smaller values than $0.35a$ we assume that the relative change of semi-major axis per event remains constant.}
\label{delta_a}
\end{figure}

We explore in this section the implications of our results for an evolutionary scenario in which a MBHB interacts with a sequence of gas clouds. In Fig.~\ref{delta_a} we show the total change of the binary semimajor axis as a function of the cloud's pericentre distance. From this figure we observe the different behaviour of $\Delta a$ for the Aligned orientation, still increasing for $r_{\rm p}=1.5a$, in contrast to the other inclinations for which the total shrinkage is approaching zero at that point. This stems from the larger capture cross-section of each MBH in corotating encounters due to the smaller velocity relative to the gas particles. This is the reason why we  run an additional simulation (A6.0) in the aligned case, as mentioned in Section~\ref{sec:simulation}.

Using the information shown in Fig.~\ref{delta_a} we construct a simple model for the evolution of a MBHB accreting clouds from different directions and with different impact parameters. If we assume a distribution of similar clouds, as the MBHB semimajor axis changes, the relative size of the cloud increases with respect to the MBHB. Therefore, our extrapolation is based on the ansatz that the important quantity is the total angular momentum of the cloud (i.e., its impact parameter) and not the spread around the mean (i.e., the relative size of the cloud). This is not necessarily true when the binary becomes more compact and then the size of the cloud with respect to it becomes bigger. However, this approximation is more accurate when the gravitational focusing of the gas is important,  e.g. when the cooling is efficient. 

\subsection{Monte Carlo evolution}
\label{sec:montecarlo}

In order to compute the evolution of a binary we need to determine the orbital parameters of the approaching clouds. For a uniform number density, clouds approach the MBHB with an impact parameter $b$ following a distribution {with a probability density given by $P(b)\propto b\dd b$, i.e. increasing with the geometrical cross-section. We can use gravitational focusing to link the impact parameter with the pericenter distance ($r_{\rm p}$) as follows:
\begin{equation}
r_{\rm p}= \frac{v_\infty^2}{2GM_c}b^2,
\end{equation}
where we used the fact that the binary is significantly more massive than the individual clouds ($M_c\ll M$).}
{Differentiating this expression we obtain $dr_{\rm p}\propto b\dd b$, or simply $P(r_{\rm p})\propto \dd r_{\rm p}$}, which means that the pericentre distance is distributed uniformly between zero and some maximum value. We compute this maximum value for each inclination using the extrapolation shown in Fig.~\ref{delta_a}.

\begin{figure}
\centering
\includegraphics[width=0.45\textwidth]{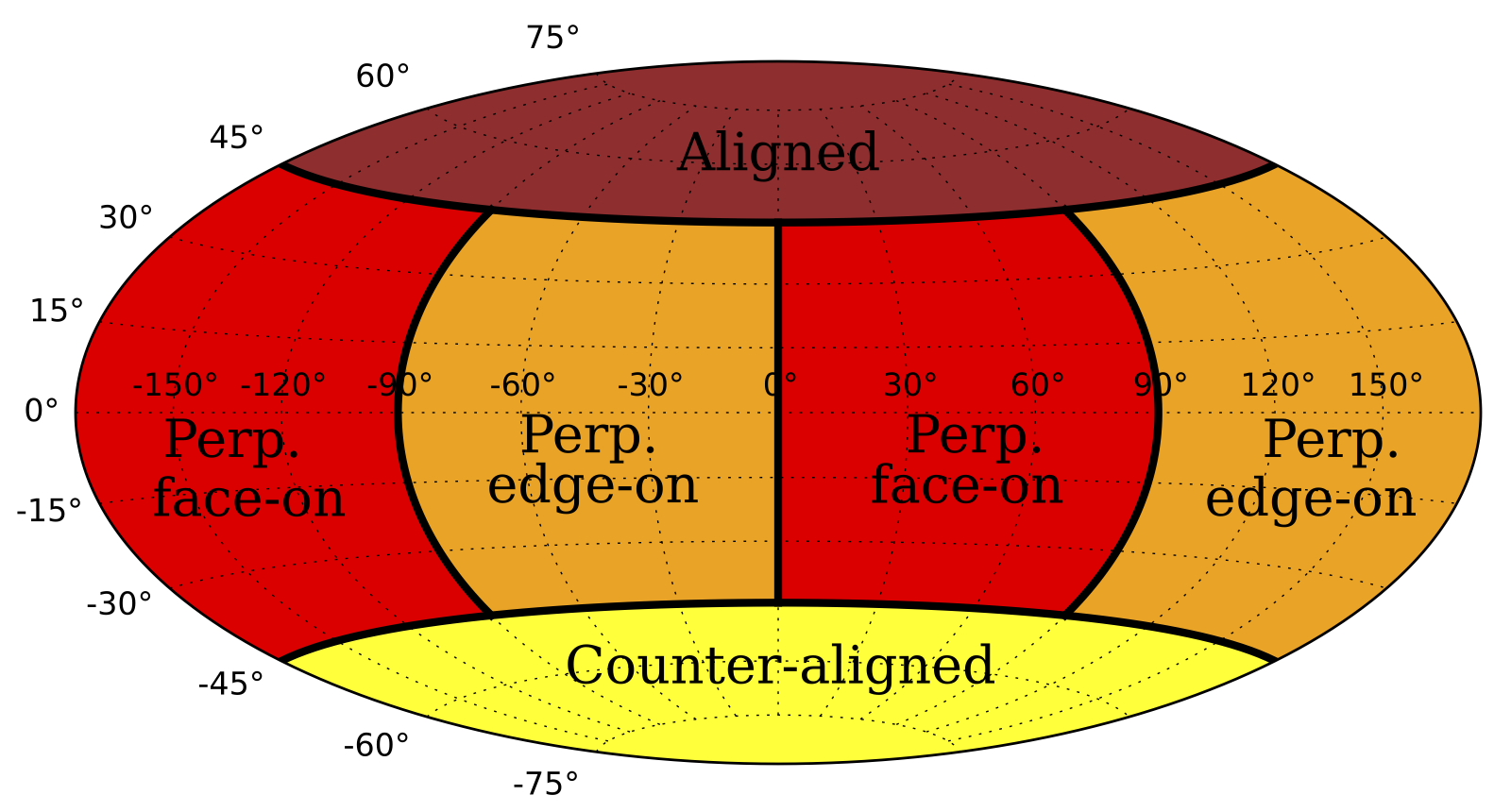}
\caption{Map of the four zones we define in the angular momentum direction distribution to determine the relative inclination of the cloud orbit with respect to the binary.}
\label{zones}
\end{figure}
\begin{figure}
\centering
\begin{picture}(230,390)
\put(0,0){\includegraphics[width=0.45\textwidth]{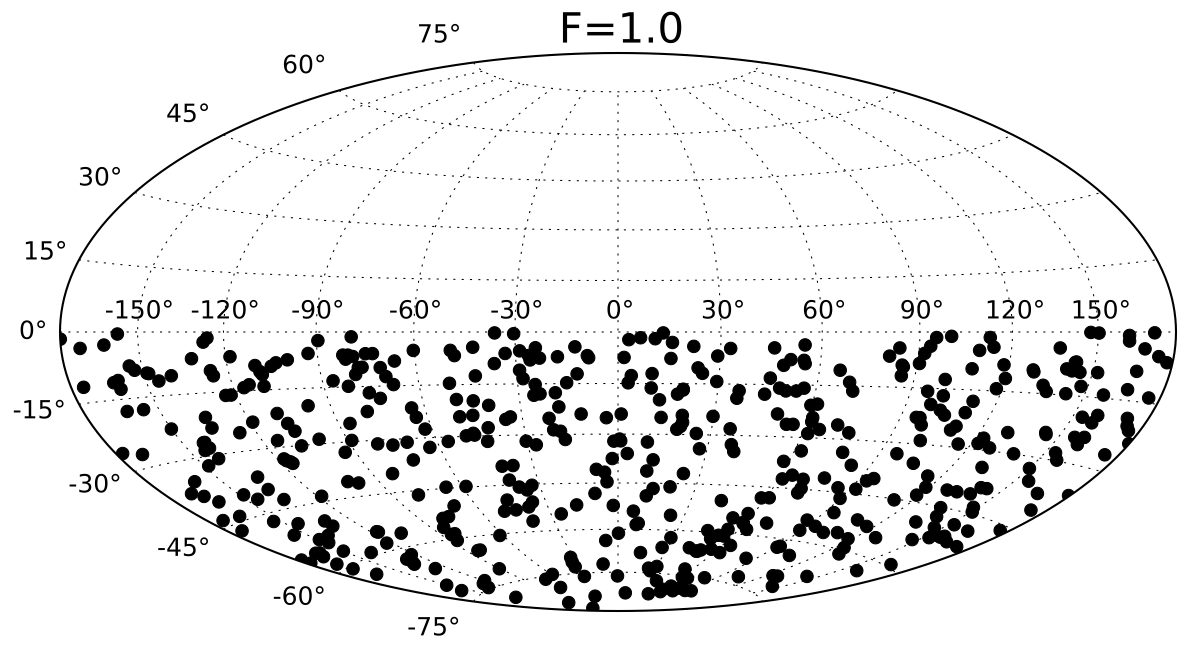}}
\put(0,130){\includegraphics[width=0.45\textwidth]{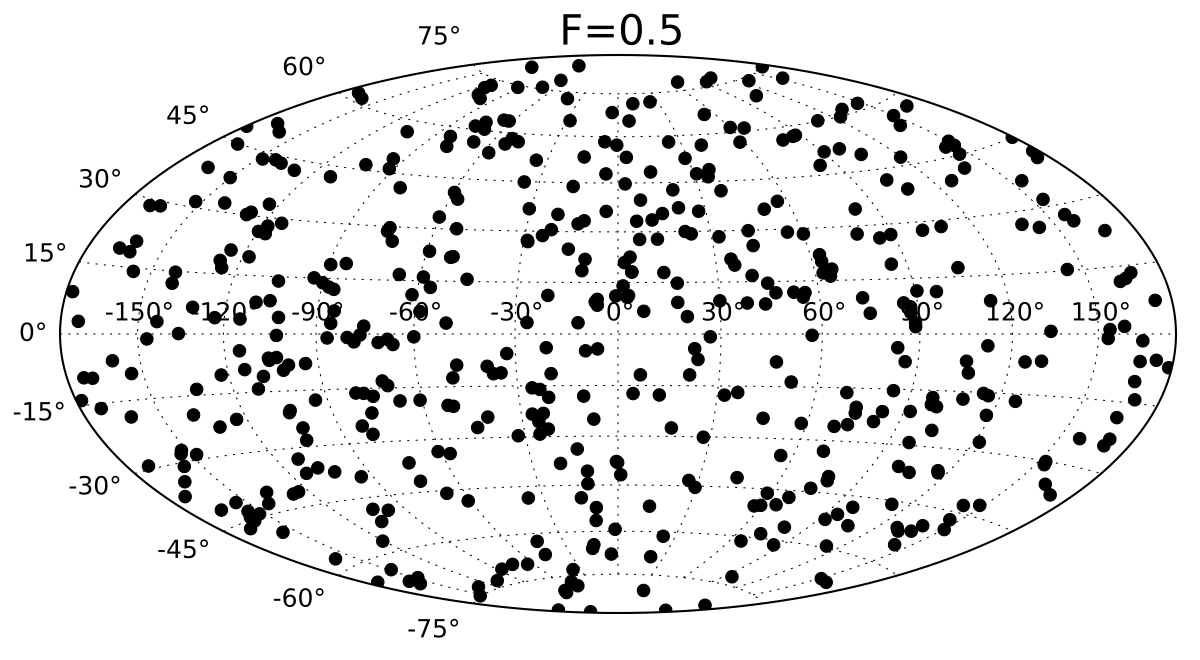}}
\put(0,260){\includegraphics[width=0.45\textwidth]{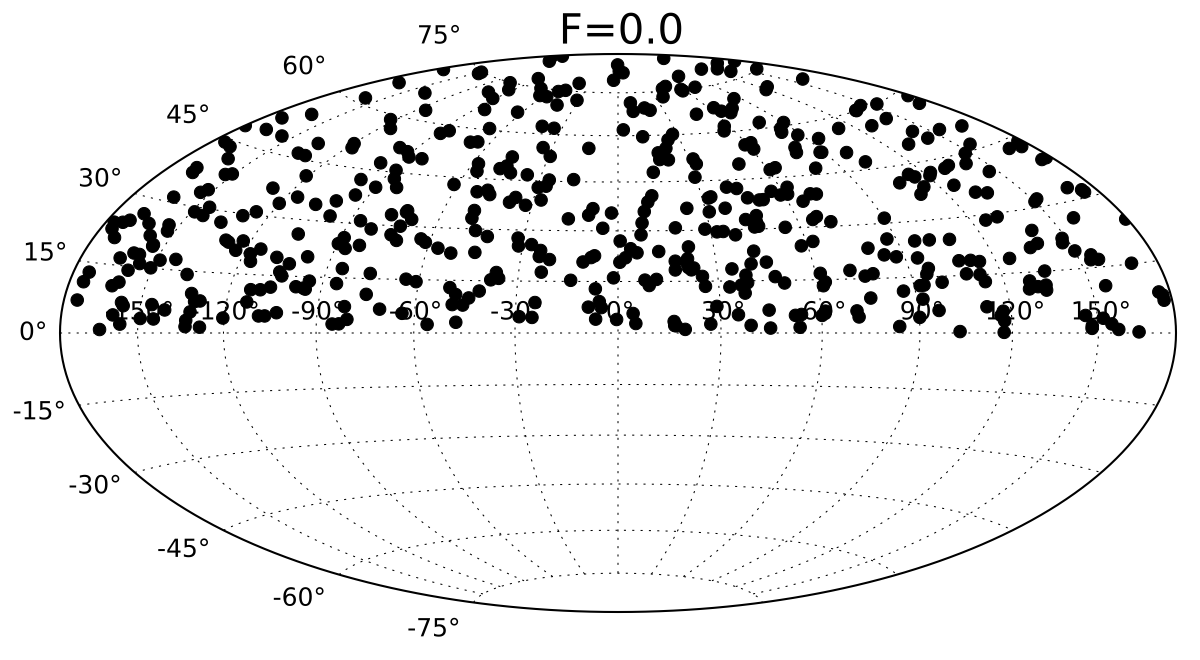}}
\end{picture}
\caption{Realisations of cloud angular momenta distributions for the different  adopted values of  $F$, as indicated at the top of each projection. The number $F$ represents the probability of having events in the `southern hemisphere'. The model $F=0.5$ is usually referred to as `chaotic accretion', where portions of gas are accreted from uniformly distributed directions around a MBH, while for $F=0.0$ all accretion events are prograde, and for $F=1.0$ they are retrograde. Figure adapted from \citet{Dotti2013}.}
\label{Fs}
\end{figure}

The impact parameter defines only the magnitude of the cloud angular momentum, not its direction. The cloud angular momentum ($\boldsymbol{L}_{\rm c}$) will point in some direction in the 2D sphere. We need to map this sphere into the four relative orientations we modelled (A, CA, PE and PF). We do so by dividing the 2D sphere in 4 different zones, each assigned to one of the orientations of our simulations (Fig.~\ref{zones}). If $\theta$ is the angle between $\boldsymbol{L}_{\rm c}$ and $\boldsymbol{L}$ (being $\boldsymbol{L}$ the MBHB angular momentum), we assign to the Aligned encounters the 2D region enclosed in $45^\circ<\theta<90^\circ$. Similarly, Counter-aligned events correspond to the region defined by $-45^\circ>\theta>-90^\circ$. The rest of the sphere maps into perpendicular encounters. In order to separate between edge-on and face-on, we note that by fixing $\boldsymbol{L}_{\rm c}$ we are still free to rotate the orbit by an azimuthal angle. Let $O$ be the origin of our coordinate system (corresponding to the binary CoM) and $\hat{x}$ a unit vector along the two planes defined by the MBHB and the cloud orbits. If $r_{\rm p}$ is the periapsis of the cloud orbit, the angle $\phi$ defined by $\hat{x}-O-r_{\rm p}$ can be used to discriminate between edge-on (PE) and face-on (PF) encounters. If $45^\circ<\phi<135^\circ$ or $225^\circ<\phi<315^\circ$, then the encounter is PF, otherwise it is PE. This translates into perpendicular encounters being evenly distributed between PE and PF, as we schematically represent in Fig.~\ref{zones}. Furthermore,  Fig.~\ref{delta_a} shows that the difference between PE and PF is minimal in terms of semimajor axis evolution. 
{In practice, with this procedure we are assigning a fixed probability to each of the four inclinations. This will be given by the fraction of the solid angle that zone subtends in the sky of the binary.}

With the mapping in hand, we just need to determine the distribution of $\boldsymbol{L}_{\rm c}$. We set the $z$ axis such as $\boldsymbol{L}=(0,0,L_z)$, with $L_z>0$, and define $F$ to be the fraction of events with $L_{{\rm c},z}<0$. Therefore, following \cite{Dotti2013}, $F$ represents the probability of having clouds coming from the southern hemisphere, where the northern hemisphere is defined by the direction of the $z$ axis (see Fig.~\ref{Fs}). Besides the constraint imposed by $F$, the events are assumed to be isotropic.  Therefore, when $F=0.5$ the events are uniformly distributed over the whole sphere around the binary \citep[known as `chaotic accretion',][]{KP06}, while for $F=0.0$ and $F=1.0$ the events are uniformly distributed over the northern and southern hemisphere, respectively. Note, however that these latter cases are different from coherent aligned/anti-aligned accretion, which would imply all $\boldsymbol{L}_{\rm c}$ along the $z$ axis. \citet{Sesana2014} linked these distributions to the morphological and kinematical properties of the host galaxies in order to explain the spin measurements of MBHs. Disc galaxies, where the gas dynamics is dominated by rotational velocity, will produce mainly coherent accretion events (prograde or retrograde depending on the binary orientation); while bulge galaxies, dominated by velocity dispersion, will tend to produce more uniformly distributed events.

\begin{figure*}
\centering 
\includegraphics[width=0.8\textwidth]{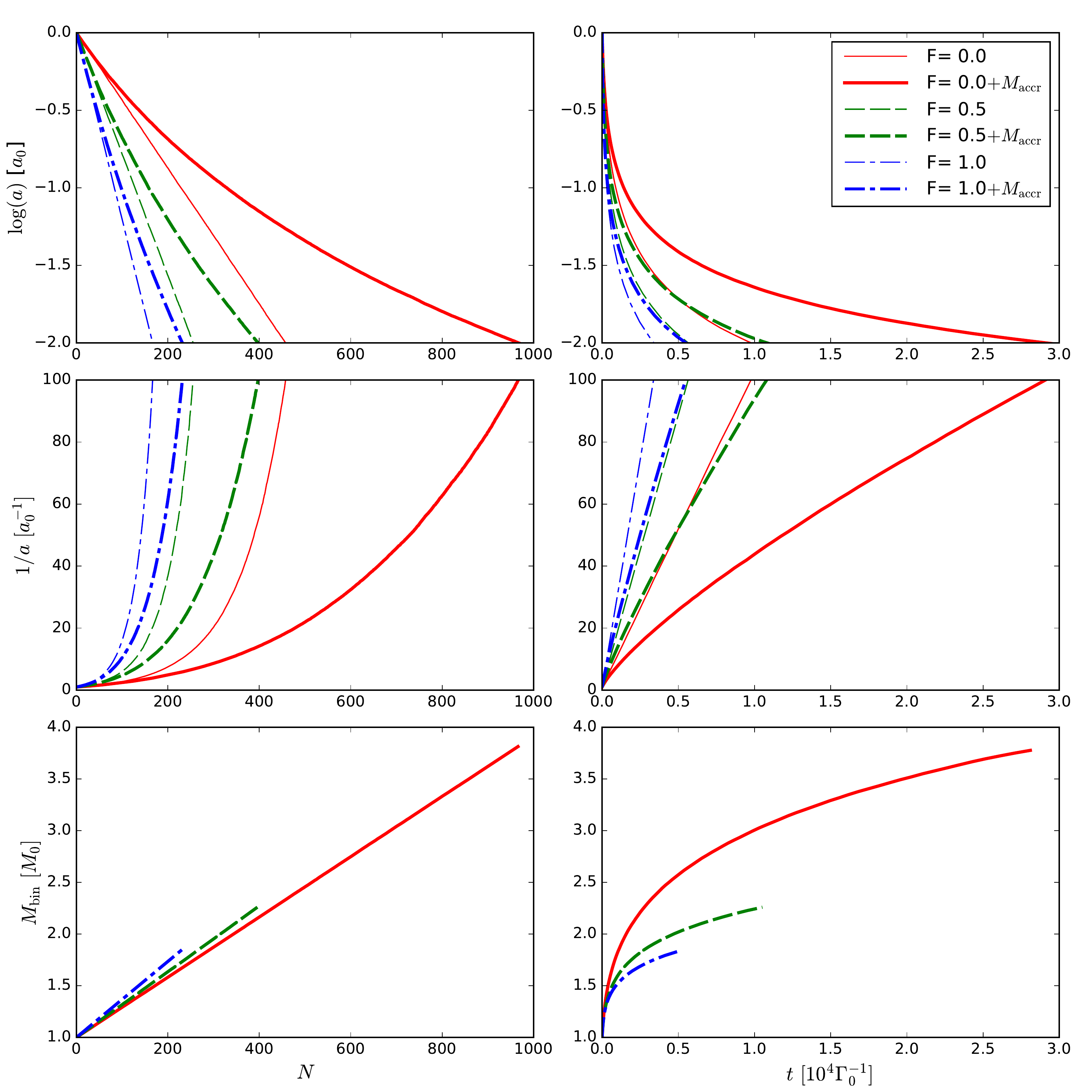}
\caption{Evolution of the binary as a function of the number of cloud encounters (left column) and as a function of time (right column). From top to bottom: semimajor axis, inverse of the semimajor axis, and binary mass. The different lines represent models with different  distributions of clouds. The thick lines show models that consider the evolution of the binary mass, while  the thin lines correspond to models that do not consider this evolution. {Each line is the result of averaging 1000 different Monte Carlo runs.}}
\label{sim_events}
\end{figure*}

Finally, using the distributions described above, we generate Monte Carlo populations of clouds with different levels of anisotropy (i.e., different $F$ values) interacting with the binary and evolving its semimajor axis according to their pericentre distance and relative orientation, as given by the curves shown in Fig.~\ref{delta_a}. The results of these models are shown in Fig.~\ref{sim_events}. {In order to suppress the stochasticity of a single run, each line is the average of 1000 Monte Carlo realisations.} On the left panel we show the evolution of the semimajor axis as a function of the number of encounters for the different distributions. To translate this into a temporal evolution we set an initial {rate ($\Gamma_0$)} at which we expect the clouds will interact with the binary. As the binary shrinks the encounters will be less frequent because of the decrease on the binary cross section. This effect can be simulated by adapting the timescale at each encounter as
\begin{equation}
\Delta T=\Gamma_0^{-1}\left(\frac{a_0}{a}\right),
\label{Tenc}
\end{equation}
which becomes longer as the semimajor axis decreases. Using this characteristic timescale we draw the cloud arrival times from a Poisson distribution. On the right panels of Fig.~\ref{sim_events} we show the results obtained for the evolution of the semimajor axis as a function of time for the different distributions. 


So far, we have not included in our description the growth of the MBHB mass. However, in order to evolve significantly its semimajor axis, the MBHB needs to interact with at least a few hundred gas clouds (see thin lines in the upper left panel of Fig.~\ref{sim_events}), which, in the long run, will imply an accreted mass comparable to the binary initial mass. We therefore need to include mass accretion in our model. The net effect will be a slowdown of the shrinking process: as we add mass to the binary, the mass ratio of the interacting clouds to the MBHB decreases, meaning that each accretion episode becomes progressively less effective in affecting the binary orbital elements. We therefore  include the accreted mass on to the binary after each encounter as follows. Using the same method we used to evolve the semimajor axis, we compute the total mass change according to each cloud's pericentre distance and relative orientation, as given by  the values measured from the simulations (see Fig.~5 of \paperI) and their extrapolation. Therefore, each accretion episode causes a change in both the MBHB semimajor axis and its total mass. In Fig.~\ref{sim_events} we show the evolution of the MBHB both when the mass growth is ignored (thin lines) and when it is properly taken into account (thick lines). As expected, the evolution of the binary orbit slows down in the latter case.

It is important to mention that even though perpendicular encounters change the MBHB inclination angle (see Fig.~\ref{perp_angle}), we do not consider this effect into the evolution of the binary angular momentum. Due to the symmetry in the azimuthal angle of all the distributions used, the net evolution of the binary orientation will be roughly zero after a significant number of interactions.

\subsection{Analytical model}
The results of our Monte Carlo runs can be used to calibrate a simple analytical model for the evolution of the MBHB. In general, this can be written as
\begin{equation}
\dv{a}{t}=\Gamma \Delta{a},
\end{equation}
where $\Gamma$ is the rate at which clouds are supplied to the binary and $\Delta{a}$ is the average relative change in semimajor axis caused by each cloud. Assuming a uniform density distribution of clouds $n$, travelling at an average speed $v$, one can write
\begin{equation}
  \Gamma=n\Sigma v=n\pi b^2 v=2\pi n \frac{GM}{v} (\chi a) = \Gamma_0\frac{a}{a_0},
  \label{eq:gamma}
\end{equation}
where we defined $\Gamma_0=2\pi n (\chi a_0) GM/v$. Here we used the fact that the geometric cross section is $\Sigma=\pi b^2$. Since $M_c\ll M$, encounters are gravitational focusing dominated, and the impact parameter $b$ is related to the maximum approach $r_{\rm p}$ through $b^2=2GMr_{\rm p}/v^2$. We consider encounters with a maximum approach $r_{\rm p,M}=\chi a$ ($\chi=3$ in our experiment). On the other hand, we showed in Section \ref{sec:angmom} that $\Delta{a}\propto \Delta{M}$, where $\Delta{M}$ is the mass accreted by the MBHB. We can therefore write
\begin{equation}
\Delta{a}=-\eta\delta \frac{M_c}{M}a,
\end{equation}
where $\delta$ represent the average mass fraction of the cloud captured (and eventually accreted) by the binary, and $\eta$ is an average efficiency coefficient. In each individual encounter, both parameters depend on the cloud inclination and impact parameter, but we are concerned here with finding their average values only. As shown in \paperI, for $r_{\rm p}<3a$ we have $\delta\approx 0.3$ (averaged over impact parameters), with little dependence on the cloud--MBHB orientation. Conversely, the angular momentum transfer efficiency strongly depends on the MBHB--cloud orientation, as shown by Equations (\ref{deltaa_ACA}) and (\ref{deltaa_PEPF}). Therefore the exact value of $\eta$ will depend on the level of anisotropy of the cloud distribution.

In the interaction process, the binary also gains mass at the same rate defined by equation (\ref{eq:gamma}), so that the evolution of the system is given by the coupled linear differential equations:
\begin{equation}
  \begin{cases}
  \dv{a}{t}=-\eta\delta M_c \frac{\Gamma_0}{a_0} \frac{a^2}{M}\\
  \dv{M}{t}=\delta M_c \frac{\Gamma_0}{a_0} a
  \end{cases},
  \label{eq:dadtdmdt}
\end{equation}
which combined trivially give the evolution of semimajor axis with mass
\begin{equation}
\dv{a}{M}=-\eta\frac{a}{M},
  \label{eq:dadm}
\end{equation}
which is immediately solved to get
\begin{equation}
  M_{\rm f}=M_0 \exp\left\{-\frac{1}{\eta}{\rm ln}\left(\frac{a_f}{a_0}\right)\right\}.
  \label{eq:Mf}
\end{equation}

The analytical evolution described by the system of equations (\ref{eq:dadtdmdt}) depends on the values of $\eta$ and $\delta$, that can be calibrated to match the results of the Monte Carlo runs shown in Fig.~\ref{sim_events}. By doing so, we obtain $(\delta,\eta)=(0.29,3.45),(0.31,5.6),(0.36,7.4)$ for $F=0.0,\,0.5,\,1.0$ respectively. 
{With these values the analytical model reproduces satisfactorily the evolution of the system observed in the Monte Carlo runs, obtaining residuals of $\lesssim 0.2\%$ for the mass evolution  and $\lesssim 3\%$ for the semimajor axis evolution.}

 We also note that by setting $\dd M/\dd t=0$ (i.e., the gas is never accreted by the MBHB), the system evolution is analogue to that of a MBHB scattering intervening stars from a uniform distribution. This is because the underlying physical description is the same: individual objects drawn from some uniform distribution transfer an amount of energy and angular momentum which is dictated by the binding energy of the binary. A major difference, however, resides in the fact that in the cloud case, the MBHB shrinking is mainly due to absorption of angular momentum from the accreted portion of the cloud, which is a strong function of the incoming cloud direction. Therefore, prograde and retrograde encounters result in different MBHB shrinking rates, and the evolution of the binary semimajor axis depends on the level of anisotropy of the cloud distribution. This is not true in the stellar scattering case, where the MBHB shrinking is mostly due to binding energy taken away by each scattered star which, contrary to the angular momentum exchange, is insensitive to its incoming direction. The net result is that in the stellar case, for different levels of anisotropy of the interacting stellar distribution, the binary semimajor axis evolution is the same but the eccentricity evolution is extremely different, as demonstrated in \cite{2011MNRAS.415L..35S}. 


\subsection{Scaling to astrophysical systems}
\begin{figure}
\centering
\includegraphics[width=0.45\textwidth]{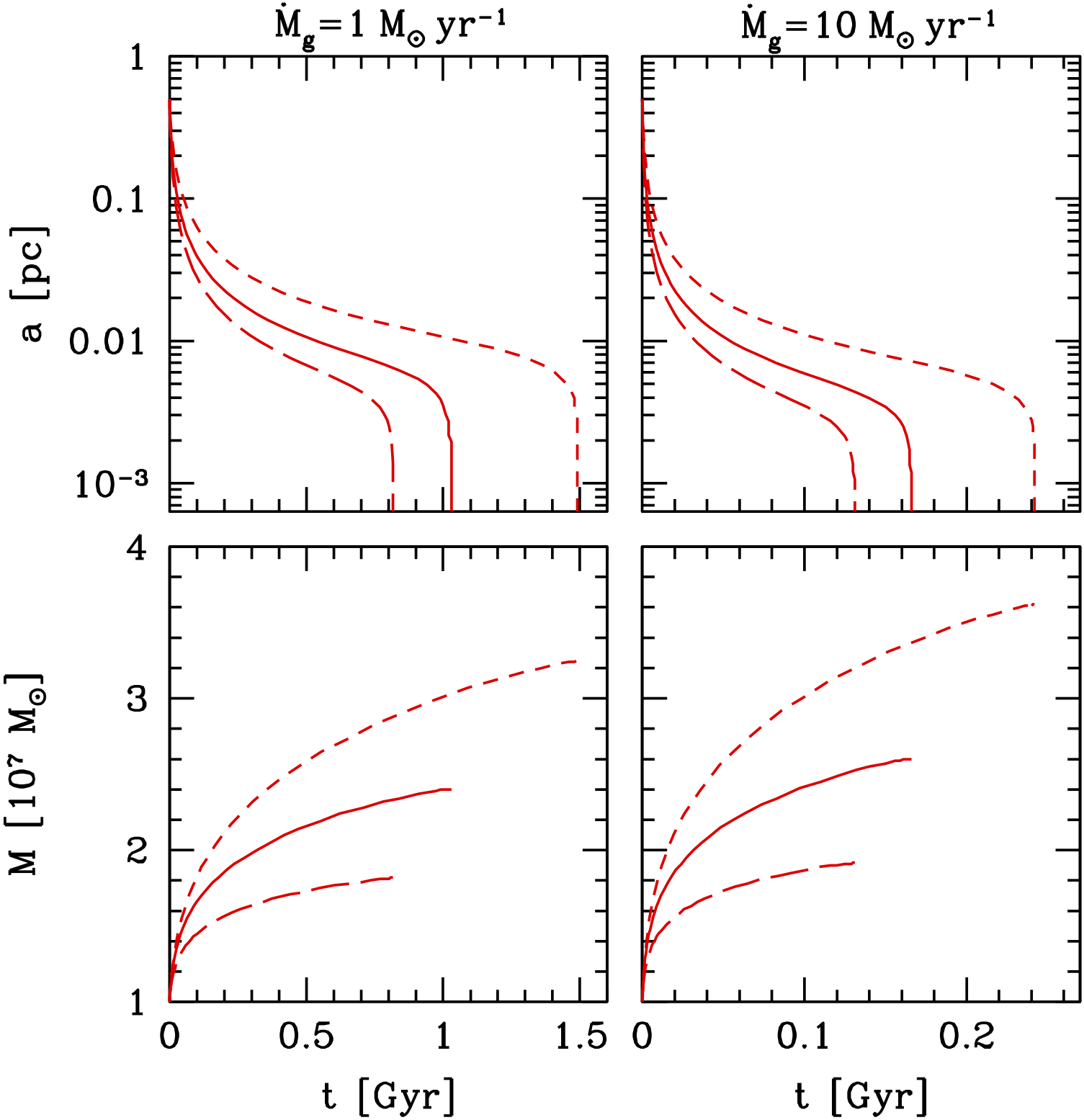}
\caption{Evolution of a binary with $M_1=M_2=5\times10^6\msun$ as a function of time for two different gas inflow rates as indicated at the top of the figure. The top panels show the semimajor axis evolution, while the bottom panels show the MBHB mass evolution. Long--dashed, solid and short--dashed curves are for $F=1.0,\,0.5,\,0.0$ respectively.  The late, fast evolution of the semimajor axes is driven by the emission of gravitational waves.}
\label{fig:evolution}
\end{figure}
To study the final fate of the binary, we generalise the analytical model introduced in the previous subsection (eq.~\ref{eq:dadtdmdt}) including the GW emission term \citep{PetersMathews1963}, to get
\begin{equation}
  \begin{cases}
  \dv{a}{t}=-\eta\delta M_c \frac{\Gamma_0}{a_0} \frac{a^2}{M}-\frac{64}{5}\frac{G^3}{c^5}\frac{M^3}{4a^3}\\
  \dv{M}{t}=\delta M_c \frac{\Gamma_0}{a_0} a
  \end{cases},
  \label{eq:dadtdmdt_gw}
\end{equation}
where we assumed circular binaries and $M_1=M_2=M/2$ throughout the process (i.e. we assume each individual MBH gets the same share of accretion). The interaction rate is connected to the physical properties of the system via $\Gamma_0=2\pi n (\chi a_0) GM/v$. One can either specify the cloud number density $n$ and infall velocity $v$, or assume a certain gas mass inflow rate $\dot{M}_{\rm g}$. As an example, we take two Milky Way like MBHs, $M_1=M_2=M/2=5\times10^6\msun$, at an initial semimajor axis of $0.5$ pc  (roughly corresponding to the hardening radius $a_h\approx GM/(4\sigma^2)$ for a Milky Way galaxy), with a mass inflow rate $\dot{M}_{\rm g}$ of either $1\msun$yr$^{-1}$ or $10\msun$yr$^{-1}$, corresponding to $\Gamma_0=10^{-5}$yr$^{-1}$ or $\Gamma_0=10^{-4}$yr$^{-1}$ (our cloud mass is $M_c=0.01 M=10^5\msun$ for $M=10^7\msun$). Such inflow rates might be typical in the central regions of relatively gas-rich post-merger galaxies. 

Results are shown in Fig.~\ref{fig:evolution}, both the binary shrinking and the mass growth. The MBHB coalesce within $\approx$1 Gyr and $\approx$0.2 Gyr for the two gas inflow rates assumed. The degree of anisotropy of the gas inflow has only a mild impact on the evolution of the system, affecting the coalescence timescale by a factor of $\approx 2$. More relevant is the impact on the mass growth. While in the $F=1.0$ case the MBHB barely doubles its mass, the mass growth is almost an e-fold larger in the $F=0.0$ case. This is because prograde cloud distribution is less efficient ($\eta\approx3.5$) than a retrograde one ($\eta\approx7$) in shrinking the MBHB, and consequently the process requires more accreted mass.

\subsection{Robustness of the model}
\label{sec:limitations}

The models described above are subject to limitations 
{arising from some of our simplifying assumptions. In this subsection, we explore the extent that some of these assumptions have on the results we present.}

\subsubsection{{Discrete orientations}}
We are collapsing the parameter space of the cloud angular momentum direction into four selected configurations (A, CA, PF, PE), with rather arbitrary boundaries. Most of the events given by any of the $F$ values we are using will be neither exactly parallel (aligned or counter-aligned) nor perpendicular (edge or face on) to the MBHB, so our approximation is bound to introduce some error in the estimate of the binary evolution. In particular, it would be important to simulate the infall of clouds with $\boldsymbol{L}_{\rm c}$ partially aligned to $\boldsymbol{L}$, since here lies the separation between events leading to shrinking (perpendicular) and expanding (aligned) MBHBs. {In order to observe the transition between these two regimes we would need to cover that parameter range with a series of simulations, which is unfortunately not feasible with our current computational capabilities.}

{Nevertheless, we can roughly estimate the uncertainty of our long-term model by using different functions to interpolate between the values obtained with the simulations.  As we discuss in Section~\ref{sec:angmomexch}, for the parallel orbits (A and CA models) a fraction of the cloud's initial angular momentum is directly added to the binary. Hence, for an arbitrarily inclined orbit, we expect that the main contribution comes from the projection of the gas velocity on to the binary plane. We propose the following function that captures this behaviour
\begin{equation}
\left(\frac{\Delta a}{a}\right)_{\rm tot}=\left(\frac{\Delta a}{a}\right)_{\parallel}\cos\beta+\left(\frac{\Delta a}{a}\right)_{\perp}\sin\beta,
\label{eq:model1}
\end{equation}
where we define the subscripts as
\begin{equation}
{\parallel}=
  \begin{cases}
  {\rm A} & \mbox{if} \quad0^\circ<\theta<90^\circ\\
  {\rm CA} & \mbox{if} \quad-90^\circ<\theta<0^\circ
  \end{cases},
\end{equation}
\begin{equation}
{\perp}=
  \begin{cases}
  {\rm PE} & \mbox{if} \quad0^\circ<\phi<90^\circ \;\mbox{or}\;-180^\circ<\phi<-90^\circ\\
  {\rm PF} & \mbox{if} \quad90^\circ<\phi<180^\circ \;\mbox{or}\;-90^\circ<\phi<0^\circ
  \end{cases},
\end{equation}
and the angle $\beta$ as
\begin{equation}
\beta=
  \begin{cases}
  \theta & \mbox{if} \quad0^\circ<\theta<90^\circ\\
 \theta + 90^\circ & \mbox{if} \quad-90^\circ<\theta<0^\circ
  \end{cases}.
\end{equation}
}

While conceptually simple, the main disadvantage of equation~\eqref{eq:model1} is that it does not reproduce a scenario where both parallel and perpendicular changes are equal. In consequence, we propose an alternative form that takes into account this fact,
\begin{equation}
\left(\frac{\Delta a}{a}\right)_{\rm tot}=\left(\frac{\Delta a}{a}\right)_{\parallel}\cos^2\beta+\left(\frac{\Delta a}{a}\right)_{\perp}\sin^2\beta,
\label{eq:model2}
\end{equation}
where we use the same definitions described above.

\begin{figure}
\centering
\includegraphics[width=0.45\textwidth]{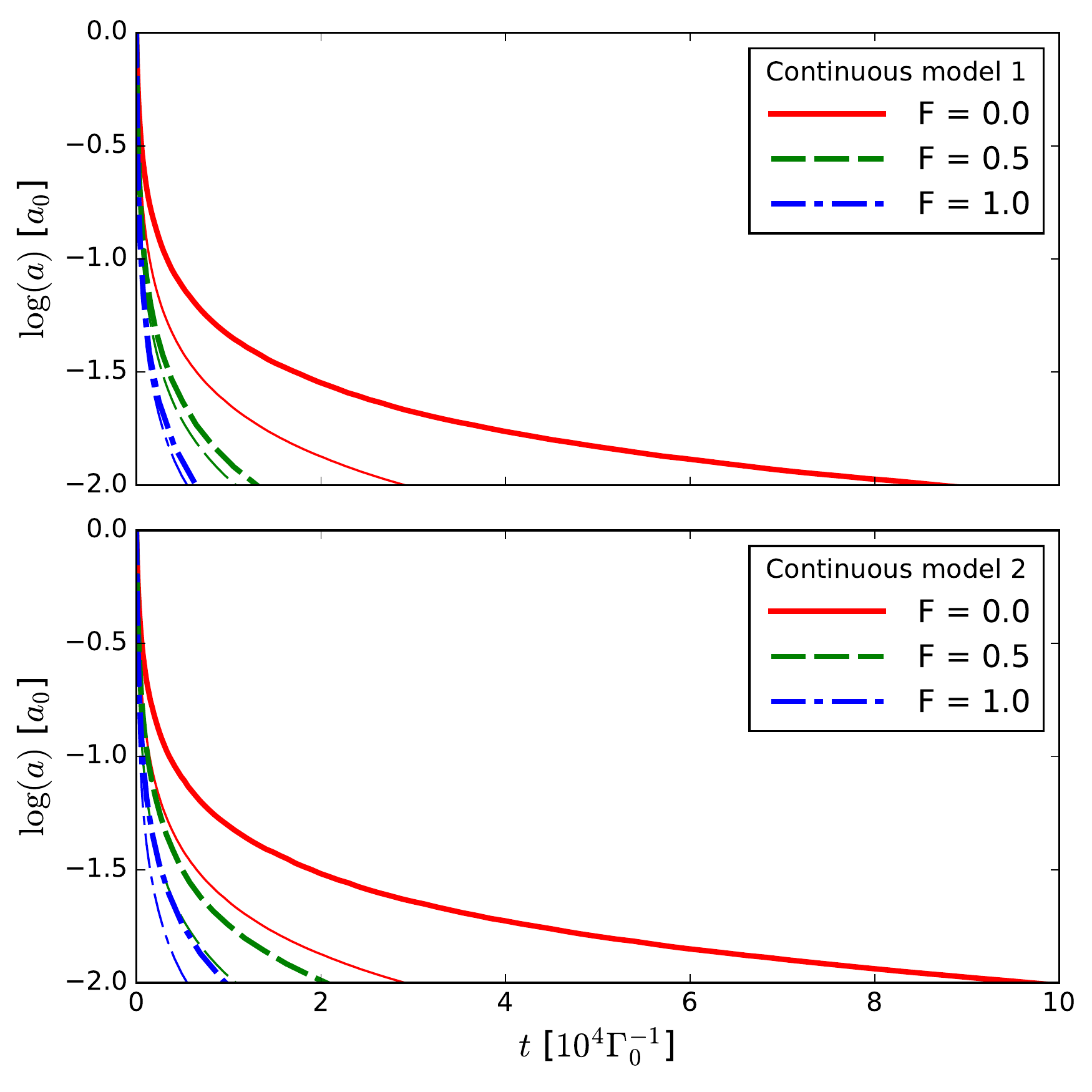}
\caption{Evolution of the binary semimajor axis as a function of time, computed with Monte Carlo runs. The different lines represent models with different distributions of clouds as indicated in the legend. The upper panel shows the result using equation~\eqref{eq:model1}, while the lower panel corresponds to equation~\eqref{eq:model2}. Both new models are depicted with thick lines, while the thin lines represent the original runs using the discrete orientations (Fig.~\ref{sim_events}).}
\label{fig:new_models}
\end{figure}

{We refer to these two equations (Eqs.~\ref{eq:model1} and \ref{eq:model2}) as `Continuous model~1' and `Continuous model~2', 
respectively. Despite the rather arbitrary nature of these interpolations, we expect they, together with the discrete model presented in \S~\ref{sec:montecarlo}, illustrate a reasonable range of possible evolutions for the binary population. We show the results obtained with the two continuous models in Fig.~\ref{fig:new_models}. We find that the timescale to reach the GW regime somewhat increases compared with the discrete scenario (Fig.~\ref{sim_events}). This occurs because the aligned configuration has a maximum impact parameter larger than the others (see Fig.~\ref{delta_a}), which means that an important fraction of the events would only expand the binary.} 
{In fact, the evolution timescale increases by $\approx3$ in the $F=0.0$ scenario, but only by a factor $\lesssim 2$ in all other cases. Although highlighting the uncertainties related to our model, these tests essentially confirm our basic MBHB evolution scenario.}

\subsubsection{Effects of the non-accreted material}
{In our Monte Carlo models (\S~\ref{sec:montecarlo}) we consider} a sequence of cloud infall events, treating them {independently} from each other. {In reality}, a large fraction of each cloud is not accreted by the binary, {and part of it will end up} forming circumbinary structures. As more clouds interact with the binary, those left-over structures can interact with each other, {cancelling out their angular momenta}, or accumulating to give rise to a 
massive circumbinary disc, that can {speed up the binary shrinking, or even reverse its transient expansion (Goicovic et al. in prep.).
  We describe here a modified Monte Carlo model that accounts for the left-over gas. In absence of simulations to calibrate the effect of the cumulative gas distribution, we make a number of assumptions, as detailed below.  }

As shown by \citet{Nix11b}, randomly-oriented events around a circular binary can produce external discs that efficiently either align or counter-align with the binary due to differential precession and dissipation within the disc. Consequently, we expect that some of the non-accreted material will result in coplanar circumbinary discs, either prograde or retrograde with respect to the binary. The condition for counter-alignment is 
\begin{equation}
\theta < 0 \quad {\rm and } \quad L_{\rm d}<2L_{\rm b},
\end{equation}
where $L_{\rm d}$ and $L_{\rm b}$ are the angular momentum of the disc and the binary, respectively \citep{Nix11a,Nix11b,Nix12}. 
{Assuming}
that all of the non-accreted material will form a disc with the original orientation of the cloud, we use eq.~\eqref{eq_Lgas} to estimate its angular momentum. We derive {a} counter-alignment condition of $\alpha\lesssim70$ for events coming from the south hemisphere ($\theta<0$). Given that we are modelling clouds with almost radial orbits and little initial angular momentum ({average} $\alpha\lesssim2$), we assume that every cloud from the south hemisphere will tend to counter-align, while events from the north hemisphere will always tend to align with the binary. This will be particularly relevant for the $F=0.0$ and $F=1.0$ distributions, where the events are somewhat coherent, thus increasing the probability of having a well defined disc. 
{The effect is most striking for the $F=0.0$ distribution, where the torques caused by the accretion of (partially) aligned clouds tend to slightly expand the binary (cf Figure~\ref{delta_a}). This is true as long as the mass of the leftover circumbinary disc is negligible. However, following many such events, a massive circumbinary disc eventually forms, which effectively transports angular momentum outwards, eventually shrinking the binary \citep[e.g.,][]{C09}.}

{For the Monte Carlo calculations of the coherent distributions ($F=0.0$ and $F=1.0$) we assume that all the non-accreted material ($\approx0.7M_c$ per event) becomes part of the disc after the transient phase. Having established a recipe for circumbinary disc growth, we now turn to estimate its effect on the binary orbital evolution. The latter has been studied analytically by  \citet{Syer1995} and \citet{Ivanov1999}, among others, as a function of the disc properties.  Here we use the results of \citet{C09}, who applied the formalism of \citet{Ivanov1999} to the case of a disc in which the transport of angular momentum is due to its own self-gravity \citep[e.g.,][]{Rice2005}, as appropriate for the large disc mass that quickly builds up.   
\citet[their eq.\ 5]{C09} find an analytical description of the semimajor axis evolution of the form 
\begin{equation}
\dot a_{\rm circumb}\sim-10^{-5}a_0\Omega_0\left(\frac{M}{M_0}\right)^{-2}\left(\frac{M_{\rm disc}}{0.2M_0}\right)^{5/2}\left(\frac{a}{a_0}\right)^{-2},
\label{eq:circumb}
\end{equation}
where $\Omega_0$ represents the binary initial orbital frequency and $M_{\rm disc}$ the disc mass.  This result is confirmed by their numerical simulations.}

{ For the modeling of the isotropic distribution ($F=0.5$), we take a different approach}. In this case, clouds are highly misaligned with respect to the binary \textit{and} between one another, hence most of the material will not experience enough precession to significantly change its inclination before the infall of a new cloud. For instance, in \paperI~we derive an alignment timescale of the order of 1000 orbits for a completely perpendicular disc. We therefore expect this gas to behave as a collection of independent precessing rings, similar to what occurs when a misaligned circumbinary disc breaks under the gravitational pull of the binary \citep{2013MNRAS.434.1946N, Aly2015}. In this scenario, the interaction between rings causes partial cancellation of angular momentum, which results in a direct plunge of a fraction of the gas into the binary, thus increasing the accretion rate. Assuming this gas delivers negligible net angular momentum to the binary, we expect a semimajor axis evolution similar to eq.~\eqref{deltaL2P}, namely,
\begin{equation}
\dot a_{\rm cancelled}=-3\dot M_{\rm cancelled}\frac{a}{M},
\label{eq:cancelled}
\end{equation}
where $\dot M_{\rm cancelled}$ is the accretion rate of the plunging material on to the binary. The interaction of these misaligned discs is a highly non-linear process, {so numerical modelling is needed to estimate}
how much material reaches the MBHs. 
{Since to our knowledge no published simulations of this multi-cloud process exists yet, we use as a proxy}
 the fiducial value of $\dot M\sim10^{-7}M_0\Omega_0$ from
 {the disc tearing models of}
  \citet[][their Fig.~5]{2013MNRAS.434.1946N}, 
{rescaling it to} 
 be proportional to the available gas surrounding the binary ($M_{\rm av,gas}$) as follows:
\begin{equation}
\dot M_{\rm cancelled} \sim 10^{-7}M_0\Omega_0 \left(\frac{M_{\rm av,gas}}{10^{-2}M_0}\right).
\label{eq:mcancelled}
\end{equation}
Since on average there is $70\%$ of the cloud's mass left after each event, $M_{\rm av,gas}$ increases with time.

\begin{figure}
\centering
\includegraphics[width=0.45\textwidth]{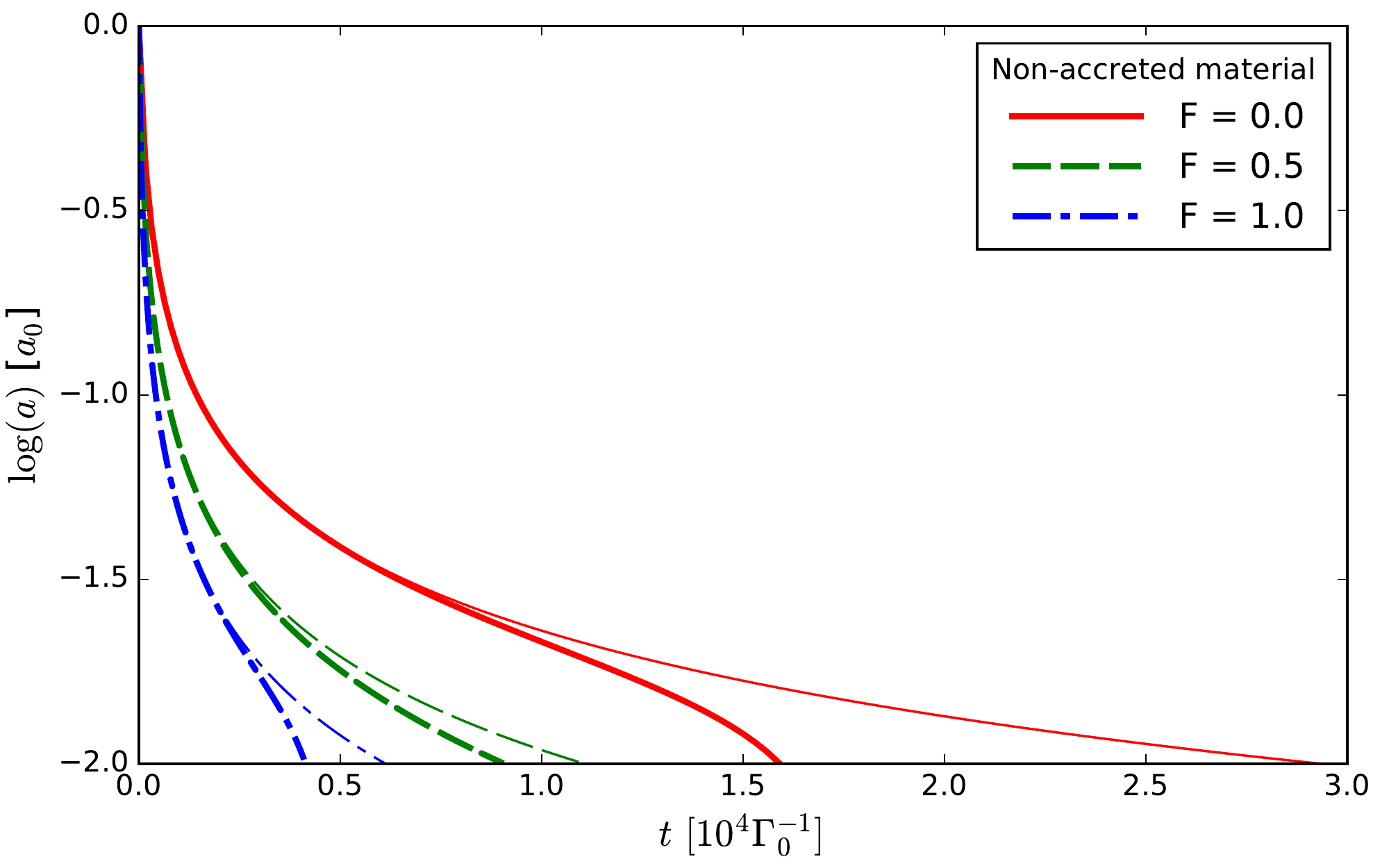}
\caption{{Evolution of the binary semimajor axis as a function of time, computed with Monte Carlo runs. The thick lines represent the model including the effects of the non-accreted material, while the thin lines show the original runs where the clouds were treated on isolation (Fig.~\ref{sim_events}). For the $F=0.0$ and $F=1.0$ distributions we model the presence of a coplanar massive circumbinary disc as an extra source of angular momentum transport, while for $F=0.5$ we are enhancing the accretion onto the binary due to cancellation of angular momentum in the gas.}}
\label{fig:non_accreted}
\end{figure}

{Putting all the pieces together, the Monte Carlo evolution of the binary described in \S~\ref{sec:montecarlo} is modified by including} an additional term $\Delta a=\dot a\Delta T$, where $\Delta T$ is given by equation~\eqref{Tenc} and $\dot a$ {(always negative)} by equation~\eqref{eq:circumb} or~\eqref{eq:cancelled} depending on the {cloud} distribution. 
{Notice the same expression is used for the  $F=0.0$ and $F=1.0$ distributions,}
since \citet{Roedig2014} showed that, despite substantial differences in the underlying physical processes, the semimajor evolution is {very similar} when the binary is surrounded by a prograde or by a retrograde disc.

The evolution of the binary obtained with this model is shown in Fig.~\ref{fig:non_accreted}.
{As expected, the MBHB shrinks faster compared {to} the original {Monte Carlo} runs due to the action of the surrounding material.
The largest difference is {obtained} for the $F=0.0$ distribution, with {an evolution almost twice as fast as in the standard model.}}

{It is worth stressing the two main assumptions made in these modified models.  First, all of the non-accreted material ($\approx 70\%$ of each cloud) is assumed to form a circumbinary structure, while this generally will not be the case.  Indeed, in \paperI~(see Table~1 there) we computed the expected mass of the forming circumbinary disc and found it to be only a small fraction of the non-accreted gas,  up to $\approx20\%$ for the largest impact parameters. We notice, however, that the interaction between gas from  successive infall events could modify this fraction. The evolution of the binary semimajor axis is proportional to either $M_{\rm disc}^{5/2}$ or $M_{\rm disc}$ in our models (cf. equations \ref{eq:circumb},~\ref{eq:cancelled}~and~\ref{eq:mcancelled}). Therefore, by including all the non accreted material in the calculation, the models provide a robust upper limit of the impact of the leftover gas in the evolution of the MBHB. Second, for the  $F=0.0$ and $F=1.0$ distributions, we assume a stable disc that effectively transport angular momentum as it grows.  Given that the MBHs more than double their own mass during the evolution (see Fig.~\ref{sim_events}), and that by construction more mass goes into forming the circumbinary disc than into accretion, the disk will eventually become more massive than the binary. Before reaching that point, it is expected that the disk will fragment \citep[e.g.,][]{Rice2005}, strongly diminishing its impact on the orbital evolution with respect to what we find in our simple model \citep{Pau2013}.}   

{
  In conclusion, the effect of any circumbinary structure forming as a consequence of the continuous infall of clouds can speed-up the evolution of the binary toward coalescence by at most a factor of two. This demonstrates that the prompt gas capture phase is {\it at least} as effective in extracting energy and angular momentum from the system than the slowly forming circumbinary structure. For the cloud infall rates explored in this study, the prompt interactions alone are enough to bring the binary orbit down from parsec separations to the gravitational wave regime on a time-scale $\simlt 1\,$Gyr. This is independent of other angular momentum transfer mechanisms that may be simultaneously at work, namely, three-body scattering or persistent torques from a larger-scale circumbinary structure.}

\section{Summary}
\label{sec:summary}
We studied the orbital evolution of an equal mass, circular MBHB interacting with an impacting gas cloud. Exploiting a suite of 13 high resolution hydrodynamical simulations, described in \paperI, we investigated the response of the MBHB orbit as a function of the cloud impact parameter and relative inclination with respect to its orbital plane. In our simulations, the cloud mass is only 1\% of the MBHB mass, so the binary orbit is expected to suffer only small changes. In order to establish if we have the accuracy to reliably measure the effect, we compared the evolution of the MBHB angular momentum to the level of angular momentum conservation during the simulations. We found that the former is appreciably larger than the numerical noise during the first few binary orbital periods, which allowed us to robustly measure the transient evolution of the orbital parameters in this phase. We therefore presented only results for the strong initial binary--cloud interaction, discarding the subsequent (much slower) secular evolution of the binary. Our main findings can be summarised as follows:
\begin{enumerate}
\item We focused our analysis in the semimajor axis evolution. We found that its total change during the strong transient interaction depends on the orbital configuration of the system, and it is closely related to the fraction of mass that gets accreted. In particular, the binary shrinks the most when interacting with counterrotating clouds which carry negative angular momentum (with respect to the MBHB orbital angular momentum) that cancels out upon accretion. Conversely, prompt accretion of corotating gas, causes the MBHB to expand, in contrast to the long-term evolution seen in persistent massive circumbinary discs.
\item Using a simple analytical model, we were able to show that, for all configurations, the evolution of the binary orbital elements is dominated by the transfer of angular momentum from the cloud to the binary through the accretion of gas during the first stages of the interaction. Considering only the angular momentum budget of the accreted material is sufficient to satisfactory reproduce all the trends for the total evolution of the MBHB angular momentum vector (either magnitude or inclination) and semimajor axis.
\item By further including the effects of gravitational slingshot from the MBHB onto the non-accreted material, and by using a more accurate estimate of the accreted angular momentum, we could reproduce the evolution of the MBHB semimajor axis and angular momentum within a few percent. This confirms that the simulations are accurate enough to capture the correct evolution of the MBHB, and that the underlying physics is understood.
\item Since the MBHBs were initially circular, any asymmetric torque (such as that exerted by an infalling cloud) should excite some eccentricity. This is in fact observed in all our simulations, however, the eccentricity growth is too small to draw any further conclusions.  We will explore this effect in forthcoming work where we plan to extend our simulation set to include initially eccentric binaries.
 \end {enumerate}
 
The resulting semimajor axis evolution as a function of the orbital configuration of the system was then used as the basis to construct a simple Monte Carlo model for evolving a MBHB interacting with a sequence of impacting clouds. We take cloud distributions from reasonable populations with different levels of anisotropy in their angular momenta, based on the studies of \citet{Dotti2013} and \citet{Sesana2014}. These distributions can be linked to the morphological and kinematical properties of the host galaxies; disc galaxies, where the gas dynamics is dominated by rotational velocity, will produce mainly coherent accretion events, while bulge galaxies, dominated by velocity dispersion, will tend to produce more uniformly distributed events. We found that the  evolution of the orbit is fastest when the distribution of clouds corresponds to mostly retrograde events ($F=1.0$), while as we transition to isotropic ($F=0.5$) and then to mainly prograde ($F=0.0$) events, the evolution progressively slows down. This is because retrograde interactions are more efficient subtracting angular momentum from the MBHB, while prograde encounters tend to slightly expand its orbit.

Finally, we used the results from the Monte Carlo realisations to calibrate a simple analytical system of coupled differential equations that captures the long term evolution of the MBHB as a function of two efficiency parameters. By scaling this analytical description to astrophysical systems, we found that typical MBHBs efficiently evolve down to the GW emission regime within a few hundred million years. This demonstrates that the interaction with individual clouds in near-radial infall is capable of efficiently shrinking the binary orbit, providing a viable solution to the final parsec problem in clumpy gas-rich environments.

We recall that our treatment is subject to a number of caveats and limitations. In particular: i) we are discretising the angular momentum space into just four configurations, which does not take into account intermediate inclinations,  and ii) we treated each MBHB--cloud interaction in isolation, neglecting the non-accreted material that will eventually accumulate, forming circumbinary structures and/or increasing accretion onto the binary. 
We test the robustness of our model by including these two effects into the calculations. We find that the timescale to reach the GW regime changes by no more than a factor of 3, which does not change significantly our conclusions. {For instance, we show that even when all the remaining material is assumed to form a massive circumbinary disc and fragmentation is neglected, its impact on the binary evolution is at most comparable to the cumulative effect of the transient accretion episodes of the individual infalling clouds.}
Additionally, the accretion rates that we measure during the transient evolution are typically super-Eddington (up to $\approx 50\dot M_{\rm edd}$, \paperI), thus suggesting that radiation pressure-driven outflows can alter the amount of material that actually reaches the MBHs. We stress however that even if all the material that crosses the accretion radius is eventually ejected from the system, the semimajor axis evolution will be mildly affected. This is because the exchange of angular momentum occurs during the capture of the material rather than the accretion itself.

 Besides these conceptual points, our model is also limited to circular, equal-mass binaries. While we can infer plausible trends of the results with the binary mass ratio, it would be interesting to see how eccentric binaries respond to clouds with different orientations and impact parameters. However, these simulations are computationally expensive, so we leave the exploration of the relevant parameter space for future work.  On the other hand, we are planning a series of simulations of MBHBs interacting with multiple clouds. This will allow us to combine the `discrete' evolution due to the prompt interaction of each individual cloud, with the secular torques eventually exerted by the circumbinary structures that we expect to form in the long term.

\section*{Acknowledgments}
FGG thanks Camilo Fontecilla and Francisco Aros for very inspiring discussions, as well as the warm hospitality of the University of Birmingham and the Albert Einstein Institute (AEI) during the development of this work.
The simulations were performed on the {\it datura} cluster at the AEI.
We acknowledge support from CONICYT-Chile through FONDECYT (1141175), Basal (PFB0609), Anillo (ACT1101), Redes (120021) and Exchange (PCCI130064) grants; and from DAAD (57055277). 
FGG is supported by CONICYT PCHA/Doctorado Nacional scholarship.  AS is supported by a University Research Fellowship of the Royal Society.
This work was completed while JC was on sabbatical leave at MPE.  JC and FGG acknowledge the kind hospitality of MPE, and funding from the Max Planck Society through a ``Partner Group'' grant.

\bibliographystyle{mnras}
\bibliography{refs}

\appendix

\section{Convergence tests with the accretion radius}
\label{sec:Ap1}

{As we discuss in Section \ref{sec:angmomexch}, the transient evolution of the binary orbital elements is dominated by the exchange of angular momentum with the accreted material. Hence, resolving the accretion process properly is a key aspect of this study. Due to numerical considerations explained in \paperI, we model each MBH with a very large sink radius compared to the Schwarzschild radius, producing unrealistically large accretion rates. }

\begin{figure}
\centering
\includegraphics[width=0.45\textwidth]{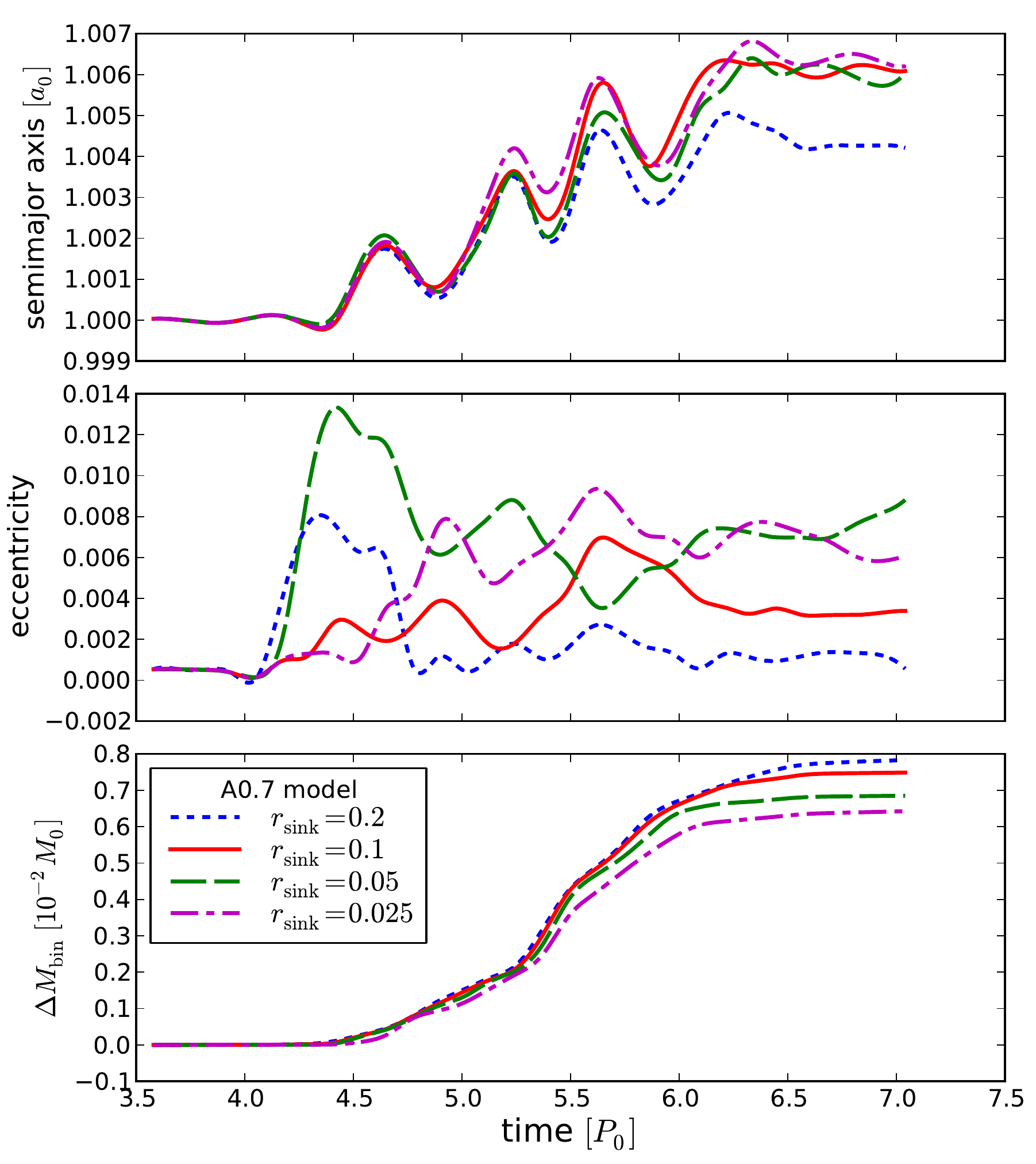}
\caption{{Evolution of the binary semimajor axis (upper panel), eccentricity (middle panel) and accreted mass (lower panels) for the CA0.7 model with different sink radii, as indicated in the legend. Recall that the solid red line represents the ``original'' system.}}
\label{fig:convergence1}
\end{figure}
\begin{figure}
\centering
\includegraphics[width=0.45\textwidth]{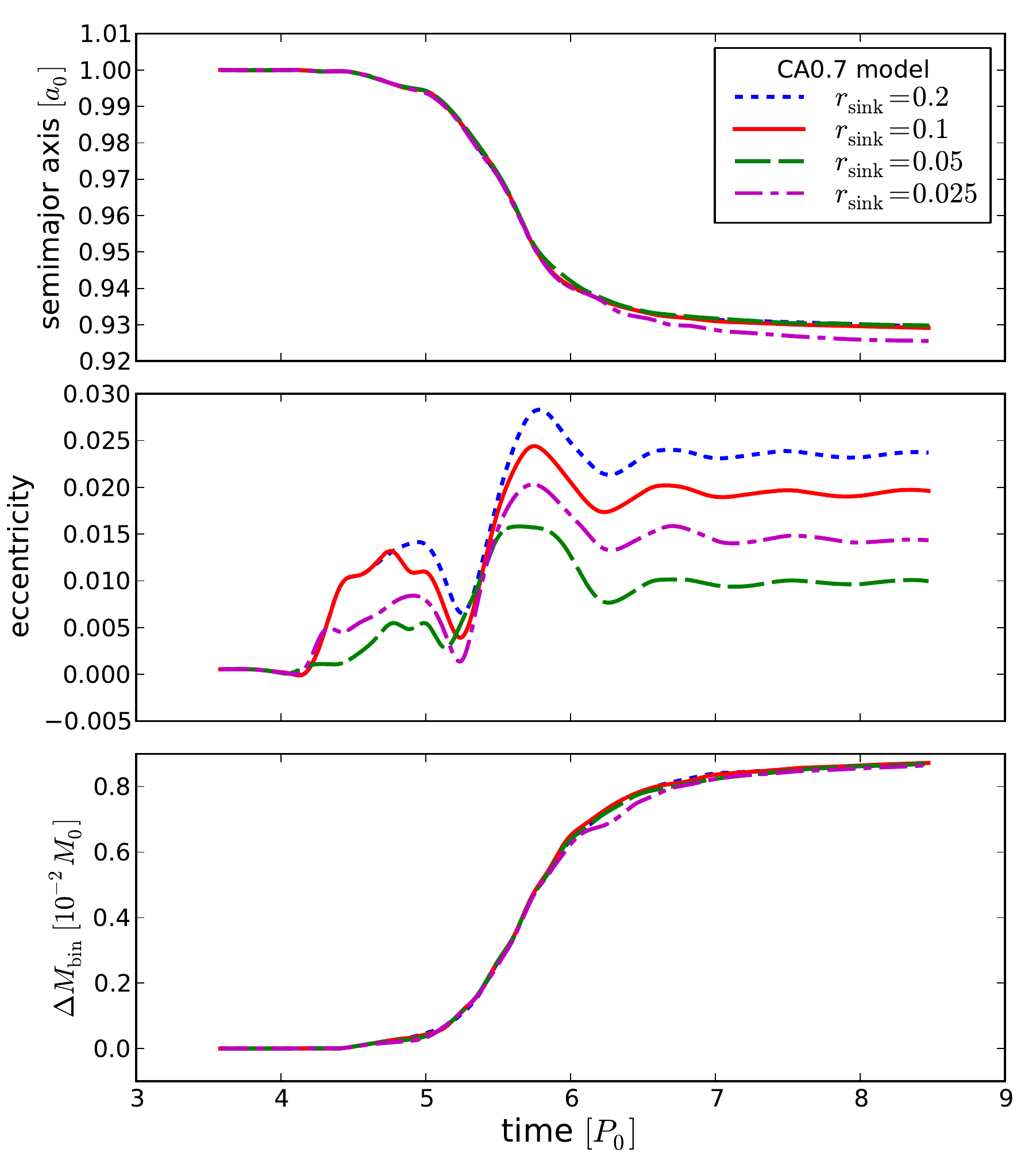}
\caption{{Same as Fig.~\ref{fig:convergence1}, but for the CA0.7 model.}}
\label{fig:convergence2}
\end{figure}

In order to establish the role of our chosen sink radius, we rerun some of the models changing only this parameter. We test two smaller values of the sink radius (0.05 and 0.025) and a larger one (0.2) 
to study the convergence of the binary orbital evolution.
We choose the A0.7 and CA0.7 models because they represent two extremes in regards to accretion: in the first case we observe the formation of very prominent mini-discs around each MBH, while in the latter the material plunges almost directly due to cancelation of angular momentum in the gas (\paperI). Additionally, we have chosen the smallest pericentre distance since the accretion is larger and thus any difference could be enhanced.
 
The binary evolution during the first orbits is shown in Figs.~\ref{fig:convergence1}~and~\ref{fig:convergence2}. Since the simulations with smaller accretion radius are more computationally expensive, we run these models only until the prompt accretion stops. Nevertheless, this is the point where the transient binary evolution also stops, as we show in Section~\ref{sec:angmom}.

For the A0.7 models (Fig.~\ref{fig:convergence1}) we observe that the semimajor axis evolution is roughly equal for radii smaller than 0.2, even though the accreted mass changes with each value of $r_{\rm p}$, as expected. This is because the bulk of angular momentum exchange occurs when the material is `captured' by the MBHs, rather than the accretion itself. For this system orientation, decreasing the accretion radius translates in the inner edge of the mini-discs being closer to the MBHs, delaying the accretion somewhat, but not changing the dynamics any further.

For the CA0.7 models (Fig.~\ref{fig:convergence2}), the difference in $\Delta a$ between the 0.2 and 0.025 simulations is $\sim 4\times 10^{-3}$, which is $\sim 6\%$ of the total change. Interestingly, the accretion seems to converge at the end, but for the smaller $r_{\rm sink}$ there is a ``transient suppression'' around $t=6P_0$, which makes the semi-major axis decrease even more respect to the other cases. This is because we are allowing some of the gas to wonder around the binary a while longer before being captured, interacting gravitationally with the MBHs.

In conclusion, the semimajor axis evolution of the binary its only mildly dependent on our choice of the sink radius. This indicates that the exchange of angular momentum between the gas and the MBHs occurs through the \textit{captured} material rather than the \textit{accreted} material, since the gas that crosses the sink radius has already lost most of its initial angular momentum during the interaction with the binary.

On the other hand, the eccentricity does not appear to converge in any of the models. This is consistent with it having always very small values, which are dominated by numerical noise instead of physical processes. Since we are not considering the eccentricity evolution in this work, this does not affect the conclusions of our study.

\bsp

\label{lastpage}

\end{document}